\documentclass[aps,footinbib, notitlepage,superscriptaddress, longbibliography,eqsecnum]{revtex4-1}
\usepackage{graphicx}
 \usepackage{amssymb}
 \usepackage{amsmath}
 \usepackage{amsfonts}
\usepackage{overpic}
\usepackage{enumerate}
\usepackage{color}
\usepackage[dvipsnames]{xcolor}
\usepackage{xspace}
\usepackage[normalsize]{subfigure}
\usepackage{hyperref}
\usepackage{bm}
\usepackage{empheq}
\usepackage[capitalize]{cleveref}

\crefname{section}{Sec.}{Secs.}

\DeclareFontFamily{OT1}{pzc}{}
\DeclareFontShape{OT1}{pzc}{m}{it}{<-> s * [1.10] pzcmi7t}{}
\DeclareMathAlphabet{\mathpzc}{OT1}{pzc}{m}{it}

\providecommand{\st}[1]{_{\text{#1}}}
\providecommand{\sfrac}[2]{#1/#2}

\providecommand{\pfrac}[2]{\left(\frac{#1}{#2}\right)}
\providecommand{\ut}[1]{^{\text{#1}}}

\def\onehalf{\frac{1}{2}}

\def\bra{\ensuremath{\langle}}
\def\ket{\ensuremath{\rangle}}

\def\const{\mathrm{const}}

\def\pd{\partial}

\def\im{\mathrm{i}}
\def\Av{\bv{A}}

\def\qv{\bv{q}}
\def\uv{\bv{u}}

\def\pv{\bv{p}}
\def\rv{\bv{r}}
\def\Rvp{\bv{R}_\parallel}
\def\rvp{\bv{r}_\parallel}

\def\Rv{\bv{R}}

\def\b0{\bv{0}}

\def\Acal{\mathcal{A}}
\def\Fcal{\mathcal{F}}

\def\Hcal{\mathcal{H}}
\def\Hc2{\Hcal^{(2)}}

\def\Jcal{\mathcal{J}}
\def\Ccal{\mathcal{C}}
\def\Dcal{\mathcal{D}}
\def\Kcal{\mathcal{K}}

\def\Mcal{\mathcal{M}}
\def\Ncal{\mathcal{N}}

\def\Ocal{\mathcal{O}}
\def\Pcal{\mathcal{P}}
\def\Qcal{\mathcal{Q}}

\def\Ucal{\mathcal{U}}
\def\Vcal{\mathcal{V}}
\def\Zcal{\mathcal{Z}}

\def\hyp13{{_1 F_3}}

\def\bcs{BCs\xspace}
\def\pbc{\ut{(p)}}
\def\Dbc{\ut{(D)}}
\def\Nbc{\ut{(N)}}
\def\NbcNZM{\ut{(N$^*$)}}
\def\izero{^{(0)}}
\def\ione{^{(1)}}

\def\chit{\tilde\chi}

\def\hscal{\mathpzc{h}}

\def\d{\mathrm{d}}

\def\ad{\ut{ad}}
\def\pdphi{{\pd\phi}}

\def\sump{\sideset{}{'}\sum}

\newcommand{\bitem}{\begin{itemize}}
\newcommand{\eitem}{\end{itemize}}
\newcommand{\benum}{\begin{enumerate}}
\newcommand{\eenum}{\end{enumerate}}
\newcommand{\btab}[1]{\begin{tabular}{#1}}
\newcommand{\etab}{\end{tabular}}
\newcommand{\beq}{\begin{equation}}
\newcommand{\eeq}{\end{equation}}
\newcommand{\beqn}{\begin{equation*}}
\newcommand{\eeqn}{\end{equation*}}

\newcommand{\bv}[1]{\mathbf{#1}}

\graphicspath{{Figs/}{figs/}}

\begin{document}
\title{Dynamics and steady states of a tracer particle in a confined critical fluid}
\author{Markus Gross}
\email{gross@is.mpg.de}
\affiliation{Max-Planck-Institut f\"{u}r Intelligente Systeme, Heisenbergstra{\ss}e 3, 70569 Stuttgart, Germany}
\affiliation{IV.\ Institut f\"{u}r Theoretische Physik, Universit\"{a}t Stuttgart, Pfaffenwaldring 57, 70569 Stuttgart, Germany}
\date{\today}

\begin{abstract}
The dynamics and the steady states of a point-like tracer particle immersed in a confined critical fluid are studied. The fluid is modeled field-theoretically in terms of an order parameter (concentration or density field) obeying dissipative or conservative equilibrium dynamics and (non-)symmetry-breaking boundary conditions.
The tracer, which represents, e.g., a colloidal particle, interacts with the fluid by locally modifying its chemical potential or its correlations.
The coupling between tracer and fluid gives rise to a nonlinear and non-Markovian tracer dynamics, which is investigated here analytically and via numerical simulations for a one-dimensional system.
From the coupled Langevin equations for the tracer-fluid system we derive an effective Fokker-Planck equation for the tracer by means of adiabatic elimination as well as perturbation theory within a weak-coupling approximation.
The effective tracer dynamics is found to be governed by a fluctuation-induced (Casimir) potential, a spatially dependent mobility, and a spatially dependent (multiplicative) noise, the characteristics of which depend on the interaction and the boundary conditions.
The steady-state distribution of the tracer is typically inhomogeneous.
Notably, when detailed balance is broken, the driving of the temporally correlated noise can induce an effective attraction of the tracer towards a boundary.
\end{abstract}

\maketitle

\section{Introduction}

Fluids near critical points are characterized by fluctuations with large correlation length and long relaxation time, which give rise to a plethora of intriguing phenomena.
A particular example is the critical Casimir force (CCF), which acts on objects immersed in a near-critical medium \cite{fisher_wall_1978,krech_casimir_1994,brankov_theory_2000,kardar_friction_1999,gambassi_casimir_2009}.
CCFs have been utilized to externally control the behavior of colloidal particles in critical solvents by, e.g., altering the solvent temperature or tuning the surface properties of the particles (see Ref.\ \cite{maciolek_collective_2018} for a review).
Except in $d=2$ spatial dimensions, where exact methods exist \cite{machta_critical_2012,nowakowski_critical_2016,vasilyev_critical_2013-1,squarcini_critical_2020}, the interaction between two colloidal particles or between a colloid and a solid wall is typically analyzed separately in the near-distance (Derjaguin) and the far-distance limit. 
In the former, an approximate parallel plate geometry is realized, allowing one to invoke known results for the CCF in a thin film \cite{krech_casimir_1994,hanke_critical_1998,schlesener_critical_2003,gambassi_critical_2009}. 
By contrast, in the far-distance limit, where the ratio between the particle radius and the distance is small, the CCF can be obtained from a ``small-sphere expansion'' of the Boltzmann weight \cite{burkhardt_casimir_1995,eisenriegler_casimir_1995,hanke_critical_1998,hasenbusch_thermodynamic_2013}. 
Recently, the non-equilibrium dynamics of colloidal particles in critical media has received increased attention \cite{gambassi_relaxation_2008,furukawa_nonequilibrium_2013}, examples including studies of drag forces \cite{demery_drag_2010,demery_thermal_2011,furukawa_nonequilibrium_2013,okamoto_drag_2013,tani_drag_2018,yabunaka_drag_2020}, aggregation \cite{furukawa_nonequilibrium_2013,magazzu_controlling_2019}, diffusion \cite{demery_perturbative_2011,demery_diffusion_2013,dean_diffusion_2011,reister_lateral_2005,reister-gottfried_diffusing_2010,camley_contributions_2012,torres-carbajal_brownian_2015}, shear flow \cite{fujitani_effective_2014,rohwer_correlations_2019}, solvent coarsening \cite{roy_solvent_2018,roy_phase_2018,gomez-solano_transient_2020}, and interplay between criticality and activity \cite{zakine_field-embedded_2018,zakine_spatial_2020}.

Here, we study the behavior of a point-like tracer particle in a confined critical fluid within a dynamical field theory (see \cref{fig_sketch}). The fluid medium is modeled in terms of a scalar order-parameter (OP) field $\phi(\rv)$ governed by non-conserved or conserved equilibrium dynamics within the Gaussian approximation \cite{hohenberg_theory_1977,tauber_critical_2014}. 
In a single-component fluid, the OP represents the deviation of the local fluid density from its critical value ($\phi\propto n-n_c$), whereas, in a binary mixture, $\phi$ correspondingly represents the concentration deviation of a certain species \cite{onuki_phase_2002}.
We distinguish between a \emph{reactive} and a \emph{passive} tracer: a reactive tracer interacts with the fluid in a way that preserves detailed balance and that thus renders a steady state in accordance with equilibrium statistical mechanics. 
A passive tracer, by contrast, is affected by the fluid but does not act back on it and hence represents a non-equilibrium system. In the present case, the passive tracer is driven by a temporally correlated noise and thus can, in fact, be regarded as a special type of ``active'' particle \cite{szamel_self-propelled_2014,farage_effective_2015,maggi_multidimensional_2015,fodor_how_2016}.
Following Refs.\ \cite{dean_diffusion_2011,demery_perturbative_2011,demery_diffusion_2013}, fluid and tracer are coupled such that, in the reactive case, one obtains either a locally enhanced mean OP or a locally suppressed OP variance (and reduced correlation length) at the tracer location. This resembles the typical behavior of a colloid in a critical fluid \cite{fisher_wall_1978,diehl_field-theoretical_1986,schlesener_critical_2003} or of a magnetic point defect in the Ising model \cite{hanke_critical_2000,wu_critical_2015}.
A reactive tracer can thus be regarded as a simplified model for a colloidal particle.
We analyze theoretically and via simulations the resulting dynamics and long-time steady states for various symmetry- and non-symmetry-breaking boundary conditions (\bcs) imposed on the OP by the confinement.
Specifically, we focus on a one-dimensional system, i.e., an interval of length $L$.
However, we also calculate equilibrium distributions in three dimensions, which turn out to be qualitatively similar to the one-dimensional case.
By comparing with the literature on CCFs, we show that the equilibrium distributions obtained here encode the static behavior of a colloid in a half-space geometry in the far-distance limit \cite{burkhardt_casimir_1995,eisenriegler_casimir_1995,hanke_critical_1998}.
The point-like representation of the tracer thus retains the essential character of the interaction with the medium, while it facilitates analytical and numerical approaches.

The present study is structured as follows: In \cref{sec_model,sec_prelim}, we introduce the model and provide the necessary preliminaries. In \cref{sec_passive,sec_reactive}, the statics and dynamics of a passive and a reactive tracer, respectively, are analyzed.
Results of numerical simulations are presented in \cref{sec_sim}, together with a physical interpretation of the observed phenomena and a discussion in the context of existing literature. \Cref{sec_sum} provides a summary of our study.
Technical details are collected in \crefrange{app_dimensions}{app_steadyst_3d}.

\section{Model}
\label{sec_model}

\begin{figure}[t]\centering
	\includegraphics[width=0.33\linewidth]{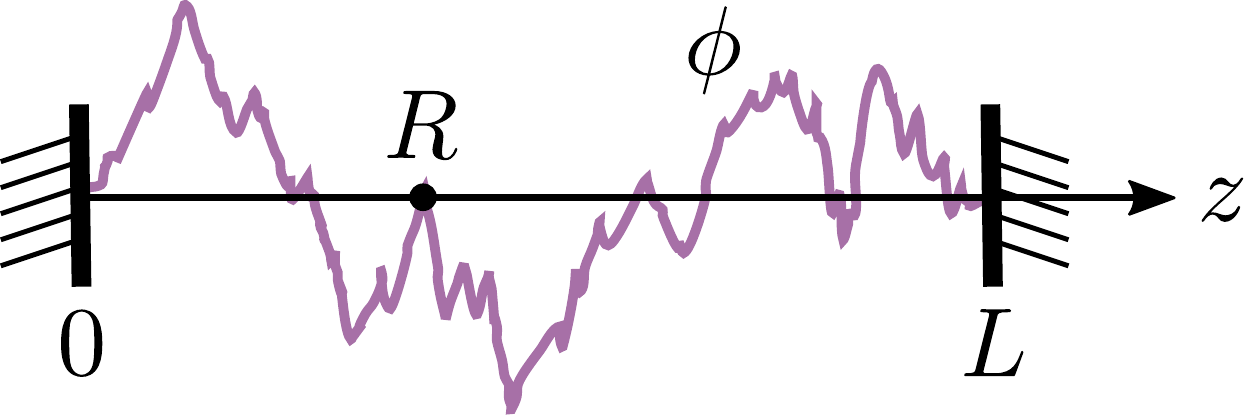} 
    \caption{Situation considered in the present study: a point-like tracer at location $R(t)$ (black dot) is immersed in a fluctuating medium described by a scalar field $\phi(z,t)$ (order parameter). The system is confined by boundaries at $z=0,L$, which exert reflective \bcs on the tracer and either periodic, Neumann, Dirichlet, or capillary \bcs on the field $\phi$ [see \cref{eq_bcs}]. } 
    \label{fig_sketch}
\end{figure}

Before specializing to the one-dimensional case, we introduce the model for arbitrary spatial dimension $d$.
We consider a system consisting of a tracer at position $\Rv(t)$ and a fluctuating OP field $\phi(\rv,t)$. The system is finite in at least one direction, which we take to be the $z$-direction, having length $L$.
In thermal equilibrium, the system is described by the joint steady-state probability distribution 
\beq P_s(\Rv,[\phi]) = \frac{1}{\Zcal_0} \exp\left(-\frac{1}{T}\Hcal(\Rv,[\phi]) \right)
\label{eq_Pss_joint}\eeq
with the Hamiltonian 
\beq\begin{split} 
\Hcal(\Rv,[\phi]) \equiv \Hcal_\phi[\phi] + \Hcal_R(\phi(\Rv))&,\\
\Hcal_\phi[\phi] \equiv  \int_V \d^d r\, \left\{ \onehalf [\nabla\phi(\rv)]^2 - h_1 \phi(\rv)[ \delta(z) + \delta(L-z)] \right\},\qquad 
&\Hcal_R(\phi(\Rv)) \equiv - h \phi(\Rv) + \onehalf c\phi(\Rv)^2,
\end{split}\label{eq_Hamilt}\eeq 
where $\Zcal_0 = \int_V\d^{d} R \int \Dcal\phi\, \exp(-\Hcal(\Rv,[\phi])/T)$ is a normalization factor and $V$ is the system volume.
The OP fulfills one of the following \bcs:
\begin{subequations}
\begin{align}
 \text{periodic:} \qquad & \phi(\{\rvp,z\})=\phi(\{\rvp,z+L\}), \label{eq_bcs_per}\\
 \text{Dirichlet:}\qquad & \phi(\{\rvp,z\in\{0,L\}\})=0, \label{eq_bcs_Dir}\\
 \text{Neumann:} \qquad & \pd_z\phi(\rvp,z)|_{z\in\{0,L\}}=0, \label{eq_bcs_Neu}\\
 \text{capillary \bcs:} \qquad & \pd_z\bra\phi(\{\rvp,z\})\ket|_{z\in\{0,L\}}=-h_1, \label{eq_bcs_cap}
\end{align}\label{eq_bcs}
\end{subequations}
\hspace{-0.1cm}with the decomposition $\rv=\{\rvp,z\equiv r_z\}$. Capillary \bcs are imposed onto the (mean) OP by the action of boundary fields of equal strength $h_1$ [see \cref{eq_cap_bcs} below] \footnote{Alternatively, \bcs can be enforced by adding suitable boundary terms to the Hamiltonian in \cref{eq_Hamilt} \cite{diehl_field-theoretical_1986}. However, since these are not needed in the present approach, we instead impose the BCs directly via \cref{eq_bcs}.}.
Tracer and OP field are coupled either linearly (with strength $h$) or quadratically (with strength $c$) \cite{dean_diffusion_2011,demery_perturbative_2011,demery_diffusion_2013}.
The coupling $h$ corresponds to a local bulk field (chemical potential), which leads to an excess OP around the tracer, resembling critical adsorption on a colloid \cite{fisher_wall_1978,schlesener_critical_2003}.
A quadratic coupling $c>0$ describes a locally reduced correlation length and results in a suppression of the OP fluctuations near the tracer. 
The dynamics of the system is described in terms of coupled Langevin equations \cite{dean_diffusion_2011,demery_perturbative_2011,demery_diffusion_2013}:
\begin{subequations}
\begin{align}
\dot \Rv(t) &= -\gamma_R \nabla_\Rv \Hcal + \sqrt{\gamma_R} \bm{\eta}(t) ,\label{eq_tracer_dyn_bare} \\
\dot\phi(\rv,t) &= -\gamma_\phi (-\nabla^2)^a \frac{\delta \Hcal}{\delta \phi(\rv)} + \sqrt{\gamma_\phi} \xi(\rv,t), 
\label{eq_field_dyn_bare}
\end{align}\label{eq_dyn_bare}
\end{subequations}
where $\gamma_R$ and $\gamma_\phi$ are kinetic coefficients and the Gaussian white noises $\bm{\eta}$ and $\xi$ have zero mean and correlations
\begin{subequations}
\begin{align}
\bra \eta_\alpha(t) \eta_\beta(t') \ket &= 2  T_R \delta_{\alpha\beta} \delta(t-t'),
\label{eq_tracer_noise} \\
\bra \xi(\rv,t)\xi(\rv',t')\ket &= 2  T_\phi (-\nabla^2)^a \delta(\rv-\rv')\delta(t-t'). \label{eq_phi_noise_correl}
\end{align}\end{subequations}
Dissipative dynamics (model A \cite{hohenberg_theory_1977}) is realized for $a=0$ and conserved dynamics (model B) for $a=1$.
The tracer is subject to reflective \bcs at $z\in \{0,L\}$.
In general, we allow different heat bath temperatures for the tracer ($T_R$) and the OP ($T_\phi$). The case $T_R\neq T_\phi$ describes an energy transfer between tracer and OP, which can be regarded as a simplified means to capture varying levels of non-equilibrium activity in the system \cite{dotsenko_two-temperature_2013,grosberg_nonequilibrium_2015,weber_binary_2016,tanaka_hot_2017, ilker_phase_2020}.
The Fokker-Planck equation (FPE) associated with \cref{eq_dyn_bare} is given by [$P\equiv P(\Rv,[\phi],t)$] \cite{gardiner_stochastic_2009,gruen_thin-film_2006}
\beq \pd_t P = \int_V \d^d r \frac{\delta}{\delta \phi(\rv)} \left[\gamma_\phi (-\nabla^2)^a \frac{\delta \Hcal}{\delta\phi(\rv)} + T_\phi \gamma_\phi (-\nabla^2)^a \frac{\delta}{\delta\phi(\rv)}\right] P + \gamma_R \nabla_\Rv\cdot\left(\nabla_\Rv \Hcal \right)P + T_R \gamma_R \nabla^2_\Rv P,
\label{eq_FPE_funct}\eeq 
which, in the equilibrium case, is solved by \cref{eq_Pss_joint}. We consider in the following also the distribution
\beq P_s(\Rv,[\phi]) = \frac{1}{\Zcal_0} \exp\left(-\frac{1}{T_\phi} \Hcal_\phi[\phi] - \frac{1}{T_R} \Hcal_R(\phi(\Rv))\right),
\label{eq_Pss_joint_T}\eeq
generalizing the one in \cref{eq_Pss_joint}.

Upon defining a rescaled time $\tilde t \equiv \gamma_R t$ (which has dimensions of $L^2/T_R$, see \cref{app_dimensions}) and setting $\Rv(t) = \tilde \Rv(\tilde t)$, $\phi(t) = \tilde \phi(\tilde t)$, $\bm{\eta}(t) = \bm{\tilde \eta}(\tilde t)$, etc., above equations take the form: 
\begin{subequations}
\begin{align}
\dot \Rv(t) &= (\nabla_\Rv \phi)(\Rv(t),t)[h - c\phi(\Rv(t),t)] + \bm{\eta}(t),  \\
\dot\phi(\rv,t) &= \chi^{-1} (-\nabla_\rv^2)^a \left[\nabla_\rv^2 \phi(\rv,t) + \zeta [h - c\phi(\Rv(t),t)] \delta(\rv-\Rv(t)) + h_1 [\delta(z) + \delta(L-z)] \right]  + \chi^{-1/2} \xi(\rv,t), \end{align}\label{eq_dynamics_genD}
\end{subequations}
\hspace{-0.11cm}where, for convenience, we dropped all tildes.
Furthermore, we have introduced the ``adiabaticity'' parameter
\beq 
\chi\equiv \sfrac{\gamma_R}{\gamma_\phi},
\label{eq_chi}\eeq 
as well as the control parameter $\zeta$, which takes the values $\zeta=0$ for a passive and $\zeta=1$ for a reactive tracer.
A tracer in thermal equilibrium with the OP field (describing, e.g., a colloidal particle) is realized for $T_R=T_\phi=T$, $\zeta=1$, and its steady state obeys \cref{eq_Pss_joint}.
A passive tracer, or one for which $T_R\neq T_\phi$, receives energy input that is not balanced by dissipation and therefore represents a driven, ``active'' particle \cite{szamel_self-propelled_2014,fodor_statistical_2018,grosberg_nonequilibrium_2015,weber_binary_2016,tanaka_hot_2017} \footnote{In Refs.\ \cite{dean_diffusion_2011,demery_perturbative_2011}, a tracer with $\zeta=1$ is called ``active''. However, in order to avoid confusion with other sources of activity, we prefer here the term ``reactive''.}.

We henceforth focus on a confined system of size $L$ in $d=1$ spatial dimensions (see \cref{fig_sketch}), such that \cref{eq_dynamics_genD} reduces to 
\begin{subequations}
\begin{align}
\dot R(t) &= (\pd_R \phi)(R(t),t)[h - c\phi(R(t),t)] + \eta(t), \label{eq_tracer_dyn} \\
\dot\phi(z,t) &= \chi^{-1} (-\pd_z^2)^a \left\{\pd_z^2 \phi(z,t) + \zeta [h - c\phi(R(t),t)] \delta(z-R(t)) + h_1 [\delta(z) + \delta(L-z)]\right\} + \chi^{-1/2} \xi(z,t). \label{eq_field_dyn}
\end{align}\label{eq_langevin_1d}
\end{subequations}
\hspace{-0.12cm}Since $L$ is the only OP-specific length scale in this system, we define the dimensionless counterpart of $\chi$ in \cref{eq_chi} as 
\beq \chit \equiv T_R L^{2a} \chi .
\label{eq_adiab_param}\eeq 
In the \emph{adiabatic limit}, $\chit\ll 1$, the dynamics of the OP field is much faster than the one of the tracer \footnote{As shown in \cref{app_adiab_elim}, the validity of the adiabatic approximation not only requires $\chit$, but instead the product $L h^2 (T_\phi/T_R) \chit$ (in the linearly coupled case) to be small.}. This will be utilized in \cref{sec_passive,sec_reactive} below to derive effective Markovian descriptions. 

The time-dependent statistics of $R$ and $\phi$ is described by a probability distribution $P(R,[\phi],t)= P(R,R_0,[\phi],t,t_0)$, where $R_0\equiv R(t_0)$ denotes the tracer position at the initial time $t_0$.  
Integrating $P$ over the fluctuations of the OP field $\phi$ renders the marginal probability distribution 
\beq \bar P(R,t) \equiv \int\Dcal\phi \, P(R,[\phi],t) ,
\label{eq_P_margin}\eeq
which, due to global conservation of probability, fulfills
\beq \int_0^L \d R \, \bar P(R,t) = 1.
\label{eq_P_cons}\eeq 
One aim of the present study is to determine the time-dependence of $\bar P(R,t)$ by deriving effective Fokker-Planck equations.
Importantly, we assume the OP to always remain in thermal equilibrium.
Accordingly, in the presence of a reactive tracer, the OP distribution $P_\phi(R,[\phi])$ is time independent and given by $P_s$ in \cref{eq_Pss_joint}. In the passive case, instead, one has [see \cref{eq_Hamilt}]
\beq P_\phi([\phi]) = \Zcal_0^{-1}\exp(-\Hcal_\phi([\phi])/T_\phi),
\label{eq_Pss_joint_pass}\eeq 
which is independent of $R$. 
The initial joint probability distribution is given by 
\beq P(R,R_0, [\phi], t= t_0, t_0) = \delta(R-R_0) P_\phi(R_0,[\phi]).
\label{eq_P_initcond}\eeq 
Note that, at times $t>t_0$, the joint distribution does not generally factorize into $R$- and $\phi$-independent parts.
For simplicity, we set $t_0=0$ and usually suppress the dependence of $P$ on $R_0$ and $t_0$.
We will frequently express the tracer probability distribution $\bar P(R,t)$ in terms of the dimensionless coordinate $\rho \equiv R/L$, which formally renders a new distribution $\bar p(\rho,t) = L \bar P(\rho L,t)$.
However, for notational simplicity we will use $\rho$ also as a shorthand notation in the expressions for $\bar P$. 

\section{Preliminaries}
\label{sec_prelim}

\subsection{Mode expansion and boundary conditions of the OP}
\label{sec_BCs}

We begin by introducing a set of eigenfunctions $\sigma_n(z)$ and eigenvalues $k_n^2$ of the operator $-\pd_z^2$ for the various \bcs [see \cref{eq_bcs}] considered in this study:
\begin{subequations}
\begin{align}
\sigma_n\pbc(z) &= \frac{1}{\sqrt{L}} \exp \left(\im k_n\pbc z\right) ,\qquad k_n\pbc = \frac{2\pi n}{L}, \qquad n=0,\pm 1, \pm 2, \ldots, \qquad  &\text{periodic \bcs}, \label{eq_eigenf_pbc} \\
\sigma_n\Dbc(z) &= \sqrt{\frac{2}{L}} \sin\left(k_n\Dbc z\right), \qquad k_n\Dbc = \frac{\pi n}{L}, \qquad n=1,2,\ldots,  \qquad &\text{Dirichlet \bcs}, \label{eq_eigenf_Dbc}\\
\sigma_n\Nbc(z) &= \sqrt{\frac{2-\delta_{n,0}}{L}} \cos\left(k_n\Nbc z\right), \qquad k_n\Nbc = \frac{\pi n}{L}, \qquad n=0,1,2,\ldots,  \qquad &\text{Neumann \bcs}. \label{eq_eigenf_Nbc}
\end{align}\label{eq_eigenspec}
\end{subequations}
\hspace{-0.12cm}The eigenfunctions are orthonormal:
\beq \int_0^L \d z\, \sigma_m(z) \sigma_n^*(z) = \delta_{m,n},
\label{eq_eigenf_ortho}
\eeq 
and complete:
\beq \sum_n \sigma_n(z) \sigma_n^*(z') = \delta(z-z').
\label{eq_eigenf_complete}\eeq
Note that complex conjugation is only relevant for periodic \bcs.
The eigenfunction expansions of the OP field and the noise take the form:
\begin{equation}
\phi(z,t) = \sum_{n} \sigma_n(z) \phi_n(t),\qquad 
\xi(z,t) = \sum_{n} \sigma_n(z) \xi_n(t) ,
\label{eq_phi_expand}
\end{equation}
which imply the inverse relations [see \cref{eq_eigenf_ortho}]:
\begin{equation}
\phi_n(t) = \int_0^L\d z\, \sigma^*_n(z) \phi(z,t),\qquad
\xi_n(t) = \int_0^L\d z\, \sigma^*_n(z) \xi(z,t).
\end{equation}
For a real-valued function such as $\phi(z,t)$, one has $\phi_{-n}(t) = \phi_n^*(t)$.
According to \cref{eq_phi_noise_correl}, the correlations of the noise modes are given by 
\beq \bra \xi_m(t) \xi_n^*(t')\ket = 2  T_\phi k_m^{2a} \delta_{m,n} \delta(t-t').
\label{eq_phi_noise_mode_correl}\eeq 
Upon inserting \cref{eq_phi_expand} into \cref{eq_field_dyn}, one obtains the dynamic equation of the OP modes:
\beq \pd_t \phi_n = \chi^{-1} k_n^{2a} \left\{ -k_n^2 \phi_n + \zeta [h - c\phi(R(t),t)]  \sigma_n^*(R(t)) + h_1 [\sigma_n^*(0) + \sigma_n^*(L)] \right\} + \chi^{-1/2} \xi_n.
\label{eq_phi_mode_eqn}\eeq
We assume the initial condition $\phi(z,t=t_i)=0$ to apply in the infinite past ($t_i = -\infty$), such that at $t=t_0=0$ the OP is equilibrated [see \cref{eq_P_initcond}] and its initial condition plays no role anymore.
The zero mode $\phi_{n=0}$, which occurs (except for Dirichlet \bcs) in the case of dissipative dynamics ($a=0$), has to be treated separately.
For all \bcs except standard Dirichlet ones [\cref{eq_bcs_Dir}], \cref{eq_langevin_1d} with $a=1$ conserves the OP globally, i.e., $\int_0^L\d z\, \phi(z,t)=\const$. 
Standard Dirichlet \bcs generally entail a non-zero flux through the boundaries, requiring to use suitable ``no-flux'' basis functions \cite{gross_first-passage_2018-1} instead in order to ensure global OP conservation. However, since this is technically involved, we do not consider conserved dynamics in conjunction with Dirichlet \bcs in the following.

Due to the equilibrium assumption, the mean OP profile $\bra\phi(z)\ket$ is time-independent and, for the \bcs specified in \cref{eq_eigenspec}, in fact, vanishes, $\bra\phi(z)\ket=0$.
A spatially inhomogeneous profile can arise here either due to a local bulk field ($h\neq 0$), representing a linearly coupled reactive tracer, or due to boundary fields ($h_1\neq 0$).
For a system with $h=0$ and boundary fields of identical strength at $z\in \{0,L\}$, the mean OP profile is given by [see \cref{app_profile_correl_lin}] 
\beq \bra\phi(z)\ket_{h_1}   =  h_1 L \left[\left( \frac{1}{2} - \frac{z}{L} \right)^2 - \frac{1}{12}\right],
\label{eq_avg_prof_h1}\eeq 
which fulfills standard capillary \bcs [see \cref{eq_bcs_cap} as well as, e.g., Ref.\ \cite{gross_critical_2016} and references therein]:
\beq \pd_z\bra \phi(z)\ket_{h_1}\big|_{z\in\{0,L\}} = \mp h_1.
\label{eq_cap_bcs}\eeq 
Capillary \bcs are applicable only in the absence of a zero mode, as otherwise $\bra\phi(z)\ket_{h_1}$ would be divergent [see \cref{app_profile_correl_lin}].
A system without a zero mode is naturally realized for conserved dynamics. 

\subsection{Solution of the OP dynamics}
\label{sec_OP_sol}
We next present exact as well as perturbative solutions of \cref{eq_phi_mode_eqn}, which will be used in the course of this work. 

\subsubsection{Solution of \cref{eq_phi_mode_eqn} for $c=0$ or $\zeta=0$}
For $c=0$ and arbitrary $\zeta$, as well as generally for $\zeta=0$ (passive tracer), the solution of \cref{eq_phi_mode_eqn} is given by: 
\begin{subequations}
\begin{align}
\phi_n(t) &= \int_{-\infty}^t \d s\, e^{- k_n^{2+2a} (t-s)/\chi} \left\{ \chi^{-1} k_n^{2a} \left[ \zeta h \sigma_n^*(R(s)) + h_1 \tau^*_n \right] + \chi^{-1/2} \xi_n(s) \right\},\qquad (c=0)  \label{eq_phi_mode_sol_n} \\
\phi_0(t) &= (t-t_i)\chi^{-1} \delta_{a,0} (\zeta h \sigma_0 + h_1 \tau_0) + \chi^{-1/2} \int_{t_i}^t \d s\, \xi_0(s) , \label{eq_phi_mode_sol_zm}
\end{align} \label{eq_phi_mode_sol}
\end{subequations}
\hspace{-0.1cm}where $\tau_n \equiv \sigma_n(0)+\sigma_n(L)$ and $\sigma_0= 1/\sqrt{L}$.
The zero mode $\phi_0$ performs a diffusive motion superimposed on a linear growth. 
Since this growth depends on the initial time $t_i$, the latter has to be kept finite in \cref{eq_phi_mode_sol_zm}, whereas in \cref{eq_phi_mode_sol_n}, it is unproblematic to set $t_i=-\infty$ in the lower integration boundary.
In the adiabatic limit $\chit\ll 1$, \cref{eq_phi_mode_sol_n} reduces, for $\zeta=0$ or $h=0$, to [see \cref{eq_phi_h_adiab_expand_n}]
\begin{align}
\phi_n\izero(t) \simeq \frac{h_1}{k_n^2}\tau_n^* + \sqrt{\chi}\frac{\xi_n(t)}{ k_n^{2+2a}}, \qquad (n\neq 0,\quad  \chit\ll 1)
\label{eq_phi_mode_adiab}
\end{align} 
while the zero mode [\cref{eq_phi_mode_sol_zm}] does not admit an expansion for small $\chit$.
We conclude that \bcs involving a zero mode are not compatible with the adiabatic limit. This is expected, since the relaxation time of a zero mode is infinite, which violates the assumption of a fast OP dynamics. Besides standard periodic or Neumann \bcs, we shall thus also consider modified variants thereof for which the zero mode is explicitly removed [see \cref{eq_phi_correl} below].
Note that the adiabatic approximation can in general not be used to calculate equilibrium variances of $\phi$ (cf.\ \cref{sec_correl_ad}).

\subsubsection{Solution of \cref{eq_phi_mode_eqn} for $c\neq 0$ and $h=0$}
\label{sec_phisol_c}

For $c\neq 0$, a perturbative solution of \cref{eq_phi_mode_eqn} in orders of $c$ can be constructed.
To this end, we formally expand the OP as $\phi(z,t) = \phi\izero + \phi\ione + \ldots$, where $\phi^{(i)}\sim \Ocal(c^i)$, and assume the noise $\xi\sim \Ocal(c^0)$ \footnote{The actual dimensionless control parameter associated with $c$ is not explicit here and will be determined later [see \cref{eq_effcoupl_c}].}. 
Inserting this expansion in \cref{eq_phi_mode_eqn} (with $h=0$) and grouping terms of the same order in $c$, we obtain
\begin{subequations}
\begin{align}
\phi\ione_n(t) &= - \zeta \chi^{-1} c k_n^{2a} \int_{-\infty}^t \d s\, e^{- k_n^{2+2a} (t-s)/\chi} \phi\izero(R(s),s) \sigma_n^*(R(s)),\qquad (n\neq 0) \label{eq_phi_c_hierarchy} \\
 \phi_0\ione(t) &= -\zeta \chi^{-1} c\, \delta_{a,0}\, (t-t_i)\,  \phi\izero(R(t),t) \sigma_0^*(R(t)), \label{eq_phi_c_hierarchy_zm}
\end{align} 
\end{subequations}
while the $\Ocal(c^0)$ solution $\phi_n\izero(t)$ is given by \cref{eq_phi_mode_sol} with $h=0$. Note that the only dependence on $h_1$ enters through $\phi_n\izero$.
Determining, analogously to \cref{app_adiab_OP}, the adiabatic limit $\chit\ll 1$ of \cref{eq_phi_c_hierarchy}, gives
\beq 
\phi\ione_n(t) \simeq -\frac{\zeta c }{k_n^2}  \phi\izero(R(t),t) \sigma_n^*(R(t)). \qquad (n\neq 0,\quad \chit\ll 1)
\label{eq_P2_c_act_phi1}\eeq

\subsection{Order-parameter correlation functions}
\label{sec_phi_correl_pass}

The (connected) equilibrium OP correlation function is defined as usual as 
\beq C_\phi(z,z',t,t') \equiv \bra \delta\phi(z,t) \delta\phi(z',t')\ket = \bra \phi(z,t) \phi(z',t')\ket - \bra\phi(z)\ket\bra\phi(z')\ket,\qquad  \delta\phi(z,t) \equiv \phi(z,t) - \bra\phi(z)\ket,
\label{eq_correl_def}\eeq 
where $\delta\phi$ denotes the fluctuating part of $\phi$.
We provide in the following expressions for $C_\phi$ within the Gaussian model and evaluate them in the adiabatic limit. 
Technical details are deferred to \cref{app_profile_correl}.

\subsubsection{Time-dependent correlation functions}

The time-dependent correlation function follows straightforwardly from the solution of the Langevin equation [\cref{eq_phi_mode_eqn}] discussed above.
In the case of a passive tracer ($\zeta=0$), \cref{eq_phi_mode_sol,eq_phi_noise_mode_correl} render the equilibrium two-time correlator 
\beq C_\phi(z,z',t,t')\big|_{\zeta=0}  =  T_\phi \sump_n \sigma_n(z) \sigma_n^*(z') \frac{e^{- k_n^{2+2a}  |t-t'|/\chi}}{k_n^2} + \frac{2  T_\phi }{L \chi} \min(t-t_i, t'-t_i) \delta_{a,0} ,
\label{eq_C_phi}\eeq 
where the prime indicates a sum excluding $n=0$ and we used \cref{eq_avg_prof_h1,eq_avg_prof_series} to identify the mean part $\bra\phi(z)\ket_{h_1}$, as required by \cref{eq_correl_def}. 
The last term in \cref{eq_C_phi} stems from the zero mode and exists only for periodic or Neumann \bcs [see \cref{eq_eigenspec}] and non-conserved dynamics ($a=0$). 
In the following, we will also need the correlation function of $\pd_z\phi$:
\beq\begin{split} C_{\pd\phi}(z,z',t-t')\big|_{\zeta=0} &\equiv  \bra \pd_z\delta\phi(z,t) \pd_{z'}\delta\phi(z',t')\ket\big|_{\zeta=0} = \pd_z\pd_{z'} C_\phi(z,z',t, t')\big|_{\zeta=0} \\
&=  T_\phi \sum_n  \tilde \sigma_n(z) \tilde \sigma_n^*(z') e^{- k_n^{2+2a}  |t-t'|/\chi},
\end{split}\label{eq_C_phi_deriv}\eeq 
where 
\beq \tilde\sigma_n(z) \equiv \frac{1}{k_n} \pd_z\sigma_n(z) = \begin{cases}
                          \im \sigma_n\pbc(z), \qquad n=\pm 1,\ldots &\qquad \text{(p)} \\
                          \sqrt{\frac{2}{L}}\cos(k_n\Dbc z) ,\qquad n=1,2,\ldots &\qquad \text{(D)}\\
                          -\sqrt{\frac{2}{L}}\sin(k_n\Nbc z) ,\qquad n=1,2,\ldots &\qquad \text{(N)} 
                         \end{cases}
\label{eq_eigenf_deriv}\eeq
and $\tilde\sigma_0\pbc(z) = \tilde\sigma_0\Nbc = 0$.
The notation in the right column indicates the \bcs of the OP [see \cref{eq_bcs}].
Since the zero mode does not contribute to $C_{\pd\phi}$, the latter is solely a function of the time difference.
The correlation function $C\pbc$ for periodic \bcs can be expressed in terms of the corresponding one for Neumann and Dirichlet \bcs \cite{gross_dynamics_2019}:
\beq  C\pbc(z-z',t-t') = \frac{1}{2}\left[ C\Nbc(z,z',t-t') + C\Dbc(z,z',t-t') \right]_{L/2},
\label{eq_C_per_ND_rel}\eeq 
where the r.h.s.\ is to be evaluated for a system of size $L/2$ instead of $L$. 

\subsubsection{Static correlation functions}

In equilibrium for $t=t'$, the (connected) OP correlation function [see \cref{eq_phi_correl_genfunc}] can be directly determined from \cref{eq_Pss_joint_T}:  
\beq C_\phi(x,y) \equiv \bra\delta\phi(x) \delta\phi(y)\ket\big|_{\zeta=0} = T_\phi \sum_n \frac{\sigma_n(x)\sigma_n^*(y) }{k_n^2} = \begin{cases} 
	  L T_\phi \left[ \frac{1}{6} - \frac{1}{L}|x-y| + \frac{1}{L^2}(x-y)^2 \right], \qquad & \text{(p$^*$)} \\
	  L T_\phi \left[ \frac{1}{L}\min( x,  y) -  \frac{1}{L^2} x  y \right], & \text{(D)} \\ 
	  L T_\phi \left[ \frac{1}{3} - \frac{1}{L}\max( x,  y) + \frac{1}{2L^2}\left( x^2 +  y^2\right) \right], \qquad & \text{(N$^*$,$\pm$)} \\ 
	  \frac{T_\phi}{L \varepsilon} + C_\phi\ut{(p$^*$,N$^*$)}(x,y). \qquad & \text{(p,N)}  
  \end{cases}\label{eq_phi_correl}
\eeq 
Here and in the following, $\pm$ refers to capillary \bcs [\cref{eq_cap_bcs}] \footnote{The notation $\pm$ indicates the sign of $h_1$ and alludes to the symmetry-breaking character of the boundary field, following the nomenclature used in boundary critical phenomena, see, e.g., Ref.\ \cite{gross_critical_2016}.}, while a boundary condition labeled by an asterisk indicates that the zero mode is removed from the set of modes in \cref{eq_eigenspec}. 
Physically, the vanishing of the zero mode can be ensured within conserved dynamics if a vanishing mean OP $\bra\phi(t_i)\ket=0$ is imposed as initial condition [see \cref{eq_phi_mode_eqn}]. 
However, in order to elucidate the effect of the conservation law, we will use (p$^*$) and (N$^*$) \bcs also in conjunction with dissipative dynamics, if this is required to obtain a well-defined model (e.g., in the adiabatic limit, see \cref{sec_correl_ad} below).
Note that, for capillary \bcs, the OP correlation function takes a Neumann form (see \cref{app_profile_correl_lin}).
As a characteristic of the Gaussian model, \cref{eq_phi_correl} holds independently from the presence of boundary or bulk fields (see \cref{app_profile_correl_lin}); in particular, it applies to an OP coupled linearly to a passive or reactive tracer. 
If the tracer is coupled quadratically, by contrast, the static correlation function differs in the passive and the reactive case (see \cref{sec_phivar_quadcoupl}).

\begin{figure}[t]\centering
    \subfigure[]{\includegraphics[width=0.32\linewidth]{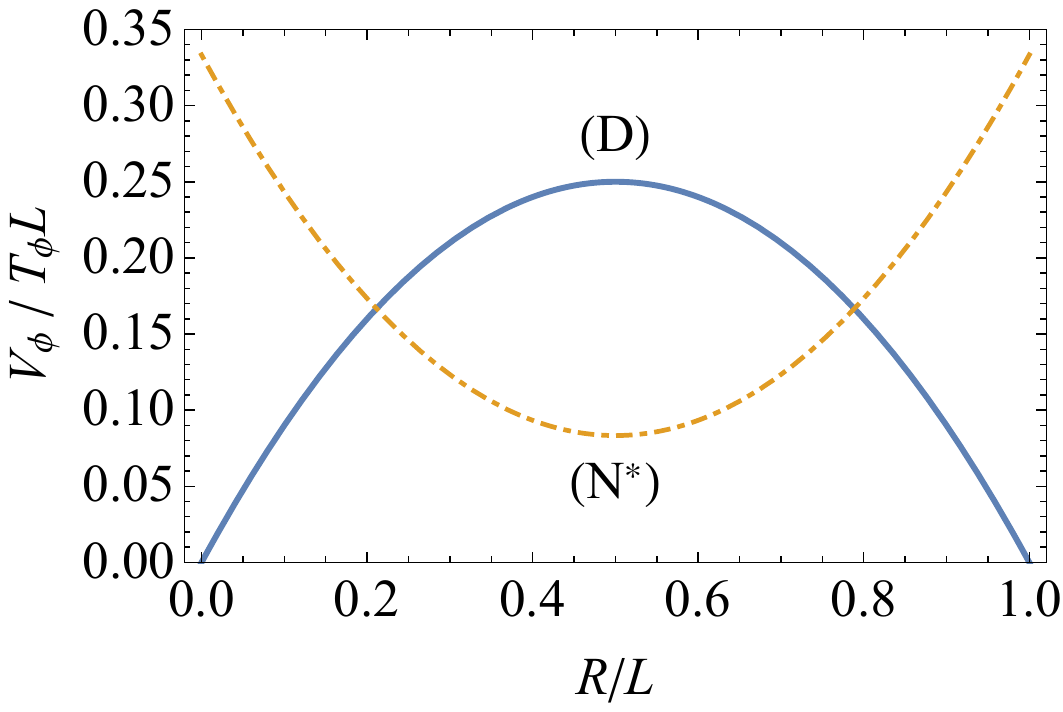} \label{fig_phiVar}}\qquad
    \subfigure[]{\includegraphics[width=0.32\linewidth]{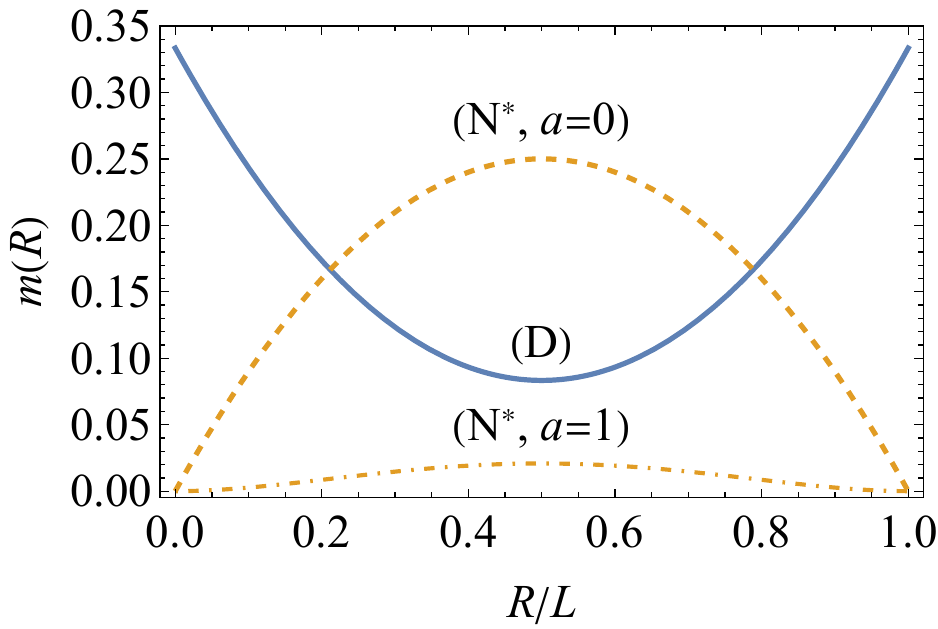} \label{fig_mFunc}}
    \caption{(a) Static variance $V_\phi(R) = \bra\delta\phi(R)^2\ket$ [\cref{eq_phiR_var}] of an OP field $\phi$ obeying Dirichlet (D) or Neumann \bcs without a zero mode (N$^*$), in the absence of a (reactive) tracer ($\zeta=0$). (b) Amplitude $m(R)$ of the correlation function $\bra [\pd_R\delta\phi(R,t)] [\pd_R\delta\phi(R,t')] \ket$ in the adiabatic limit [see \cref{eq_m_func_spec}].
    }    
\end{figure}

For $x=y=R$, \cref{eq_phi_correl} reduces to the variance ($\rho\equiv R/L$)
\beq V_\phi(R) \equiv C_\phi(R,R) = \bra\delta\phi(R)^2\ket\big|_{\zeta=0}  =  T_\phi \sump_n \frac{|\sigma_n(R)|^2}{k_n^2} 
= \begin{cases}
   T_\phi L \times \frac{1}{12}, \qquad & \text{(p$^*$)}\\
    T_\phi L \left( \rho - \rho^2 \right), & \text{(D)}\\
    T_\phi L \left( \frac{1}{3} - \rho + \rho^2 \right), & \text{(N$^*$,$\pm$)}
  \end{cases}
\label{eq_phiR_var}\eeq
which turns out to be a central quantity for the tracer dynamics and is illustrated in \cref{fig_phiVar}.
It is useful to remark that $V_\phi(R) = \bra \phi\izero(R)^2\ket - \bra\phi\ket_{h_1}^2$, which follows from \cref{eq_phi_mode_sol,eq_avg_prof_series} (with $\zeta=0$).
Note that we defined $V_\phi$ only for \bcs without a zero mode, as the variance is infinite otherwise.

\subsubsection{Adiabatic approximation}
\label{sec_correl_ad}

The adiabatic limit of $C_\phi$ can be obtained by inserting \cref{eq_phi_mode_adiab} into \cref{eq_correl_def}:
\beq C\ad_\phi(z,z',t-t')\big|_{\zeta=0} = 2 T_\phi \chi \sump_n \frac{\sigma_n(z) \sigma_n^*(z')}{k_n^{4+2a}} \delta(t-t') = L^{3} \chit_\phi \Ccal(z/L, z'/L) \delta(t-t') ,
\label{eq_C_phi_adiab}\eeq 
with [cf.\ \cref{eq_adiab_param}]
\beq \chit_\phi \equiv T_\phi L^{2a} \chi 
\label{eq_adiab_param_phi}\eeq 
and 
\beq \Ccal(\hat z, \hat z') = 
\begin{cases}
	\frac{1}{6} \left[ | \hat z-\hat z'| ^3-| \hat z+\hat z'| ^3+2 \hat z \hat z' \left(\hat z^2+\hat z'^2+2\right) \right], \qquad &\text{(D)}\\
	\frac{1}{6} \left[| \hat z-\hat z'| ^3+| \hat z+\hat z'| ^3-3 \hat z^2 \hat z'^2-\frac{1}{2}\hat z^4 -2 \hat z^2-\frac{1}{2}\hat z'^4 -2 \hat z'^2 \right] +\frac{1}{45}. \qquad &\text{(N$^*$, $a=0$)}
\end{cases}
\eeq
The corresponding expression for Neumann \bcs with $a=1$ is of similar polynomial form, but rather lengthy and not stated here.
The expression for periodic \bcs can be constructed by means of \cref{eq_C_per_ND_rel}.
Since the relaxation time of a zero mode is divergent [see also \cref{eq_C_phi}], the actual limit $\chit,\chit_\phi \to 0$ does not exist for standard periodic or Neumann \bcs. 
We remark that $\int_{-\infty}^\infty \d t\, C_\phi(z,z',t-t') = \int_{-\infty}^\infty \d t\, C\ad_\phi(z,z',t-t')$, i.e., the term on the right-hand side of \cref{eq_C_phi_adiab} multiplying the $\delta$-function corresponds to the time integral of the correlation function.

Using \cref{eq_C_phi_adiab}, one obtains 
\beq\begin{split} C\ad_{\pd\phi}(z,z',t-t')\big|_{\zeta=0} &= 2 T_\phi \chi \sump_n \frac{\tilde \sigma_n(z) \tilde\sigma_n^*(z')}{k_n^{2+2a}} \delta(t-t')  \\
 &= L\chit_\phi \delta(t-t')\times
  \begin{cases}
  	- \hat z -\hat z' - |\hat z -\hat z'| + \hat z^2 + \hat z'^2 + \frac{2}{3},  &\text{(D)} \\
  	\hat z +\hat z' -  |\hat z -\hat z'| - 2\hat z \hat z', \qquad &\text{(N$^*$, $a=0$)} \\
  	\frac{1}{3} \hat z \hat z'(2+\hat z^2 + \hat z'^2) + \frac{1}{6} |\hat z -  \hat z'|^3 - \frac{1}{6}(\hat z +\hat z')^3 , & \text{(N$^*$, $a=1$)}
  \end{cases} \label{eq_C_phi_deriv_ad}
\end{split}\eeq 
for the expression of $C_{\pd\phi}$ in the adiabatic limit.
We introduce a function $m(R)$ by writing
\beq C\ad_\pdphi(R,R,t-t')\big|_{\zeta=0} = 2 L \chit_\phi m(R) \delta(t-t'),\qquad m(R) \equiv \frac{1}{L^{1+2a}} \sump_n \frac{|\tilde\sigma_n(R)|^2}{k_n^{2+2a}}.
\label{eq_m_func}\eeq 
For the various \bcs, one has (see \cref{fig_mFunc}): 
\begin{subequations}
\begin{align}
m(R)\big|_{a=0} &= \begin{cases} 
		  \frac{1}{3}, \qquad &\text{(p$^*$)} \\
                  \frac{1}{3} - \rho + \rho^2 , & \text{(D)} \\
                  \rho - \rho^2, & \text{(N$^*$,$\pm$)}
                \end{cases} \\
\intertext{and}
m(R)\big|_{a=1} &= \begin{cases} 
		  \frac{1}{45}, \qquad &\text{(p)} \\
                  \frac{1}{3} \rho^2 - \frac{2}{3}\rho^3 + \frac{1}{3} \rho^4 . & \text{(N$^*$,$\pm$)}
                \end{cases}
\end{align}\label{eq_m_func_spec}
\end{subequations}
\hspace{-0.1cm}Note that $m$ is dimensionless and we generally suppress the dependence on $L$.
It turns out that $m$ encodes the modification of the tracer mobility due to the OP fluctuations.

\subsection{Effective Langevin equation for the tracer}
\label{sec_eff_Lang}

\begin{figure}[t]\centering
\subfigure[]{\includegraphics[width=0.32\linewidth]{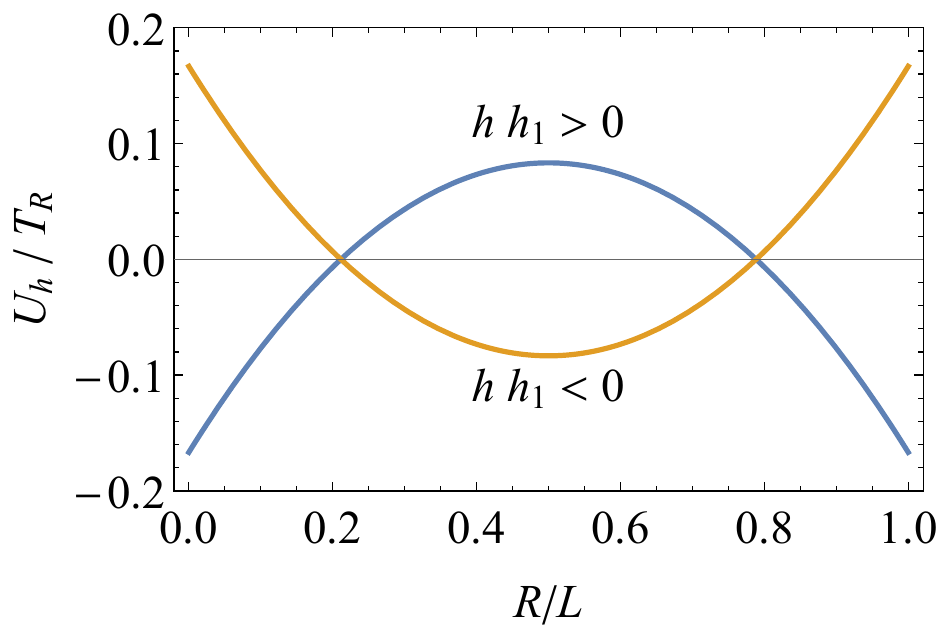} \label{fig_effpot_h} } \qquad 
    \subfigure[]{\includegraphics[width=0.31\linewidth]{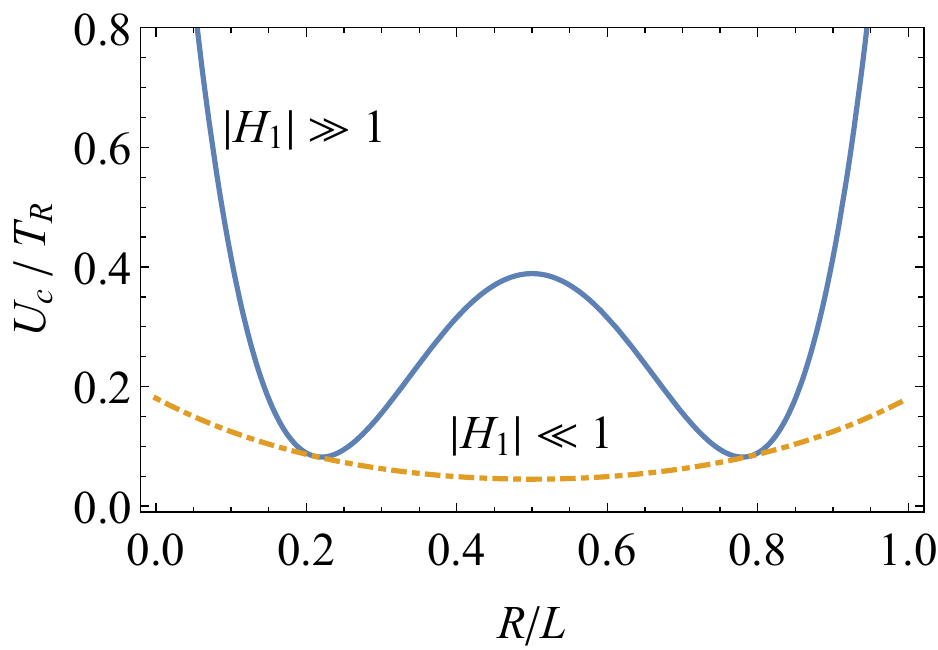} }
    \caption{Effective potentials $U_h$ and $U_c$ [\cref{eq_effpot}], which govern the the effective dynamics of a (passive) tracer [see \cref{eq_tracer_dyn_simpl}] in the presence of boundary fields $h_1$. The tracer is coupled either (a) linearly or (b) quadratically to the OP field. In (b), we have defined the dimensionless parameter $H_1 = h_1 \sqrt{L/T_\phi}$ [see also \cref{eq_effcoupl_h1}]. In the limit $|H_1|\ll 1$, $U_c$ becomes independent of $h_1$ and reduces to $U_c(R)\simeq (c/2) V_\phi\NbcNZM(R)$ [see \cref{eq_phiR_var,fig_phiVar}]. For Dirichlet \bcs, one accordingly has $U_c(R)=(c/2)V_\phi\Dbc(R)$.
    }
    \label{fig_effpot}
\end{figure}

The Langevin equation in \cref{eq_tracer_dyn} can be rewritten in a more compact form:
\beq \dot R(t) = \Xi_h(R(t),t) + \Xi_c(R(t),t) + \eta(t),\qquad \Xi_h(z,t) \equiv h \pd_z\phi(z,t),\qquad \Xi_c(z,t) \equiv -c \phi(z,t) \pd_z\phi(z,t) = -\frac{c}{2} \pd_z \phi^2(z,t),
\label{eq_tracer_dyn_simpl}\eeq 
wherein $\Xi_h$ and $\Xi_c$ represent effective forcing terms.
In the case of a passive tracer, the only source of inhomogeneity for the OP are boundary fields $h_1\neq 0$, which allows us to decompose the OP as $\phi(z) = \bra\phi(z)\ket_{h_1} + \delta\phi(z)$, with the mean equilibrium profile $\bra\phi(z)\ket_{h_1}$ stated in \cref{eq_avg_prof_h1} and $\bra\delta\phi(z)\ket=0$.
Accordingly, the mean parts of $\Xi_h$ and $\Xi_c$ give rise to (time-independent) effective potentials $U_h$ and $U_c$: 
\begin{subequations}
\begin{align}
\bra\Xi_h(z)\ket &= -\pd_z U_h(z),\qquad U_h(z) \equiv -h \bra\phi(z)\ket_{h_1}, \label{eq_Pi_h_effpot} \\
\bra\Xi_c(z)\ket &= -\pd_z U_c(z),\qquad U_c(z) \equiv \frac{c}{2} \left[V_\phi(z) + \bra\phi(z)\ket_{h_1}^2 \right], \label{eq_Pi_c_effpot} 
\end{align}\label{eq_effpot}
\end{subequations}
\hspace{-0.1cm}where $V_\phi$ is the fluctuation variance stated in \cref{eq_phiR_var}. 
It turns out that the effective potentials $U_{h,c}$ (see \cref{fig_effpot}) determine the dynamics and the steady state of a passive tracer [see \cref{eq_FP_pass_h_adiab,eq_FP_pass_c_adiab} below], as well as of a reactive tracer for weak couplings $h$, $c$ [see \cref{eq_FP_act_h_adiab,eq_FP_act_c_adiab} below].
In particular, the expression for $U_c$ in \cref{eq_Pi_c_effpot} also applies if $h_1=0$, in which case $V_\phi$ is given by \cref{eq_phiR_var} for the various \bcs.
Accordingly, a linearly coupled tracer is attracted to (repelled from) the boundary if $h$ and $h_1$ have the same (opposite) sign. A quadratically coupled tracer, instead, is generally repelled from the boundaries with $h_1\neq 0$, which, interestingly, can lead to the emergence of two distinct minima in $U_c$ provided $|h_1|$ is large. This behavior reflects the critical Casimir interactions between wall and tracer and is further discussed in \cref{sec_discuss_react}.

By subtracting the mean parts from $\Xi_h$ and $\Xi_c$, we obtain new noises
\begin{subequations}
\begin{align}
\Pi_h(z,t) &\equiv \Xi_h(z,t) - \bra \Xi_h(z)\ket, \label{eq_Pi_h_effnoise} \\
\Pi_c(z,t) &\equiv \Xi_c(z,t) - \bra\Xi_c(z)\ket \label{eq_Pi_c_effnoise} 
\end{align}\label{eq_effnoise}
\end{subequations}
\hspace{-0.1cm}having vanishing means and correlations are given by [see \cref{eq_C_phi,eq_C_phi_deriv}] 
\begin{subequations}
\begin{align}
\bra \Pi_h(z,t)\Pi_h(z',t')\ket &= h^2 C_{\pd\phi}(z,z',t-t')\label{eq_Pi_h_correl},\\
\bra \Pi_c(z,t) \Pi_c(z',t')\ket &= c^2 \Big\{ \pd_{z}\pd_{z'} \bra\phi(z)\ket_{h_1} \bra\phi(z')\ket_{h_1} C_\phi(z,z',t-t') \nonumber  \\ &\qquad + \left[\pd_{z}C_{\phi}(z,z',t-t')\right] \left[\pd_{z'}C_{\phi}(z,z',t-t')\right] + C_\phi(z,z',t,t') C_{\pd\phi}(z,z',t-t') \Big\}, \label{eq_Pi_c_correl}
\end{align}
\end{subequations}
while $\bra\Pi_{h,c}(R(t),t)\, \eta(t)\ket=0$.
We emphasize that $\Pi_c$, which is essentially the square of the Gaussian process $\phi$, is itself not Gaussian \cite{miguel_theory_1981,luczka_non-Markovian_1995}. 
A Gaussian approximation can be made by assuming a small coupling $c$, such that higher-order cumulants of $\Pi_c$ become negligible.

Due to the non-linear and non-Markovian nature of \cref{eq_tracer_dyn_simpl}, its time-dependent solution can in general not be determined exactly. 
In the following, we thus solve \cref{eq_tracer_dyn_simpl} perturbatively, either by assuming a weak coupling or by applying an adiabatic approximation. The latter approach amounts to replacing the noises $\Pi_h$ and $\Pi_c$ by suitable Markovian approximations.

\section{Passive tracer}
\label{sec_passive}

We begin by discussing a passive tracer, which is described by \cref{eq_langevin_1d} with $\zeta=0$. Since the forcing term [first term on the r.h.s.\ of \cref{eq_tracer_dyn}] is not balanced by a corresponding dissipation term in \cref{eq_field_dyn}, a passive tracer represents a driven, non-equilibrium system. We separately analyze the cases of a linear ($h\neq 0$, $c=0$) and a quadratic coupling ($h=0$, $c\neq 0$) between tracer and OP field.

\subsection{Linear tracer-field coupling}
\label{sec_passive_linear}

\subsubsection{Adiabatic approximation}
\label{sec_pass_lin_adiab}

We apply the adiabatic elimination procedure described in Refs.\ \cite{theiss_systematic_1985,theiss_remarks_1985} in order to integrate out the fast dynamics of $\phi$ from \cref{eq_langevin_1d}, which results in the following FPE for the effective tracer probability distribution $\bar P$ (see \cref{app_adiab_elim}):
\beq \pd_t \bar P(R,t) = - \pd_R \left[\mu(R)  \bar P(R,t) \right] + \pd_R^2 \left[ D(R) \bar P(R,t) \right] ,
\label{eq_FP_pass_h_adiab}\eeq 
with drift and diffusion coefficients given by [see \cref{eq_adiab_drift_diffus_p} with $\zeta=0$] 
\begin{subequations}
\begin{align}
\mu(R) &\equiv  -U_h'(R)   + \onehalf D'(R), \label{eq_FP_pass_h_drift} \\
D(R) &\equiv  T_R \left[ 1 + \chit \kappa_h m(R) \right] , \label{eq_FP_pass_h_diff}
\end{align}\label{eq_FP_pass_h_coeffs}
\end{subequations}
\hspace{-0.15cm}where the prime denotes a derivative, $\chit$, $m(R)$, and $U_h(R)\propto h_1$ are stated in Eqs.\ \eqref{eq_adiab_param}, \eqref{eq_m_func_spec}, and \eqref{eq_Pi_h_effpot}, respectively, and we defined the effective dimensionless coupling constant
\beq \kappa_h \equiv \frac{L h^2 T_\phi}{T_R^2}.
\label{eq_effcoupl_h}\eeq
The term $D'(R)/2$ in \cref{eq_FP_pass_h_drift} represents a spurious drift \cite{gardiner_stochastic_2009}, while the prefactor of $U_h'(R)$ represents a mobility, which is unity here, i.e., in physical units [see \cref{eq_tracer_dyn_bare}], the effective and bare mobilities of a passive tracer are identical.
As revealed by a comparison of the characteristic relaxation rates of the OP and the tracer, the adiabatic approximation requires not only $\chit\ll 1$, but, in fact, $\chit\kappa_h\ll 1$ [see \cref{eq_adiab_valid}].

The steady state solution of \cref{eq_FP_pass_h_adiab} (with vanishing flux at the boundaries) is given by [see also \cref{eq_app_Pss}]
\beq \bar P\st{s}(R) = \frac{1}{L \Zcal} \frac{1}{\sqrt{1 + \chit\kappa_h \, m(R)}} \exp\left[ -\int_0^R \d z \frac{U_h'(z)}{D(z)} \right].
\label{eq_Pss_h_pass_adiab}\eeq 
The exponential term is absent for $h_1=0$. In this case, the normalization constant $\Zcal$ evaluates for $a=0$ (non-conserved OP dynamics) to $\Zcal\pbc = \sfrac{1}{\sqrt{1+\chit\kappa_h \, m\pbc}}$, $\Zcal\Dbc = \ln\left(\frac{12+7\chit\kappa_h +4\sqrt{3\chit\kappa_h(3+\chit\kappa_h)}}{12+\chit\kappa_h}\right)/ \sqrt{\chit\kappa_h}$, and $\Zcal^{(\text{N}^*)} = 2\,\mathrm{arccot}\left(2/\sqrt{\chit\kappa_h}\right)/\sqrt{\chit\kappa_h}$; for $a=1$ and generally for $h_1\neq 0$, $\Zcal$ has to be calculated numerically. We recall that standard Dirichlet \bcs are not compatible with global probability conservation ($a=1$). 
In the extreme adiabatic limit $\chit= 0$, \cref{eq_Pss_h_pass_adiab} reduces to 
\beq \bar P\st{s}(R) = \tilde\Zcal^{-1}\exp\left(-U_h(R)/T_R\right),\qquad (\chit=0)
\label{eq_Pss_h_pass_adiab0}\eeq 
with $\tilde\Zcal = \exp[-h_1/(12 T_R)] \sqrt{\pi T_R/h_1}\, \mathrm{erfi}(\sqrt{h_1/T}/2)$.
\Cref{eq_Pss_h_pass_adiab0} is what one obtain by neglecting the noise $\Pi_h$ [see \cref{eq_Pi_h_effnoise}] in \cref{eq_tracer_dyn_simpl}. Note that $\bar P_s$ in \cref{eq_Pss_h_pass_adiab0} is non-trivial only for capillary \bcs, as otherwise $\bar P_s(R)=1/L$ for $\chit= 0$.

It is interesting to contrast the above results  to a ``naive'' derivation of the adiabatic limit of \cref{eq_tracer_dyn_simpl}: by using \cref{eq_phi_mode_adiab} in the expression for $\Pi_h$ in \cref{eq_Pi_h_effnoise}, we obtain, to leading order in $\chi$, a Gaussian Markovian (white) noise
\beq \Pi_h(R(t),t) \simeq h \sqrt{\chi} \sump_n \frac{1}{k_n^{1+2a} } \tilde\sigma_n(R(t)) \xi_n(t) ,
\label{eq_tracer_pass_ad_noise}\eeq 
with correlations given by \cref{eq_m_func,eq_Pi_h_correl}, i.e.,
\beq \bra \Pi_h(R(t),t) \Pi_h(R(t'),t')\ket = 2 T_R \chit \kappa_h  m(R) \delta(t-t'),
\label{eq_tracer_pass_ad_noise_correl} 
\eeq 
which are consistent with \cref{eq_FP_pass_h_diff}.
However, $\Pi_h$ is also a multiplicative noise and thus requires specifying a stochastic integration rule \cite{gardiner_stochastic_2009,volpe_effective_2016}. 
The form of the spurious drift in \cref{eq_FP_pass_h_drift} suggests a \emph{S}tratonovich convention, such that the Langevin equation associated with \cref{eq_FP_pass_h_adiab} takes the form (see \cref{app_spurious})
\beq  
\pd_t R = -U_h'(R) + \sqrt{D(R)} \overset{\mathrm{S}}{\circ} \theta,\qquad \bra\theta(t)\theta(t')\ket = 2\delta(t-t'),
\label{eq_lang_pass_h}\eeq 
where $\theta$ is a Gaussian white noise of zero mean.

\subsubsection{Weak-coupling approximation}
\label{sec_passive_h1_weakcpl}

In a complementary approach to the adiabatic approximation [\cref{sec_pass_lin_adiab}], we determine here a perturbative solution in terms of $h$ for $\bar P(R,t)$ [\cref{eq_P_margin}].
As it turns out, one can thereby capture certain non-Markovian effects neglected in the adiabatic limit.
To this end, we regard the first term on the r.h.s.\ of \cref{eq_tracer_dyn} as a generic time-dependent force, which allows us to set up the following FPE for the ``reduced'' probability density $\hat P(R,t)=P(R,R_0,t,t_0)$ of the tracer \cite{gardiner_stochastic_2009,dean_diffusion_2011,demery_perturbative_2011}:
\beq \pd_t \hat P(R,t) = - \pd_R \hat J(R,t), 
\label{eq_FP}\eeq 
with the flux  
\beq \hat J(R,t)\equiv- T_R \pd_R \hat P(R,t) +  [h-c\phi(R,t)] [\pd_R \phi(R,t)] \hat P(R,t).
\label{eq_flux}\eeq 
Due to the reflective \bcs for the tracer, one has 
\beq \hat J\big|_{R\in \{0,L\}} = 0,
\label{eq_P_noflux}\eeq 
consistent with \cref{eq_P_cons}.
Note that the reduced distribution $\hat P$ is distinct from $\bar P$ in \cref{eq_P_margin}, since both $\hat P$ and $\hat J$ depend implicitly on the OP field $\phi(z,t)$, which will be specified below in terms of its correlations [see \cref{sec_phi_correl_pass}] \footnote{Note that the reduced distribution $\hat P$ is also different from the full distribution $P$ introduced in \cref{eq_FPE_funct}.}.
Averaging over the OP fluctuations described by \cref{eq_Pss_joint_pass}, relevant for the passive case, yields an approximation for $\bar P$:
\beq \bar P(R,t) \simeq \bra\hat P(R,t)\ket_\phi.
\label{eq_Pred_avg}\eeq 

We now take $c=0$ and $h_1=0$ and formally decompose the reduced distribution as $\hat P=\hat P_0 + \hat P_1 + \hat P_2+\ldots$, assuming $\hat P_i\sim \Ocal(h^i)$. (Note that the actual dimensionless control parameter is $\kappa_h$ [\cref{eq_effcoupl_h}], which can be made explicit by appropriate rescaling.)
Due to vanishing boundary fields, we have $\bra\phi\ket=0$. Accordingly, since \cref{eq_FP} (with $c=0$) is linear in $\phi$, the first non-trivial influence of $\phi$ arises at $\Ocal(h^2)$.
Inserting the expansion of $\hat P$ into \cref{eq_FP} and collecting terms of the same order in $h$, renders
\begin{subequations}
\begin{align}
(\pd_t - T_R\pd_R^2) \hat P_0 &= \delta(R-R_0)\delta(t-t_0), \label{eq_FP_P0} \\
(\pd_t - T_R\pd_R^2) \hat P_1 &= -h \, \pd_R \left[ (\pd_R \phi(R,t)) \hat P_0(R,t) \right], \label{eq_FP_P1} \\
(\pd_t - T_R\pd_R^2) \hat P_2 &= -h \, \pd_R \left[ (\pd_R \phi(R,t)) \hat P_1(R,t) \right]. \label{eq_FP_P2}
\end{align}\label{eq_FP_pass_h_iter}
\end{subequations}
\hspace{-0.13cm}The solution of \cref{eq_FP_P0} fulfilling the initial and the no-flux conditions of \cref{eq_P_noflux,eq_P_initcond} can be expressed in terms of Neumann modes [see \cref{eq_eigenf_Nbc}] as
\beq\begin{split} 
\hat P_0(R,R_0, t,t_0) &= \frac{1}{L} + \frac{2}{L} \sum_{n=1}^\infty \cos(k_n\Nbc R) \cos(k_n\Nbc R_0) \exp(-T_R (k_n\Nbc)^2 (t-t_0)) \\
&= \frac{1}{L} \sum_{n=-\infty}^\infty \cos(k_n\Nbc R) \cos(k_n\Nbc R_0) \exp(-T_R (k_n\Nbc)^2 (t-t_0)).
\end{split}\label{eq_P0_sol}\eeq 
Since $\bar P = \hat P_0$ [see \cref{eq_Pred_avg}] already validates \cref{eq_P_cons} at all $t$, we require 
\beq \int_0^L \d R\, \hat P_{i\geq 1}(R,t)=0.
\label{eq_P_i_cons}\eeq 
The solution of the inhomogeneous Fokker-Planck equation in \cref{eq_FP_P1} can be stated in terms of the associated Green's function $G$ (determined below) as
\beq \hat P_1(R,R_0, t,t_0) = -h  \int_0^L \d y \int_{t_0}^t \d s\, G(R, y, t-s) \pd_y\left[ (\pd_y \phi(y,s)) \hat P_0(y,R_0, s, t_0) \right] .
\label{eq_P1_h}\eeq 
We assume no-flux \bcs [\cref{eq_P_noflux}] to apply also to $\hat P_1$. Together with \cref{eq_P_i_cons}, this implies that the Green's function takes the same form as $\hat P_0$ in \cref{eq_P0_sol} except for the absence of the term with $n=0$:
\beq G(R,R_0,t-t_0) = \frac{2}{L} \sum_{n=1}^\infty \cos(k_n\Nbc R) \cos(k_n\Nbc R_0) \exp\left[-T_R (k_n\Nbc)^2 (t-t_0)\right].
\label{eq_Green_func}\eeq 
We remark that the completeness relation in \cref{eq_eigenf_complete} implies 
\beq G(R,R_0,t\to t_0) = \delta(R-R_0) - \frac{1}{L}.
\eeq
The solution of \cref{eq_FP_P2} follows in an analogous way:
\beq \hat P_2(R,R_0,t,t_0) = h^2   \int_0^L \d y \int_0^L \d z \int_{t_0}^t \d s \int_{t_0}^s \d u\, G(R,y,t-s) \pd_y \left\{ G(y,z, s-u) \pd_z \left[ (\pd_y \phi(y,s)) (\pd_z\phi(z,u)) \hat P_0(z,R_0,u,t_0) \right] \right\}.
\eeq 

Upon averaging over the fluctuations of $\phi$ [see \cref{eq_Pred_avg}], we obtain the distributions 
\begin{subequations}
\begin{align}
\bar P_1 &= 0, \label{eq_P1_sol} \\
\bar P_2(R,R_0,t,t_0) &= h^2  \int_0^L \d y \int_0^L \d z \int_{t_0}^t \d s \int_{t_0}^s \d u\, G(R,y,t-s) \pd_y \left\{ G(y,z, s-u) \pd_z \left[ C_{\pd\phi}(y,z,s-u) P_0(z,R_0,u,t_0) \right] \right\}, \label{eq_P2_sol}
\end{align}
\end{subequations}
with $C_{\pd\phi}$ defined in \cref{eq_C_phi_deriv}.
By means of \cref{eq_C_per_ND_rel}, the expression of \cref{eq_P2_sol} for periodic \bcs on $\phi$ can be obtained from the corresponding expressions for Neumann and Dirichlet \bcs as 
\beq \bar P_2\pbc(R,R_0,t,t_0) = \frac{1}{2} \left[\bar P_2\Nbc(R,R_0,t,t_0) + \bar P_2\Dbc(R,R_0,t,t_0) \right]_{L/2}.
\label{eq_P2avg_pbc}\eeq 
In order to evaluate \cref{eq_P2_sol}, we use \cref{eq_P0_sol,eq_Green_func} and subsequently perform the spatial and temporal integrations. For Neumann \bcs, the resulting expression is convergent and simplifies to three infinite sums over the mode indices, which can be efficiently evaluated numerically owing to the exponential term.
For Dirichlet \bcs, instead, one is left with four infinite sums [which can be reduced to three sums in the late-time limit, see \cref{eq_P2_latetime_D}] and $\bar P_2$ grows logarithmically with the summation cutoff. 
This divergence appears to be an artifact of the perturbative approach, as it turns out that the shape of $\bar P$ still qualitatively describes the numerical results (see \cref{sec_sim}).

For simplicity, we henceforth set $t_0=0$ [see \cref{eq_P_initcond}].
In order to determine the late time limit of $\bar P_2$, we reorder the time integration variables in \cref{eq_P2_sol} to obtain
\beq \bar P_2(R,R_0,t) = h^2  \int_0^L \d y \int_0^L \d z \int_0^t \d v \int_0^{t-v} \d w\, G(R,y,v) \pd_y \left\{ G(y,z,w) \pd_z \left[ C_{\pd\phi}(y,z,w) P_0(z,R_0,t-v-w) \right]\right\}.
\eeq 
Owing to their exponential time dependence, the functions $G$ [\cref{eq_Green_func}] and $C_{\pd\phi}$ [\cref{eq_C_phi_deriv}] are essentially nonzero only for times $v,w\ll  k\st{min}^{-2-2a} \sim L^{2+2a}$ \footnote{More precisely, we require $v,w\ll \min(k\st{min}^{-2}/T_R, k\st{min}^{-2+2a} \chi)$}, where $k\st{min}$ is the minimal wavenumber in the system and we recall that $a=0$ ($a=1$) for dissipative (conserved) OP dynamics.
When considering the limit $t\to\infty$, the integrand thus gives substantial contributions only for $v\ll t$ and $w\ll t-v$. Accordingly, we may replace $P_0$ [\cref{eq_P0_sol}] by its late-time limit,
\beq \bar P_0|_{t\to\infty} = \frac{1}{L},
\label{eq_P0_latetime}\eeq 
which is independent of position.
This allows us to extend the upper limit of the $w$ integral from $t-v$ to $t$, resulting in 
\beq \bar P_2(R,R_0,t\to\infty) = \frac{h^2}{L}  \int_0^L \d y \int_0^L \d z \int_0^\infty \d v \int_0^{\infty} \d w\, G(R,y,v) \pd_y \left\{ G(y,z,w) \pd_z \left[ C_{\pd\phi}(y,z,w)  \right]\right\}.
\label{eq_P2_latetime_gen}\eeq 
The time and space integrals can be performed explicitly, rendering, for Neumann \bcs on the OP: 
\beq\begin{split}
\bar P_2\Nbc(R,R_0,t\to\infty) &= \frac{\kappa_h}{2 L \pi^2 } \sum_{l=1}^\infty \frac{\cos(2 \pi l R/L)}{l^2 (1+\frac{\pi^{2a}}{L^{2a} T_R \chi} l^{2a})},
\label{eq_P2_latetime_Neu}\end{split}\eeq 
with the effective dimensionless coupling constant $\kappa_h$ defined in \cref{eq_effcoupl_h}.
The weak-coupling approximation applies to the regime $\kappa_h \ll 1$. 
For Dirichlet \bcs (which are relevant only for $a=0$), instead, we obtain
\begin{multline} 
\bar P_2\Dbc(R,R_0,t\to\infty) = \frac{\kappa_h}{L \pi^4 } \sum_{l=1}^N \sum_{m=1}^N  \sum_{k=1}^N \left[ \frac{k}{k-m} \Jcal_{k-m} + \frac{k}{k+m}\Jcal_{k+m}\right] \\ \times \left[\frac{k-m}{k-m-l} \Jcal_{k-m-l} + \frac{k-m}{k-m+l} \Jcal_{k-m+l} + \frac{k+m}{k+m-l} \Jcal_{k+m-l} + \frac{k+m}{k+m+l} \Jcal_{k+m+l}\right] \frac{\cos( \pi l R/L)}{l^2 [m^2+ k^2/(\chi T_R) ]},
\label{eq_P2_latetime_D}\end{multline} 
where the function $\Jcal_n$ arises from the spatial integrals and is given by 
\beq \Jcal_n = \begin{cases} \displaystyle
            1, &\qquad \text{$n$ odd} \\
            0, &\qquad \text{$n$ even}
           \end{cases}
\eeq 
and $N$ is a cutoff on the number of modes.
While the sum in \cref{eq_P2_latetime_Neu} can be evaluated analytically, we resort to a numerical calculation of \cref{eq_P2_latetime_D}. 
In total, combining \cref{eq_P0_latetime,eq_P2_latetime_Neu,eq_P2_latetime_D}, we obtain the steady-state distributions
\begin{subequations}
\begin{align}
\bar P_s\Nbc(R) &= \frac{1}{L} + 
 \begin{cases} \displaystyle \frac{\kappa_h}{2 L [1+1/(\chi T_R)]} \left(\frac{1}{6}- \rho + \rho^2\right), \qquad & (a=0) \\ \displaystyle
  \frac{\kappa_h }{2 L } \left[\frac{1}{6}- \rho + \rho^2 +\frac{\pi^{2a}}{2 \chit} - \frac{\pi^{a}}{2 \sqrt{\chit}} \frac{\cosh\left( (1-2\rho)\chit^{1/2}/\pi^{a}\right)}{\sinh(\chit^{1/2}/\pi^{a})} \right], & (a=1)
 \end{cases}
  \label{eq_Pss_h_pass_dom_N} \\
\bar P_s\Dbc(R) &= \frac{1}{L} + \frac{\kappa_h }{L} \sum_{l=1}^N \beta_l(N) \cos(2 \pi l R/L) , 
\label{eq_Pss_h_pass_dom_D}\end{align}\label{eq_Pss_h_pass_dom}
\end{subequations}
\hspace{-0.1cm}with $\rho\equiv R/L$ and $\chit$ given in \cref{eq_adiab_param}. The numbers $\beta_l(N)$ have to be determined numerically from \cref{eq_P2_latetime_D} and are found to decay exponentially with increasing $l$ (for large $l$) and grow logarithmically with $N$.
The corresponding distribution for periodic \bcs can be obtained via \cref{eq_P2avg_pbc}.
We remark that, for $\chit\to 0$, the distributions in \cref{eq_Pss_h_pass_dom} become uniform, $\bar P_s\to 1/L$. This can alternatively be shown by inserting the adiabatic approximation for $C_{\pd\phi}$ reported in \cref{eq_C_phi_deriv_ad} into \cref{eq_P2_latetime_gen}.

\subsubsection{Discussion}

\begin{figure}[t]\centering
    \subfigure[]{\includegraphics[width=0.33\linewidth]{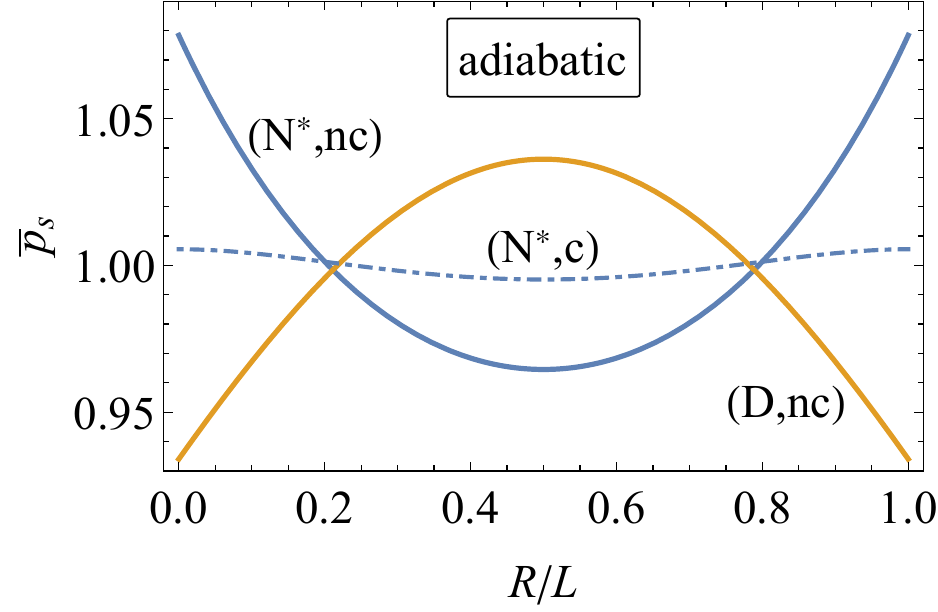} \label{fig_Pss_pass_h_ad} } \qquad
    \subfigure[]{\includegraphics[width=0.34\linewidth]{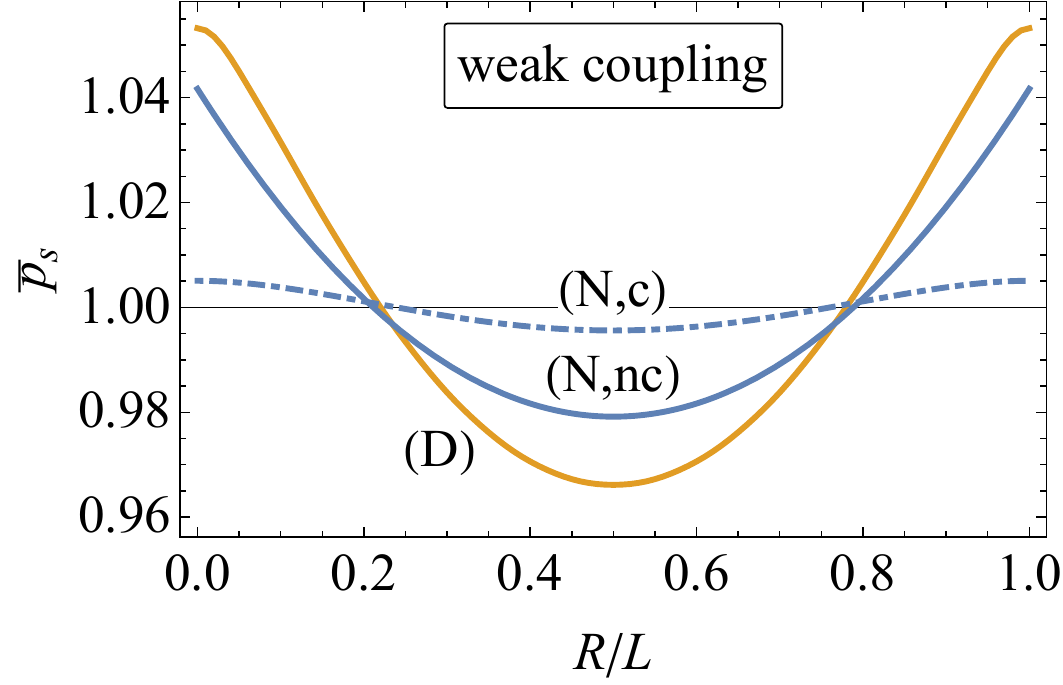} \label{fig_Pss_pass_h_wc} }
    \caption{Steady-state distribution $\bar p_s = L \bar P_s$ of a linearly coupled passive tracer obtained within (a) the adiabatic approximation [\cref{eq_Pss_h_pass_adiab}] and (b) the weak-coupling approximation [\cref{eq_Pss_h_pass_dom}]. The effective coupling constants [see \cref{eq_adiab_param,eq_effcoupl_h}] are set to $\tilde \chi=\kappa_h=1$ for illustrative purposes.  The labels indicate the boundary conditions [see \cref{eq_eigenspec}] as well as the \emph{n}on-\emph{c}onserved or \emph{c}onserved character of the OP dynamics [see \cref{eq_field_dyn_bare}]. In the weak-coupling approximation, the amplitude of $\bar P_s\Dbc$ depends on the mode number $N$ and diverges logarithmically with $N$. A value of $N=25$ is used to evaluate $\bar P_s\Dbc$ in panel (b).}
    \label{fig_Pss_pass_h}
\end{figure}

\begin{figure}[t]\centering
    \subfigure[]{\includegraphics[width=0.32\linewidth]{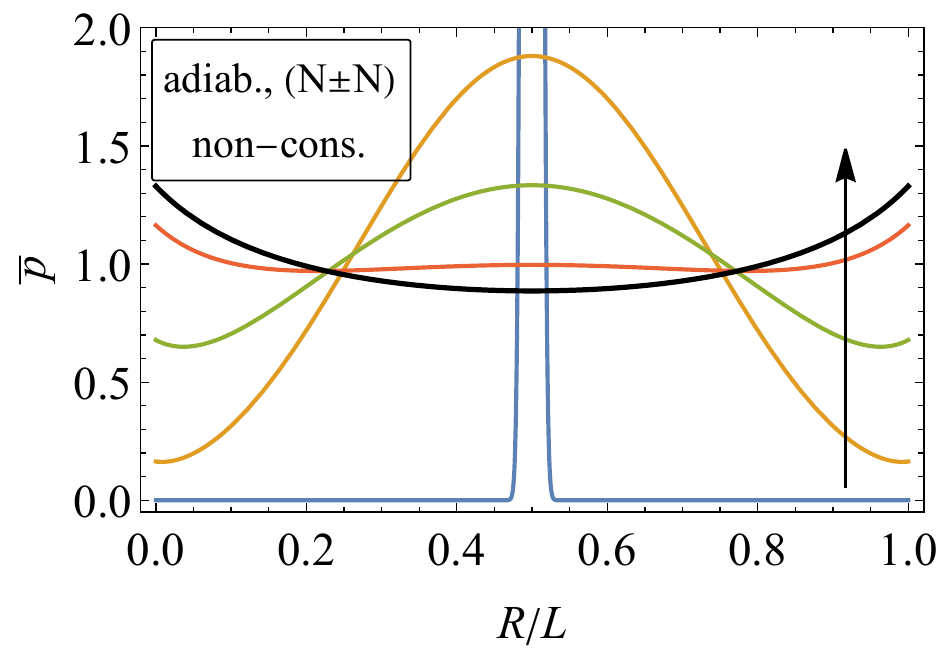} \label{fig_P_pass_lin_NhN}}\qquad
    \subfigure[]{\includegraphics[width=0.32\linewidth]{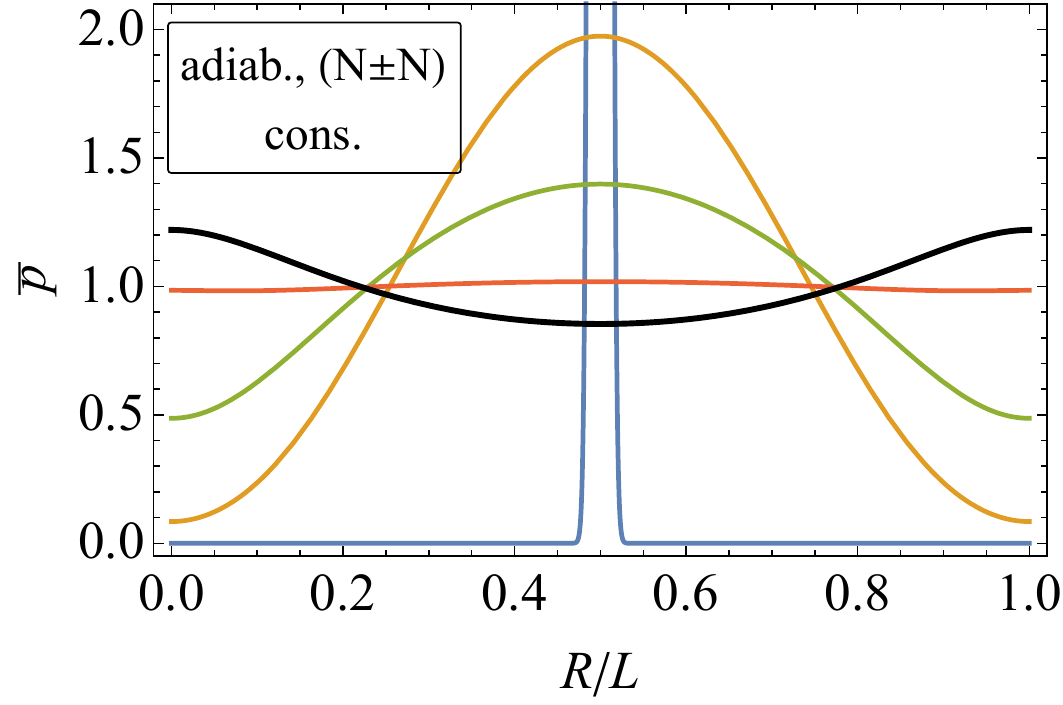} } 
    \caption{Time evolution of the probability density $\bar p = L\bar P(R,R_0=L/2,t,t_0=0)$ [solid curves, corresponding to various times, with initial condition $P(R,R_0,t_0,t_0)=\delta(R-R_0)$] for a linearly coupled passive tracer in the adiabatic limit, obtained from the FPE in \cref{eq_FP_pass_h_adiab}. The OP is subject to Neumann \bcs and follows non-conserved (a) or conserved dynamics (b). For illustrative purposes, we use a value $\chit \kappa_h=5$ in (a) and 50 in (b). In both panels, the curves correspond to times $t T_R/L^2 = 10^{-5}, 10^{-2}, 2\times 10^{-2}, 4\times 10^{-2}$ (in the direction of the arrow) as indicated by the color coding. The thick black lines represent the steady-state distributions $\bar P_s(R)$ given in \cref{eq_Pss_h_pass_adiab}. }
    \label{fig_P_pass_lin}
\end{figure}

\Cref{fig_Pss_pass_h} illustrates the steady-state distribution $\bar P_s(R)$  for a linearly coupled passive tracer obtained within the adiabatic [\cref{eq_Pss_h_pass_adiab}, panel (a)] and the weak-coupling approximation [\cref{eq_Pss_h_pass_dom}, panel (b)]. 
Here and in the following, we use a notation such as (D$\pm$D) to indicate a tracer linearly coupled to a OP field obeying Dirichlet \bcs. 
For an OP subject to Neumann \bcs, both the adiabatic and the weak-coupling approximation yield an increased occupation probability at the boundaries and a reduction at the center of the system.
For Dirichlet \bcs, in contrast, the two approaches predict opposite behaviors.
Interestingly, numerical simulations indicate that both approximations describe certain aspects of the actual tracer distribution (see \cref{sec_sim} for further discussion).
In particular, the attraction of the tracer towards the wall is generically expected for a confined non-Markovian process. 
In the case of capillary \bcs, the steady-state distribution [see \cref{eq_Pss_h_pass_adiab0}] is governed by the effective potential $U_h$ [\cref{eq_Pi_h_effpot}], implying attraction (repulsion) between wall and tracer if $h$ and $h_1$ have equal (opposite) signs [see \cref{fig_effpot_h}].

\Cref{fig_P_pass_lin} illustrates the time evolution of the tracer distribution $\bar P(R,t)$ in the adiabatic limit, as obtained from \cref{eq_FP_pass_h_adiab} for Neumann \bcs of the OP. The time evolution of $\bar P$ is found to be slightly faster for dissipative (a) than for conserved dynamics (b). The dissipative dynamics for Dirichlet \bcs follows analogously, but does not exhibit any new features and is thus not shown. 

If periodic \bcs are imposed on the OP, \cref{eq_FP_pass_h_coeffs} reduces to $\mu=0$ and $D=T_R$, such that \cref{eq_FP_pass_h_adiab} describes a simple diffusion process subject to reflective \bcs, which has $\bar P_s=1/L$ as steady-state solution. However, numerical simulations (not shown) reveal this to be an artifact of the present order of the adiabatic approximation. In fact, within the weak-coupling approximation, \cref{eq_P2avg_pbc} implies that the behavior of the tracer is similar for periodic, Neumann or Dirichlet \bcs.
The adiabatic and the weak-coupling approximations need not necessarily agree in the limit $\chit\ll 1$, because, in the former approach the field $\phi$ is interpreted as a noise, whereas in the latter, it is formally regarded as a time-dependent potential.

\subsection{Quadratic tracer-field coupling}
\label{sec_passive_quadr}

\subsubsection{Adiabatic approximation}
\label{sec_passive_quadr_ad}

In the case of a quadratically coupled passive tracer, the Langevin equation in \cref{eq_tracer_dyn_simpl} (with $h=0$) reduces to 
\beq \dot R(t) =  - \pd_R U_c(R(t)) + \Pi_c(R(t),t) + \eta(t),
\label{eq_langevin_pass_c}\eeq 
where the effective potential $U_c$ and the noise $\Pi_c$ are reported in \cref{eq_Pi_c_effpot,eq_Pi_c_effnoise}, respectively.
We apply the adiabatic elimination procedure described in Refs.\ \cite{stratonovich_topics_1963,gardiner_stochastic_2009} in order to obtain an effective FPE for the tracer distribution $\bar P(R,t)$.
We furthermore assume the coupling $c$ to be small, such that $\Pi_c$ can be approximated as a Gaussian Markovian white noise with correlation (see \cref{sec_pass_c_adiab})
\beq \bra \Pi_c(R(t),t) \Pi_c(R(t'),t')\ket \simeq 2\Pcal(R)\delta(t-t'),
\label{eq_correl_adiab_noise}\eeq 
where the amplitude $\Pcal\sim \Ocal(c^2 \chi)$ is specified in \cref{eq_correl_adiab_modesum}.
Within these approximations, the FPE associated with \cref{eq_langevin_pass_c} follows as (see \S 4.8 in Ref.\ \cite{stratonovich_topics_1963}) 
\beq \pd_t \bar P(R,t) = - \pd_R \left[\mu(R) \bar P(R,t) \right] + \pd_R^2 \left[ D(R) \bar P(R,t) \right],
\label{eq_FP_pass_c_adiab}\eeq 
with the effective drift and diffusion coefficients
\begin{subequations}
\begin{align}
\mu(R) &\equiv -\pd_R U_c(R) + \onehalf \pd_R \Pcal(R), \label{eq_FPE_pass_c_driftCoeff} \\
D(R) &\equiv T_R+\Pcal(R) . \label{eq_FPE_pass_c_diffCoeff} 
\end{align}\label{eq_FPE_pass_c_coeffs}
\end{subequations}
The spurious drift term $(1/2)\Pcal'(R)$ in \cref{eq_FPE_pass_c_driftCoeff} indicates a  Stratonovich character of the noise $\Pi_c$ in \cref{eq_langevin_pass_c} \cite{gardiner_stochastic_2009}.
The steady-state distribution resulting from \cref{eq_FP_pass_c_adiab} is given by [see also \cref{eq_app_Pss}]
\beq \bar P_s(R) = \frac{1}{\Zcal} e^{-\Vcal(R)},\qquad 
\Vcal(R) \equiv \ln D(R) - \int_0^R \d z \frac{\mu(z)}{D(z)} \simeq  \frac{c}{2 T_R} \left[ V_\phi(R) + \bra\phi(R)\ket_{h_1}^2\right] + \Ocal(c^2 \chi)  = \frac{U_c(R)}{T_R},
\label{eq_Pss_pass_c_ad}\eeq 
where we expanded the potential $\Vcal$ up to $\Ocal(c \chi^0)$ and disregarded any $R$-independent constants. 
Specializing $\bar P_s$ to the various \bcs results in ($\rho\equiv R/L$)
\beq \bar P_s(R)  = \begin{cases} 
		     \frac{1}{L}, \qquad & \text{(p$^*$)} \\ 
                    \frac{\sqrt{\kappa_c} } {\sqrt{2\pi}\, L\, \mathrm{erfi}\left(\frac{\sqrt{\kappa_c}}{2\sqrt{2}}\right)} \exp\left[ \frac{\kappa_c}{2} \left(\frac{1}{2}- \rho\right)^2 \right], \qquad & \text{(D)}  \\ 
                    \frac{\sqrt{\kappa_c} }{\sqrt{2\pi}\, L\, \mathrm{erf}\left(\frac{\sqrt{\kappa_c}}{2\sqrt{2}}\right)} \exp\left[ - \frac{\kappa_c}{2} \left(\frac{1}{2}- \rho \right)^2 \right],  \qquad & \text{(N$^*$)} \\
                    \frac{1}{\Zcal} \exp\left[-\sfrac{U_c(R)}{T_R} \right], \qquad &\text{($\pm$)}
                   \end{cases}
\label{eq_Pss_pass_c_avg}\eeq 
with the dimensionless effective coupling 
\beq \kappa_c \equiv \frac{c L T_\phi}{T_R}.
\label{eq_effcoupl_c}\eeq 
The above assumption of a weak coupling implies that $\bar P_s$ in \cref{eq_Pss_pass_c_avg} is correct only to $\Ocal(\kappa_c)$ [see also \cref{eq_Pss_pass_c} below].
Note that \cref{eq_Pss_pass_c_avg} is independent of the parameter $a$ [see \cref{eq_field_dyn_bare}], as the effect of the conservation law enters at $\Ocal(\kappa_c^2)$ [see \cref{eq_correl_adiab_modesum}].

\subsubsection{Weak-coupling approximation}
\label{sec_passive_quadr_weakcpl}

Here, analogously to \cref{sec_passive_h1_weakcpl}, we develop a perturbative solution of the tracer dynamics by assuming a weak coupling $c$ [the actual dimensionless control parameter being $\kappa_c$, see \cref{eq_effcoupl_c}].
We thus consider \cref{eq_FP} with $h=0$ and formally expand the reduced tracer distribution as $\hat P = \hat P_0 + \hat P_1 + \hat P_2 + \ldots$, with $\hat P_i \sim \Ocal(c^i)$.
Upon inserting this expansion into \cref{eq_FP}, one obtains the hierarchy 
\begin{subequations}
\begin{align}
(\pd_t - T_R\pd_R^2) \hat P_0 &= \delta(R-R_0)\delta(t-t_0), \label{eq_FP_c_P0} \\
(\pd_t - T_R\pd_R^2) \hat P_1 &= c \, \pd_R \left[\phi(R,t) (\pd_R \phi(R,t)) \hat P_0(R,t) \right] ,\label{eq_FP_c_P1} \\
(\pd_t - T_R\pd_R^2) \hat P_2 &= c \, \pd_R \left[\phi(R,t) (\pd_R \phi(R,t)) \hat P_1(R,t) \right] .\label{eq_FP_c_P2}
\end{align}
\end{subequations}
The solution for $\hat P_0$ is given in \cref{eq_P0_sol}.
For the leading correction, we obtain, analogously to \cref{eq_P1_h}:
\beq \hat P_1(R,R_0, t,t_0) = c  \int_0^L \d y \int_{t_0}^t \d s\, G(R, y, t-s) \pd_y\left[ \phi(y,s) (\pd_y \phi(y,s)) \hat P_0(y,R_0, s, t_0) \right] ,
\label{eq_P1_c}\eeq 
with the Green function $G$ reported in \cref{eq_Green_func}. The $\Ocal(c^2)$ contribution to $\hat P$ follows along the same lines from \cref{eq_FP_c_P2}.
Averaging over the fluctuations of $\phi$ [see \cref{eq_Pred_avg}] and using their time-translation invariance, renders  
\begin{subequations}
\begin{align}
\bar P_1(R,R_0, t,t_0) &= \onehalf c \int_0^L \d y \int_{t_0}^t \d s\, G(R, y, t-s) \pd_y\left[  P_0(y,R_0, s, t_0) \pd_y U_c(y) \right], \label{eq_P1_c_pass_avg} \\
\bar P_2(R,R_0, t,t_0) &= \int_0^L \d y \int_0^L \d z \int_{t_0}^t \d s \int_{t_0}^t \d u \, G(R, y, t-s) \pd_y\left\{ G(y,z,s-u) \pd_z \left[ \left\bra \Xi_c(y,s)\Xi_c(z,u)\right\ket P_0(z,R_0, u, t_0) \right]\right\}, \label{eq_P2_c_pass_avg}
\end{align}\label{eq_P_c_pass_avg}
\end{subequations}
\hspace{-0.1cm}with the effective potential $U_c$ stated in \cref{eq_Pi_c_effpot} and $\bra\Xi_c(y,s) \Xi_c(z,u)\ket = \bra\Pi_c(y,s)\Pi_c(z,u)\ket + \bra\Xi_c(y)\ket\bra\Xi_c(z)\ket$ [see \cref{eq_Pi_c_effnoise,eq_Pi_c_correl}].

While \cref{eq_P_c_pass_avg} has to be evaluated numerically in the general case, the steady-state distribution $\bar P_s$ up to $\Ocal(c)$ can be readily determined from \cref{eq_FP_c_P0,eq_FP_c_P1}: the former yields $\bar P_{\text{s},0} = \sfrac{1}{L}$, whereas the latter reduces, after averaging over the field, to [see \cref{eq_Pi_c_effpot}]
\beq 
T_R\pd_R^2 \bar P_{\text{s},1} = -\frac{1}{L}  \pd_R^2  U_c(R) .
\label{eq_FP_c_P1_ss} 
\eeq 
Upon imposing the no-flux condition [\cref{eq_P_noflux}], we obtain
\beq \bar P\st{s} = \frac{1}{L} \left[ \alpha - \frac{U_c(R)}{T_R} \right]  = \begin{cases} 
		     \frac{1}{L}, \qquad & \text{(p)} \\ 
             \frac{1}{L}\left[1 + \frac{\kappa_c}{2} \left(\frac{1}{6} - \rho + \rho^2 \right)\right], \qquad & \text{(D)}  \\ 
             \frac{1}{L}\left[1 - \frac{\kappa_c}{2} \left(\frac{1}{6} - \rho + \rho^2 \right)\right], \qquad & \text{(N$^*$)} \\
             \frac{1}{L}\left\{1 + \frac{\kappa_c H_1^2}{360} - \frac{\kappa_c}{2} \left(\frac{1}{6} - \rho + \rho^2 \right)\left[1+ H_1^2 \left(\frac{1}{6} - \rho + \rho^2 \right) \right]\right\}, \qquad & \text{($\pm$)}
                   \end{cases}
\label{eq_Pss_pass_c}\eeq 
where the integration constant $\alpha$ is fixed via \cref{eq_P_i_cons} and we introduced the dimensionless coupling 
\beq H_1\equiv h_1\sqrt{\frac{L}{T_\phi}} .
\label{eq_effcoupl_h1}\eeq
These expressions agree with the ones obtained from the long-time limit of \cref{eq_P1_c_pass_avg} (not shown) and from the expansion of \cref{eq_Pss_pass_c_avg} to $\Ocal(\kappa_c)$.
A dependence of $\bar P\st{s}$ on $\chi$ and $a$, i.e., the conservation law [see \cref{eq_field_dyn_bare}], is introduced at $\Ocal(\kappa_c^2)$ via the correlation function of $\Xi_c$ [\cref{eq_Pi_c_correl}] present in \cref{eq_P2_c_pass_avg}.

\subsubsection{Discussion}
\label{sec_discuss_pass_quad}

\begin{figure}[t]\centering
    \includegraphics[width=0.32\linewidth]{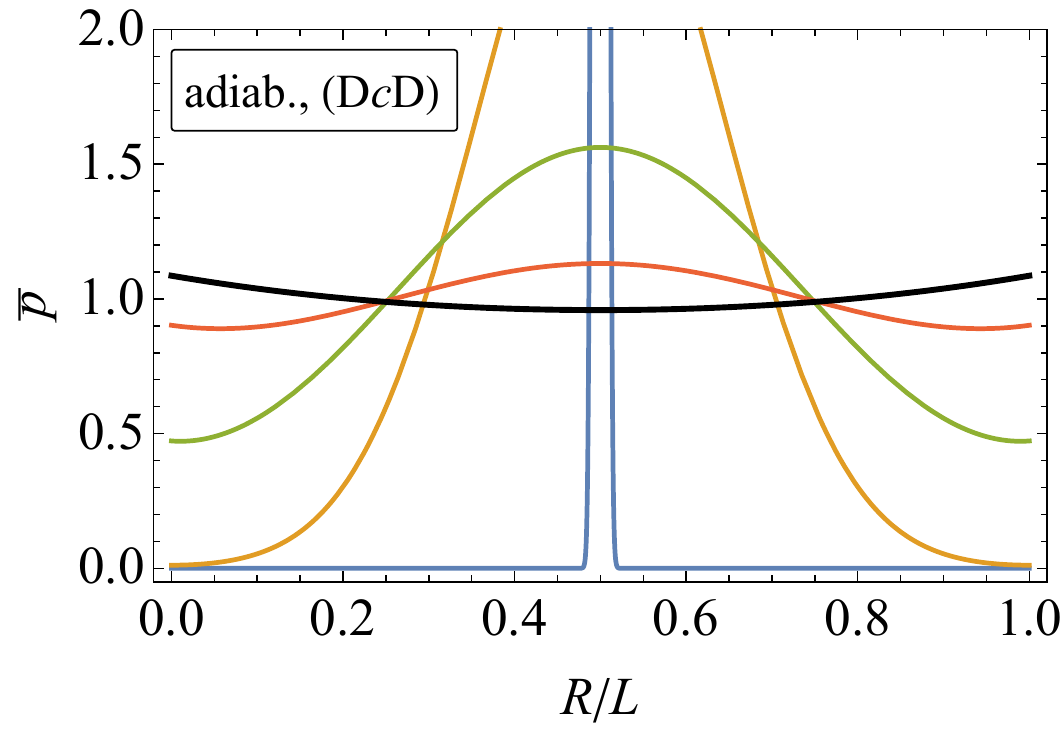} 
    \caption{Time evolution of the probability distribution $\bar P(R,R_0=L/2,t,t_0=0)$ [solid curves, corresponding to various times] at $\Ocal(c \chi^0)$ of a passive tracer in the adiabatic limit coupled quadratically to an OP with Dirichlet \bcs. The distributions are obtained by numerically solving the FPE in \cref{eq_FP_pass_c_adiab} with an initial condition $P(R,R_0,t_0,t_0)=\delta(R-R_0)$ and using an effective coupling $\kappa_c=1$ [\cref{eq_effcoupl_c}]. The thick black line represents the steady-state distribution $\bar P_s\Dbc(R)$ given in \cref{eq_Pss_pass_c_avg}, which to $\Ocal(\kappa_c)$ coincides with the weak-coupling expression in \cref{eq_Pss_pass_c}. By contrast, if the OP fulfills Neumann \bcs, the steady-state probability is enhanced in the center of the system and reduced at the boundaries [see \cref{eq_Pss_pass_c_ad} and \cref{fig_effpot}(b)].}
    \label{fig_P_pass_quadr}
\end{figure}

If non-symmetry-breaking \bcs are imposed on the OP, in both the adiabatic [\cref{eq_FP_pass_c_adiab}] and the weak-coupling approximation [\cref{eq_FP_c_P1}] the tracer follows, to leading order in $\kappa_c$ and $\chit$, a Brownian motion in a quadratic potential $U_c(R)$ subject to reflective \bcs [see \cref{eq_phiR_var,eq_tracer_dyn_simpl,fig_phiVar}]. For Neumann \bcs, this is, in fact, a confined Ornstein-Uhlenbeck process.
While the time-dependent solution for such processes can in principle be determined analytically \cite{linetsky_transition_2005}, the resulting expression is rather involved and we instead present in \cref{fig_P_pass_quadr} a numerical solution of \cref{eq_FP_pass_c_adiab}. 
Since the behavior of (N$c$N) and (D$c$D) is qualitatively similar, only the latter case is shown in the figure.

According to \cref{eq_Pss_pass_c_ad,eq_phiR_var}, at late times the occupation probability of the tracer is enhanced in the center (at the boundaries) of the system if the OP is subject to Neumann (Dirichlet) \bcs.
Notably, the dynamics of $\bar P$ in the case (D$c$D) is similar to the one for (N$\pm$N) [see \cref{fig_P_pass_lin_NhN}]; analogously, (N$c$N) is similar to (D$\pm$D). This is a consequence of the formal similarity between the expressions for $V_\phi$ [\cref{eq_phiR_var}] and $m(R)$ [\cref{eq_m_func}], which enter the FPEs in \cref{eq_FP_pass_h_adiab} and \cref{eq_FP_pass_c_adiab}, respectively.
We remark that, up to the order considered in \cref{eq_FP_c_P1_ss}, effects of the interaction between the tracer and the field at different times are neglected. In \cref{eq_FP_pass_c_adiab}, this is a direct consequence of the Markovian assumption for $\Pi_c$.
The uniform distribution for periodic \bcs predicted by both the adiabatic and the weak-coupling approach [see \cref{eq_Pss_pass_c_avg,eq_Pss_pass_c}] are considered to be an artifact of the low order of the approximations used here, as numerical simulations (not shown) reveal a non-uniform steady-state distribution. 

For capillary \bcs, the tracer behaves to leading order as a Brownian particle in an effective potential $U_c$ [see \cref{eq_Pi_c_effpot,fig_effpot}]. The latter is a fourth-order polynomial and shows a crossover from a unimodal shape ($|H_1|\ll 1$) with a minimum at the center of the system, to a bimodal shape ($|H_1|\gg 1$) with minima at 
\beq \frac{R_\pm}{L} = \frac{1}{2}\pm \frac{\sqrt{1-6/H_1^2}}{2\sqrt{3}}.\qquad (|H_1|\gg 1)
\label{eq_Pss_c_pass_h1_peakpos}\eeq

\section{Reactive tracer}
\label{sec_reactive}

We now turn to the discussion of a reactive tracer, which is described by \cref{eq_langevin_1d} with $\zeta=1$. 
Since \cref{eq_langevin_1d} satisfies in this case a fluctuation-dissipation relation for all degrees of freedom \cite{ma_modern_1976,onuki_phase_2002,tauber_critical_2014}, the resulting joint steady-state distribution for $\phi$ and $R$ is provided by \cref{eq_Pss_joint,eq_Hamilt}.
A reactive tracer represents a simplified model for a colloid in a fluid.
While previous studies of the equilibrium behavior of a single confined colloid considered spatially resolved particles in a half-space \cite{burkhardt_casimir_1995,eisenriegler_casimir_1995,hanke_critical_1998,schlesener_critical_2003,gambassi_critical_2009,hasenbusch_thermodynamic_2013} or in a slit geometry \cite{vasilyev_nonadditive_2018,kondrat_probing_2018,vasilyev_bridging_2020}, we focus here on a point-like particle under strong confinement. 
The confinement induces long-ranged interactions between the colloid and both boundaries (walls).
Further connections to previous studies are discussed in \cref{sec_discuss_react}.

In \cref{sec_react_equil}, we determine the equilibrium distribution of the tracer for various couplings and OP \bcs. 
Besides being interesting in its own right, the equilibrium distributions provide a means to independently check the steady-state solutions obtained from the effective FPEs derived in \cref{sec_react_dynamics}. While we focus here on a one-dimensional system, equilibrium distributions of a tracer in a three-dimensional slit are presented in \cref{app_steadyst_3d}.

\subsection{Equilibrium tracer distribution}
\label{sec_react_equil}

The equilibrium (steady-state) probability distribution for the tracer follows by marginalizing the joint distribution in \cref{eq_Pss_joint_T} over the OP field $\phi$ [see \cref{eq_P_margin}]:
\beq \bar P_s(R) = \int\Dcal\phi \, P_s(R,[\phi]) = \frac{1}{\Zcal} \int \Dcal\phi \exp\left\{ -\frac{1}{T_\phi} \left[\int_0^L \d z  \onehalf \left(\pd_z \phi\right)^2 - h_1 \left(\phi(0) + \phi(L)\right) \right] - \frac{1}{T_R} \left[ - h\phi(R) + \onehalf c \phi^2(R) \right] \right\},
\label{eq_Pss_pathint}\eeq 
where the normalization factor $\Zcal$ is determined by the requirement $\int_0^L \d R\, \bar P_s(R) = 1$ [see \cref{eq_P_cons}]. For generality, we keep here different temperatures for the tracer ($T_R$) and the field ($T_\phi$), although they essentially only rescale the coupling parameters.
Boundary fields $h_1$ give rise to an inhomogeneous mean profile $\bra\phi(z)\ket$ of the OP field [see \cref{eq_avg_prof_h1}]. 

Owing to the Gaussian nature of \cref{eq_Pss_pathint}, we can integrate out the field degrees-of-freedom by inserting the orthonormal transformation in \cref{eq_phi_expand}, which induces a unit Jacobian. 
The actual integration is performed over the modes $\phi_n$ with the aid of the following standard result for multidimensional Gaussian integration \cite{binney_theory_1992,le_bellac_quantum_1991}:
\beq \int \Dcal \phi \exp\left(-\onehalf \sum_{ij} \phi_i \Gamma_{i,j} \phi_j^* + \sum_i K_i^*\phi_i\right) = \frac{(2\pi)^{N/2}}{(\det \boldsymbol{\Gamma})^{1/2}}\exp\left(\onehalf \sum_{i,j} K_i \Gamma_{ij}^{-1} K_j^* \right),
\label{eq_gaussint}\eeq 
where $\boldsymbol{\Gamma}$ is a $N\times N$ matrix and $K_i$ is a given field.
Complex conjugation is relevant only for periodic \bcs, in which case the modes $\phi_n$ and $K_n$ are complex-valued with $\phi_{-n}=\phi_n^*$ (analogously for $K_n$) and the integration measure has to be suitably chosen \cite{binney_theory_1992,uzunov_introduction_1993}.
A possible zero mode, occurring for periodic and Neumann \bcs, can be regularized by replacing the vanishing eigenvalue $k_0$ [see \cref{eq_eigenspec}] by a nonzero parameter $\varepsilon$ \footnote{In the context of a Landau-Ginzburg model, a natural regularization is provided by a finite correlation length $\xi\sim \varepsilon^{-1/2}$, which can be achieved by adding the term $\varepsilon \phi^2$ to the integrand in \cref{eq_Hamilt}.}, which is set to zero in the end of the calculation [see also \cref{sec_matrix}].

\subsubsection{System without boundary fields}

\begin{figure}[t]\centering
    \subfigure[]{\includegraphics[width=0.31\linewidth]{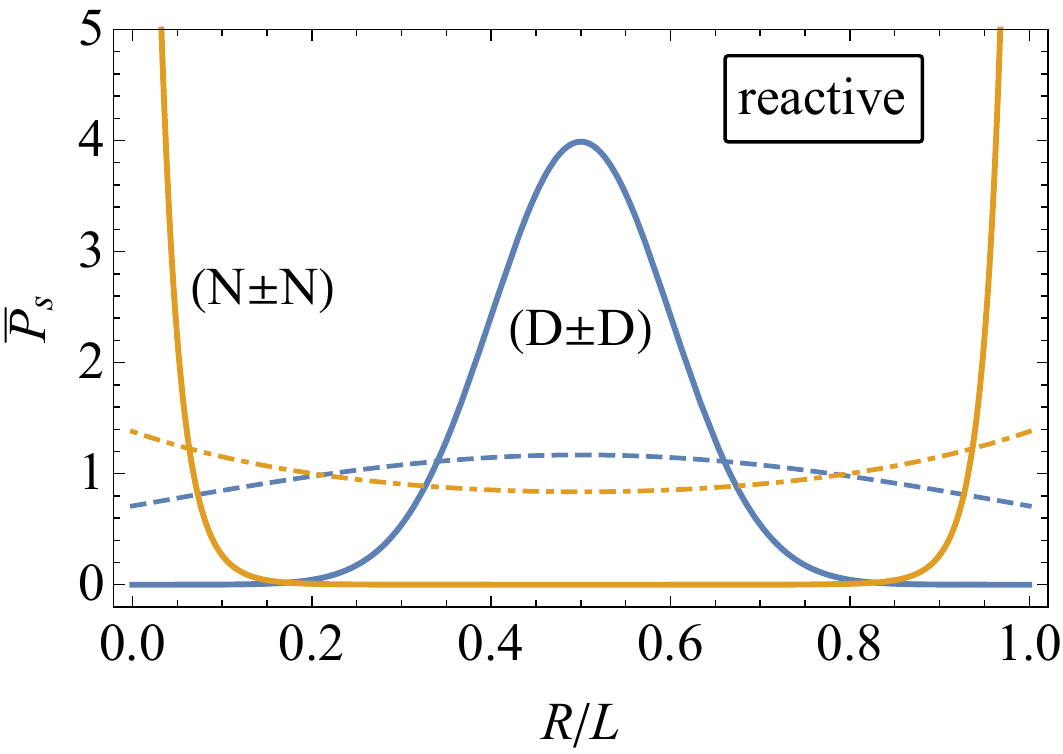} \label{fig_Pss_act_h}}\qquad
    \subfigure[]{\includegraphics[width=0.32\linewidth]{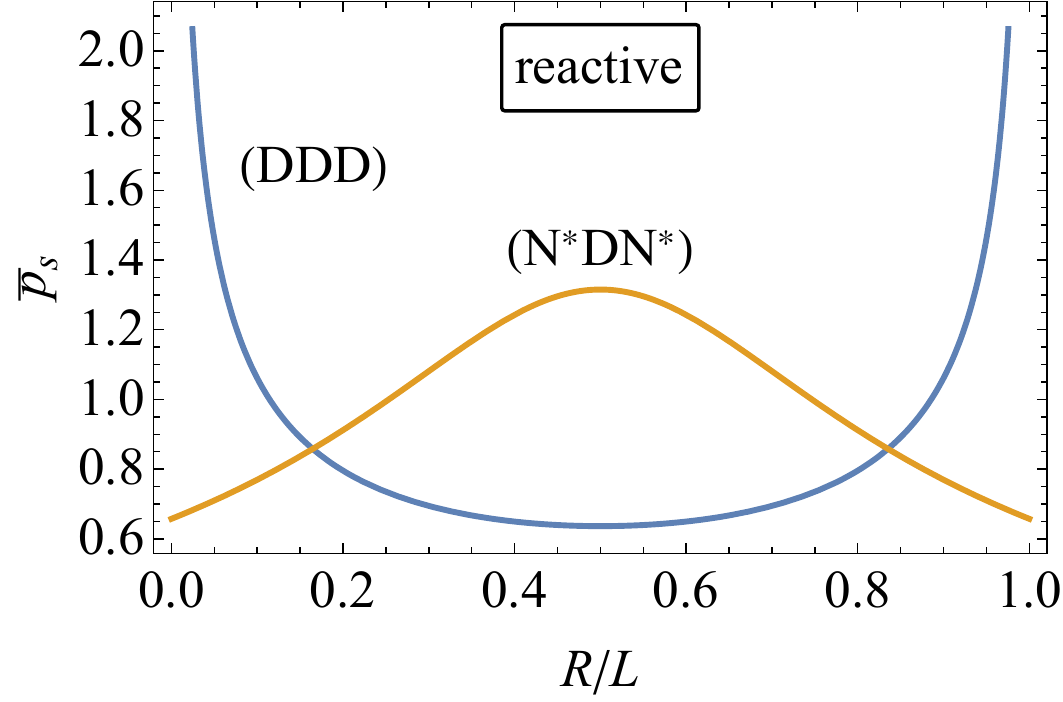} \label{fig_Pss_act_c}} 
    \caption{Equilibrium probability distribution $\bar P_s(R)$ of a reactive tracer in a finite interval (length $L$) coupled (a) linearly [\cref{eq_Pss_h_act}] or (b) quadratically [\cref{eq_Pss_c_act_largeC}] to an OP field subject to Dirichlet (D) or Neumann (N) \bcs. In (a), the solid (dashed) curves correspond to an effective coupling of $\kappa_h=10$ ($\kappa_h=1$) [see \cref{eq_effcoupl_h}]. In (b), a coupling $c\to\infty$ is used, which effectively imposes a Dirichlet boundary condition $\phi(R)=0$ at the location of the tracer. Neumann \bcs without a zero mode are denoted by (N$^*$). }
    \label{fig_Pss_act}
\end{figure}

We consider first a system without boundary fields, i.e., $h_1=0$.
In the case of a \emph{linear} coupling between OP and tracer (i.e., $h\neq 0$, $c=0$), \cref{eq_Pss_pathint} evaluates to 
\beq \bar P_s(R)\big|_{c=0} = \frac{1}{\Zcal}\exp\left[ \frac{ h^2 }{2 T_R^2} V_\phi(R) \right]  = \begin{cases}
	  \sfrac{1}{L}, \qquad & \text{(p)}\\ \displaystyle
	  \frac{\sqrt{\kappa_h}}{L \sqrt{2\pi}\, \mathrm{erf}\left(\frac{\sqrt{\kappa_h}}{2\sqrt{2}}\right)} \exp\left[- \frac{\kappa_h}{2} \left(\frac{1}{2} -  \rho\right)^2\right],  \qquad & \text{(D)}\\ \displaystyle
	   \frac{\sqrt{\kappa_h}}{L \sqrt{2\pi }\, \mathrm{erfi}\left(\frac{\sqrt{\kappa_h}}{2\sqrt{2}}\right)} \exp\left[ \frac{\kappa_h }{2}\left(\frac{1}{2} - \rho\right)^2\right], \qquad & \text{(N)}
	  \end{cases}
\label{eq_Pss_h_act}\eeq 
where $V_\phi(R)=\bra\phi^2\ket$ is reported in \cref{eq_phiR_var} and $\kappa_h$ in \cref{eq_effcoupl_h}.
The zero mode of the OP provides an $R$-independent constant in the exponential and thus plays no role for the distribution [this is different if $c\neq 0$, see \cref{eq_Pss_c_act_det} below].
In the limit $\kappa_h \to 0$, \cref{eq_Pss_h_act} reduces to $\bar P_s(R)\simeq 1/L$, as expected.
In the limit $\kappa_h \to\infty$, instead, using the asymptotic behavior $\mathrm{erfi}(x\to\infty) \simeq \sfrac{\exp(x^2)}{\sqrt{\pi} x}$, the tracer becomes highly localized for Neumann or Dirichlet \bcs on the OP:
\beq \bar P_s(R)\big|_{c=0,\kappa_h \to\infty} = \begin{cases}
                               \sfrac{1}{L}, \qquad & \text{(p)}\\ \displaystyle
                               \delta(R-L/2), \qquad & \text{(D)}\\ \displaystyle
                               \frac{1}{2}\left[ \delta(R) + \delta(L-R) \right].  \qquad & \text{(N)}                           
                              \end{cases}
\eeq 

\Cref{eq_Pss_h_act} is illustrated in \cref{fig_Pss_act}(a). 
If the OP is subject to Dirichlet \bcs, the probability of the tracer is enhanced at the center of the system and reduced at the boundaries, while the trends are opposite in the case of Neumann \bcs. 
These behaviors are qualitatively similar to a passive tracer in the adiabatic limit (see \cref{fig_Pss_pass_h_ad}). 
This can be intuitively explained based on the dynamical coupling $\pd_R\phi(R)$ [see \cref{eq_tracer_dyn}], which acts as an effective noise [see \cref{eq_tracer_dyn_simpl}] in the Langevin equation: for Neumann (Dirichlet) \bcs, this noise is strongest at the center (boundaries) and thus drives the tracer towards the boundaries (center). 
Note, however, that this effect can be overwhelmed by non-Markovian contributions, responsible for the generally strong attraction of a passive tracer towards the boundaries (see discussion in \cref{sec_sim}).

For a tracer \emph{quadratically} coupled to the OP (i.e., $c\neq 0$, $h=0$), \cref{eq_Pss_pathint} evaluates to 
\beq \bar P_s(R)\big|_{h=0} = \frac{1}{\Zcal} \left[\det \boldsymbol{\Gamma}(R) \right]^{-1/2}  = \begin{cases} \displaystyle
                               \frac{1}{L}, \qquad &\text{(p),(N)} \\ \displaystyle
                               \frac{1}{\Zcal \left[1+ \frac{c}{T_R} V_\phi(R) \right]^{1/2} }, & \text{(D),(N$^*$)} 
                  \end{cases}
\label{eq_Pss_c_act_det}\eeq 
where the matrix $\boldsymbol{\Gamma}$ is given by  \footnote{The complex conjugation is only relevant for periodic \bcs and follows by noting that $\sum_{n=-\infty}^\infty \sigma_n \phi_n =\sum_{n=-\infty}^\infty \sigma_{-n} \phi_{-n} = \sum_{n=-\infty}^\infty \sigma_n^* \phi_n^*$, see \cref{eq_eigenf_pbc}.}
\beq \Gamma_{n,m}(R) \equiv \frac{1}{T_\phi} k_n^2 \delta_{n,m} + \frac{c}{ T_R} \sigma_n(R)\sigma_m^*(R)
\label{eq_Gamma_mat}\eeq 
and the continuum limit of its determinant is calculated in \cref{sec_matrix}.
Upon normalization, \cref{eq_Pss_c_act_det} results in
\beq \bar P_s(R)\big|_{h=0} = \begin{cases} \displaystyle
                               \frac{1}{L}, \qquad &\text{(p),(N)} \\ \displaystyle
                               \frac{1}{2 L\, \mathrm{arctan}(\sqrt{\kappa_c}/2)\, \sqrt{\kappa_c^{-1} +\rho -  \rho^2}}, & \text{(D)} \\ \displaystyle
                               \frac{1}{L \sqrt{\kappa_c^{-1} + 1/3 - \rho + \rho^2 }\, \ln\left( \frac{12 + 7\kappa_c  + 4 \sqrt{3 \kappa_c (3+ \kappa_c )}}{12 + \kappa_c } \right) }, & \text{(N$^*$)} 
                              \end{cases}
\label{eq_Pss_c_act}\eeq
with the dimensionless coupling constant $\kappa_c$ in \cref{eq_effcoupl_c}. 
In the limit $\kappa_c \to\infty$, the tracer effectively imposes Dirichlet \bcs on the OP field, such that \cref{eq_Pss_c_act}, becoming independent of $\kappa_c$, reduces to
\beq \bar P_s(R)\big|_{h=0,\kappa_c\to\infty} 
= \begin{cases} \displaystyle
    \frac{1}{L}, \qquad &\text{(p),(N)} \\ \displaystyle
    \frac{1}{\pi L \sqrt{\rho (1-\rho)}}, & \text{(D)} \\ \displaystyle
     \frac{1}{\mathrm{arcosh(7)}\, L \sqrt{1/3- \rho + \rho^2}}. & \text{(N$^*$)} 
  \end{cases}
\label{eq_Pss_c_act_largeC}\eeq 
Accordingly, if the OP is subject to Dirichlet (Neumann) \bcs, a quadratically coupled tracer is most likely to be found at the boundaries (center) of the system, as illustrated in \cref{fig_Pss_act}(b). This behavior is similar to the passive case [see \cref{eq_Pss_pass_c_avg,fig_P_pass_quadr}]

\subsubsection{System with boundary fields}
\label{sec_surface_fields_eq}

\begin{figure}[t]\centering
    \subfigure[]{\includegraphics[width=0.315\linewidth]{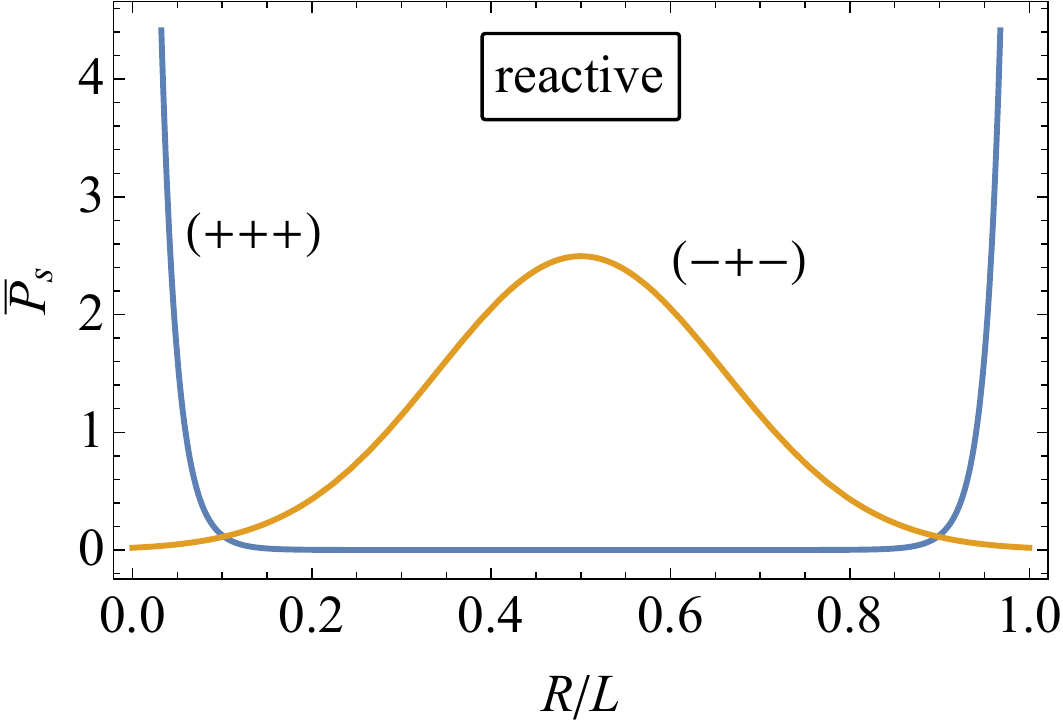} \label{fig_Pss_act_h1_h} }\qquad
    \subfigure[]{\includegraphics[width=0.315\linewidth]{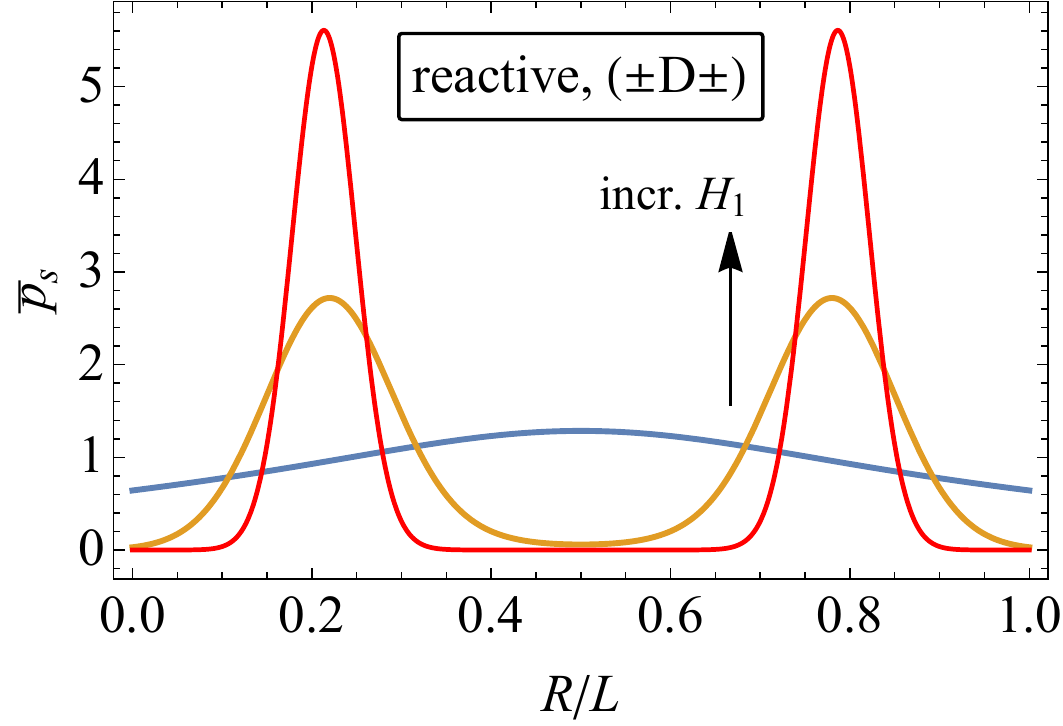} } 
    \caption{Equilibrium probability distribution $\bar P_s(R)$ of a reactive tracer in a finite interval (length $L$) coupled to an OP field subject to two boundary fields of equal strength $h_1$. In (a), tracer and OP field are coupled linearly (with coupling $h$) and $\bar P_s(R)$ is given by \cref{eq_Pss_h_act_h1_ppp,eq_Pss_h_act_h1_mpm}, corresponding to $h_1/h>0$ and $h_1/h<0$, respectively. In (b), tracer and OP field are coupled quadratically, with a coupling $c\to \infty$, implying Dirichlet \bcs (D) at the tracer location [see \cref{eq_Pss_c_act_h1_largeC}]. Upon increasing the value of the dimensionless boundary field strength $H_1$ [\cref{eq_effcoupl_h1}], $\bar P_s$ crosses over from a single- to a double-peaked shape [see \cref{eq_Pss_c_act_h1_peakpos}]. }
    \label{fig_Pss_act_h1}
\end{figure}

We assume now boundary fields of equal strength $h_1$ to act on the OP.
Upon performing in \cref{eq_Pss_pathint} the Gaussian integration over $\phi$ and omitting all $R$-independent constants, one obtains the following equilibrium distribution for a linearly coupled tracer ($c=0$):
\beq\begin{split} \bar P_s(R)\big|_{c=0} &= \frac{1}{\Zcal} \exp\left\{ \frac{T_\phi}{2 } \sum_{n=1} \frac{1}{k_n^2} \left[ \tilde h^2 \sigma_n^2(R)  + 2 \tilde h \tilde h_1 \sigma_n(R)\big( \sigma_n(0) +\sigma_n(L) \big)  \right] \right\} \\
&= \frac{1}{\Zcal} \exp\left\{ \frac{1}{2 } \tilde h^2 V_\phi\NbcNZM(R)  +  \tilde h \tilde h_1 \left[ C_\phi\NbcNZM(0,R) + C_\phi\NbcNZM(L,R) \right] \right\} 
\end{split}\label{eq_Pss_h_act_h1_base}\eeq 
with $\tilde h\equiv h/T_R$, $\tilde h_1 \equiv h_1/T_\phi$, and $C_\phi\NbcNZM$ and $V_\phi\NbcNZM$ reported in \cref{eq_phi_correl,eq_phiR_var}, respectively.
Note that the term in the last square brackets essentially simplifies to $\bra\phi(R)\ket_{h_1}$ [see \cref{eq_avg_prof_h1_correl}], such that this part of the distribution resembles \cref{eq_Pss_h_pass_adiab0} of the passive case.
The normalized tracer probability distribution is given by (with $\rho\equiv R/L$)
\beq \bar P_s(R) = \frac{\sqrt{\kappa_{h_1}}}{L \sqrt{2\pi}\, \mathrm{erfi}\left( \frac{\sqrt{\kappa_{h_1}}}{2\sqrt{2}} \right)} \exp\left[ \frac{\kappa_{h_1}}{2} \left(\frac{1}{2} - \rho\right)^2\right], \qquad \text{($+$$+$$+$),($-$$-$$-$)}
\label{eq_Pss_h_act_h1_ppp}\eeq
with the dimensionless effective coupling 
\beq \kappa_{h_1} \equiv T_\phi L \tilde h (\tilde h+2 \tilde h_1),
\label{eq_effcoupl_hh1}\eeq 
which generalizes \cref{eq_effcoupl_h}. 
We use a notation such as (+$-$+) to indicate the presence of two boundary fields with $h_1>0$ and a tracer field $h<0$.
When $\tilde h$ and $\tilde h_1$ are such that $\kappa_{h_1}>0$, $\bar P_s(R)$ is largest at the boundaries [see \cref{fig_Pss_act_h1_h}]. This attractive interaction between wall and tracer is, in fact, characteristic for an effective $(++)$ boundary condition between wall and tracer (see \cref{sec_discuss_react}). For large values of $\kappa_{h_1}$, one has $\bar P_s(R) \sim \exp \left(-T_\phi L \tilde h \tilde h_1 (1 - R/L)R/L\right)$, which implies, in particular, that $\bar P_s\simeq \onehalf\left[\delta(R) + \delta(L-R)\right]$ in the limit $h_1\to\infty$.

The case $\kappa_{h_1}<0$ corresponds to an effective $(-+)$ boundary condition between wall and tracer and applies if $h$ and $h_1$ have opposite signs and fulfill $|h|<2|h_1|$.
In this case, one obtains
\beq \bar P_s(R) = \frac{\sqrt{\kappa_{h_1}'}}{L \sqrt{2\pi}\, \mathrm{erf}\left( \frac{\sqrt{\kappa_{h_1}'}}{2\sqrt{2}} \right)} \exp\left[- \frac{\kappa_{h_1}'}{2} \left(\frac{1}{2} - \rho\right)^2\right], \qquad \text{($-$$+$$-$),($+$$-$$+$)}
\label{eq_Pss_h_act_h1_mpm}\eeq
with $\kappa_{h_1}' \equiv T_\phi L |\tilde h|(2 |\tilde h_1| - |\tilde h|)$.
This distribution has its maximum at $L/2$, indicating a repulsion between wall and tracer [see \cref{fig_Pss_act_h1_h}].
In fact, in the limit $\kappa_{h_1}'\to\infty$, the tracer is strongly localized at the center, $\bar P_s(R)\simeq \delta(R-L/2)$.
If the two boundary fields have opposite signs, we obtain asymmetric distributions, which, for sufficiently large $|h_1|$, are monotonous and have a maximum at one boundary.

In the case of a quadratically coupled tracer with non-vanishing boundary fields (i.e., $h=0$, $h_1\neq 0$), \cref{eq_Pss_pathint} yields
\beq\begin{split} \bar P_s(R)\big|_{h=0} = \frac{1}{\tilde \Zcal} \left[\det \boldsymbol{\Gamma}(R) \right]^{-1/2}  \exp\left\{ \frac{h_1^2 }{2 T_\phi^2} \sum_{n,m} \left[ \sigma_n(0) + \sigma_n(L) \right] \Gamma(R)^{-1}_{n,m} \left[ \sigma_m(0) + \sigma_m(L) \right]  \right\}, \\
\end{split}\label{eq_Pss_c_act_h1_mat}\eeq 
where the determinant and the exponential of the matrix $\boldsymbol{\Gamma}$ [\cref{eq_Gamma_mat}] are calculated in \cref{sec_inv} [see \cref{eq_actC_det_res,eq_actC_Gamma_correl}]. 
Taking, as required in this case, $\sigma_n$ to be Neumann modes without a zero mode, one obtains 
\beq \bar P_s(R)\big|_{h=0} = \frac{1}{\Zcal} \frac{1}{L \sqrt{1 + \kappa_c \left( \frac{1}{3} - \rho + \rho^2 \right)}} \exp\left[ -\frac{1}{2} H_1^2 \kappa_c \frac{\left( \frac{1}{6} - \rho + \rho^2 \right)^2}{1 + \kappa_c \left( \frac{1}{3} - \rho + \rho^2 \right)}  \right], \qquad \text{($\pm c \pm$)}
\label{eq_Pss_c_act_h1}\eeq 
where the dimensionless couplings $\kappa_c$ and $H_1$ are reported in \cref{eq_effcoupl_c,eq_effcoupl_h1}, respectively, and the normalization factor $\Zcal=\tilde \Zcal/L$ has to be calculated numerically.
If a zero mode is present, one obtains a flat distribution, $\bar P_s(R)|_{h=0} = 1/L$, instead [see \cref{eq_sherman_morrison_inv_zeromode}].
In the limit $\kappa_c \to\infty$, the OP obeys Dirichlet \bcs at the tracer location and the distribution in \cref{eq_Pss_c_act_h1} becomes independent of $\kappa_c$, reducing to
\beq \bar P_s(R)\big|_{h=0,\kappa_c\to\infty} = \frac{1}{\hat\Zcal}  \frac{1}{L \sqrt{ \frac{1}{3} - \rho + \rho^2 }} \exp\left[ - \frac{1}{2} H_1^2  \frac{\left( \frac{1}{6} - \rho + \rho^2 \right)^2}{ \frac{1}{3} - \rho + \rho^2 } \right].   \qquad \text{($\pm$D$\pm$)} 
\label{eq_Pss_c_act_h1_largeC}\eeq
Upon increasing the parameter $H_1$, the distributions in \cref{eq_Pss_c_act_h1,eq_Pss_c_act_h1_largeC} show a cross-over [see \cref{fig_Pss_act_h1}(b)] from a single- to a double-peaked shape with peaks located at 
\beq \frac{R_\pm}{L} = \onehalf \pm \frac{1}{2\sqrt{3}}. \qquad (|H_1|\gg 1)
\label{eq_Pss_c_act_h1_peakpos}\eeq 
In the limit $H_1\to\infty$, \cref{eq_Pss_c_pass_h1_peakpos,eq_Pss_c_act_h1_peakpos} coincide and $\bar P_s$ reduces to a sum of two Dirac-$\delta$ functions at $R_\pm$.

\subsubsection{Many tracers}
\label{sec_equil_manypart}

\begin{figure}[t]\centering
    \subfigure[]{\includegraphics[width=0.29\linewidth]{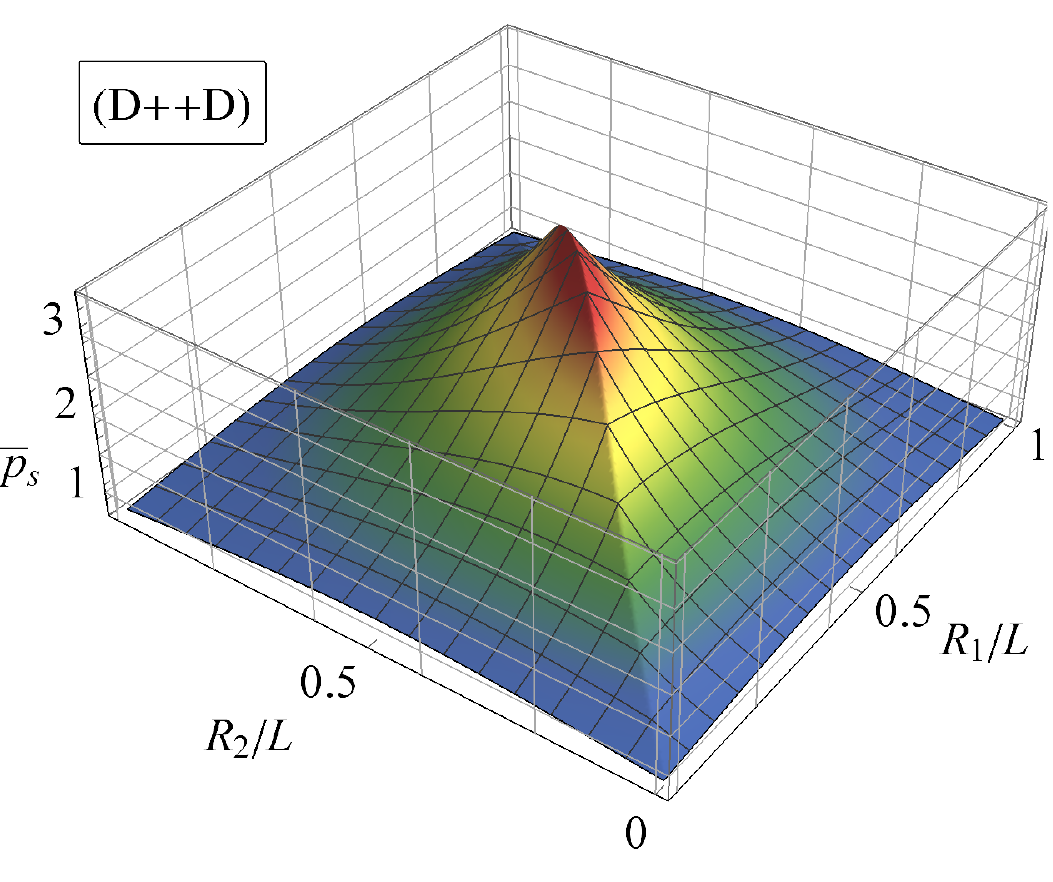} }\qquad
    \subfigure[]{\includegraphics[width=0.29\linewidth]{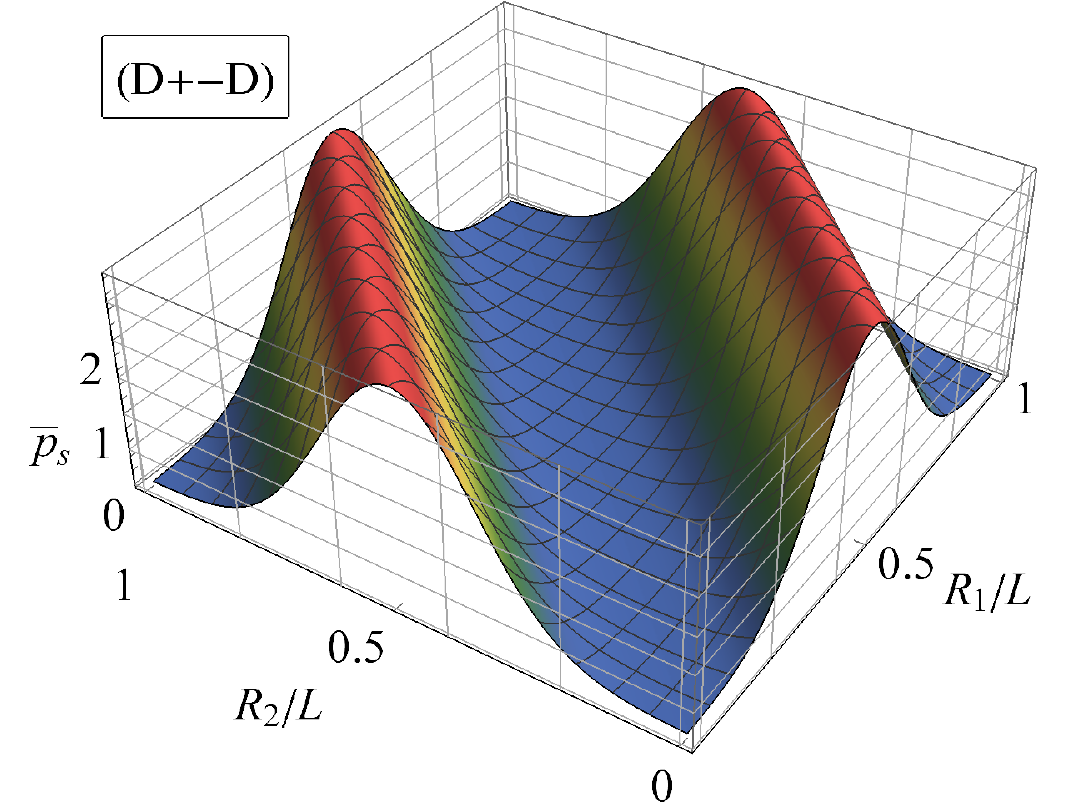} } 
    \subfigure[]{\includegraphics[width=0.29\linewidth]{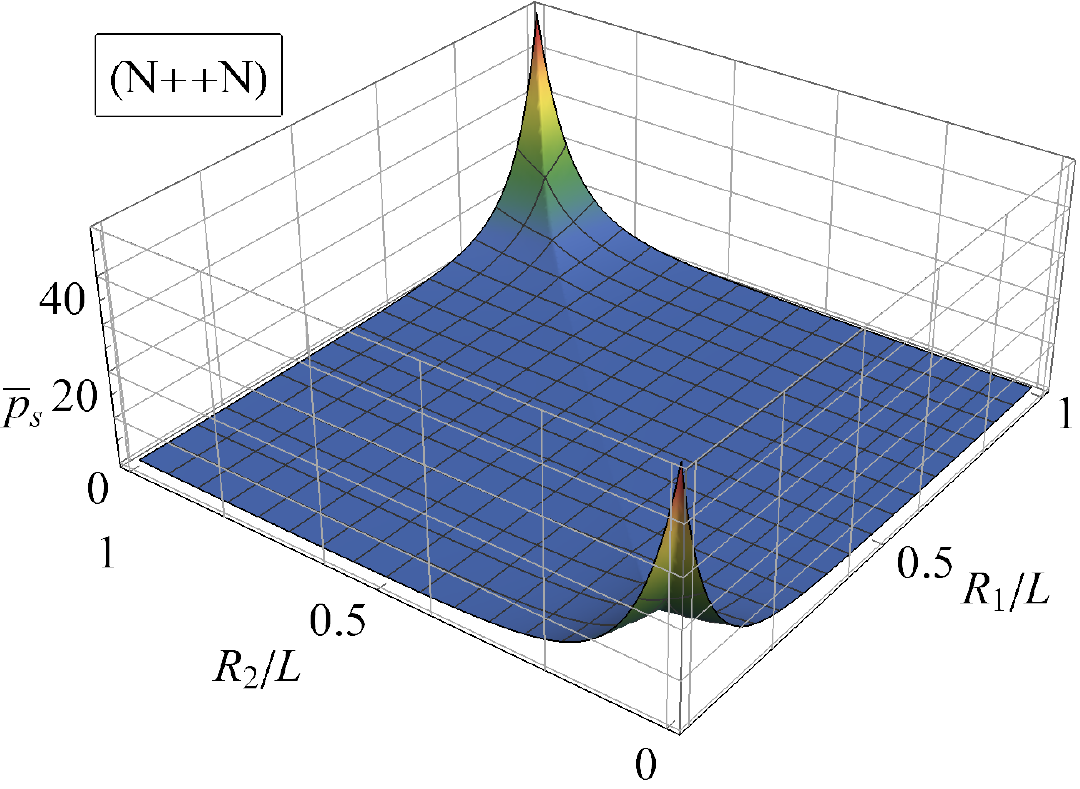} } \qquad
    \subfigure[]{\includegraphics[width=0.29\linewidth]{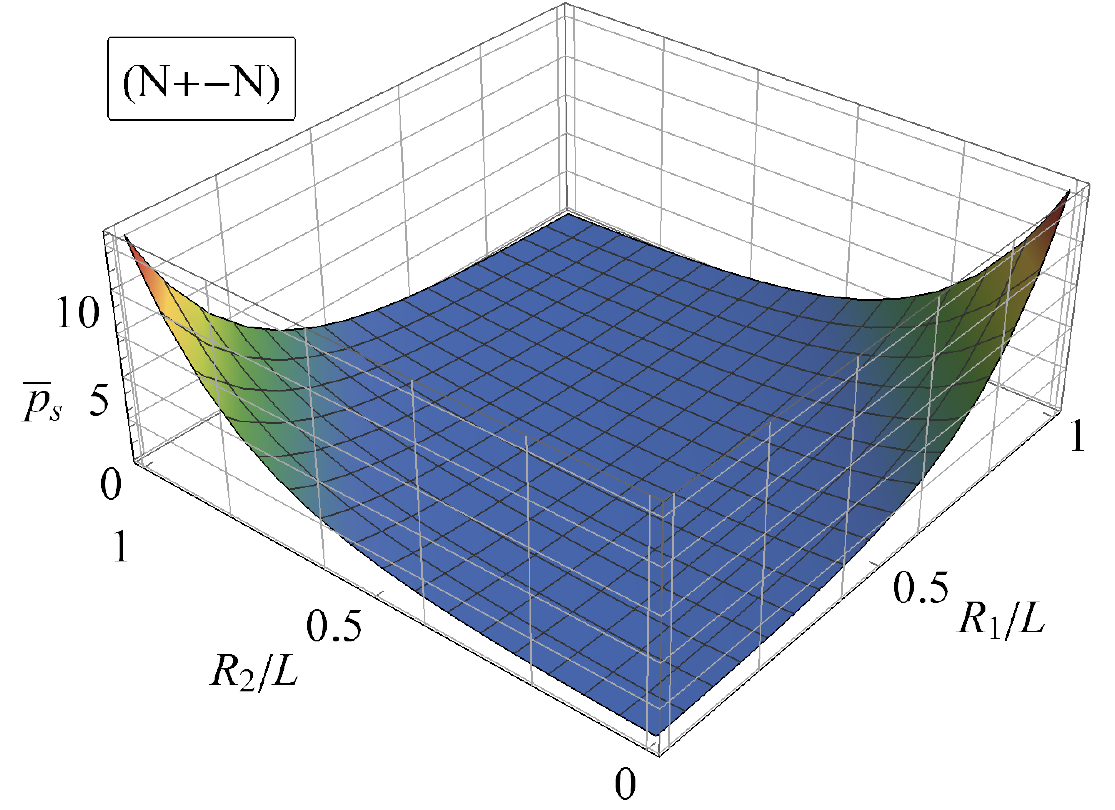} } 
    \subfigure[]{\includegraphics[width=0.29\linewidth]{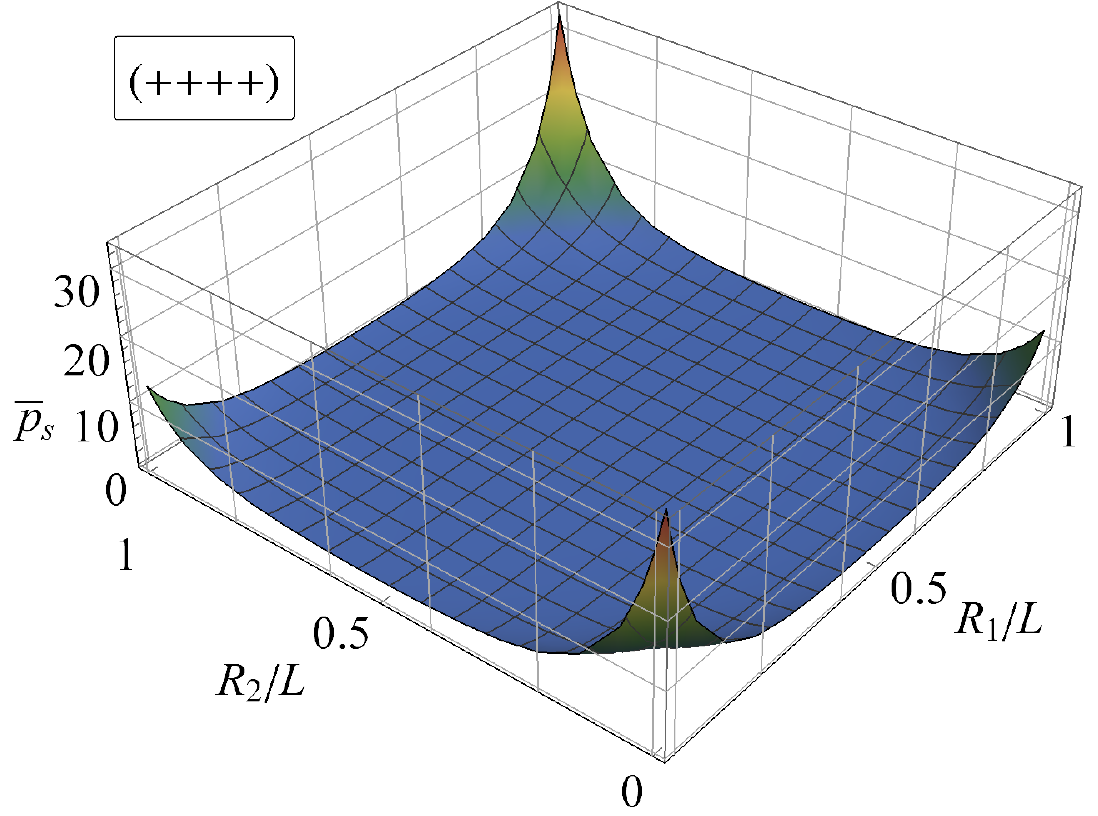} } \qquad
    \subfigure[]{\includegraphics[width=0.32\linewidth]{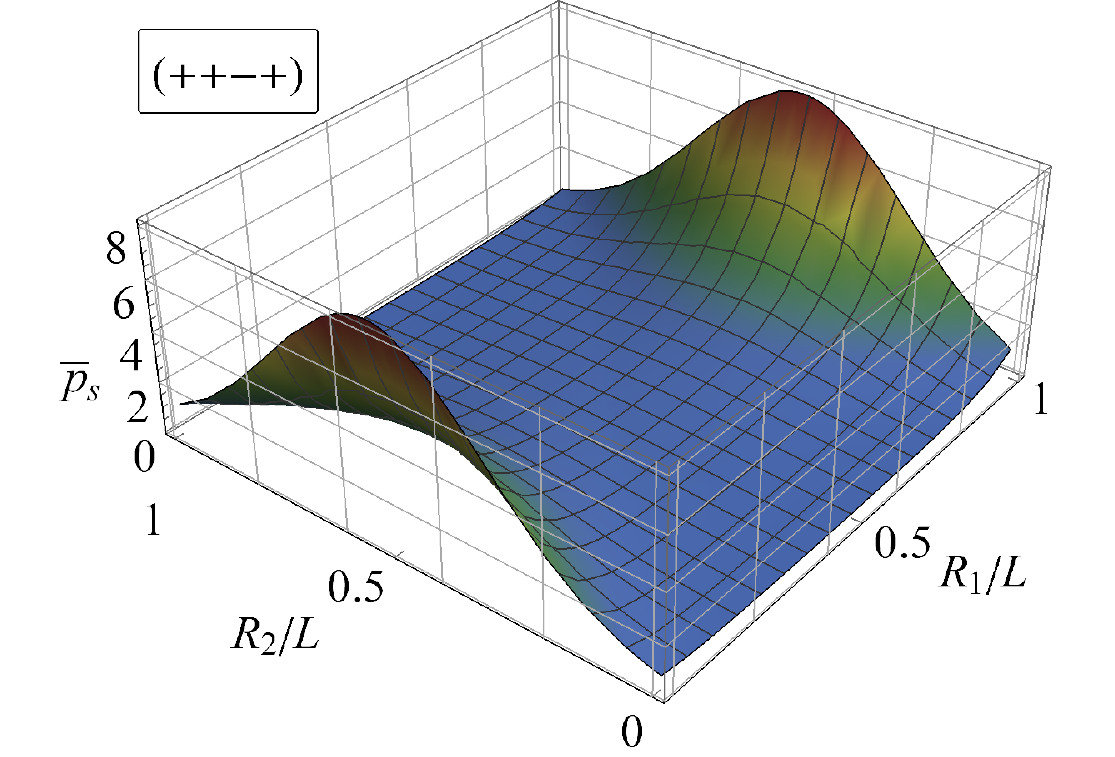} } 
\caption{Joint equilibrium probability distribution $\bar p_s(\rho_1, \rho_2) = L^2\bar P_s(R_1=\rho_1 L,R_2 = \rho_2 L)$ [\cref{eq_Pss_h_act_D_twop,eq_Pss_h_act_h1_twop}] of two reactive tracers (with positions $R_1$, $R_2$) confined to an interval $[0,L]$ and coupled linearly to a critical OP field. Up to a sign (indicated in the captions), the value of the coupling $h$ is the same for the two tracers. The OP field obeys (a,b) Dirichlet, (c,d) Neumann, or (e,f) capillary \bcs [see \cref{eq_cap_bcs}]. }
    \label{fig_Pss_act_twop}
\end{figure}

The above results can be straightforwardly generalized to $N_p>1$ tracers in the system.
To this end, the terms in the square brackets in \cref{eq_Pss_pathint} are replaced according to $h\phi(R)\to \sum_{j=1}^{N_p} \hscal_j \phi(R_j)$ and $c\phi^2(R)\to  \sum_{j=1}^{N_p} c_j \phi^2(R_j) $, where $\hscal_j$ and $c_j$ denote the coupling constants pertaining to the $j$th tracer.
As before, boundary fields of equal strength $\tilde h_1 = h_1/T_\phi$ act on the OP $\phi$.
In the case of a linear coupling between each tracer and the OP, \cref{eq_Pss_h_act_h1_base} generalizes to
\beq\begin{split} \bar P_s(R_1, R_2)\big|_{c=0} &= \frac{1}{\tilde \Zcal} \exp\left\{ \frac{T_\phi}{2} \sum_{n=1} \frac{1}{k_n^2}\left[  \sum_{j=1}^{N_p} \tilde \hscal_j \sigma_n(R_j)  + \tilde h_1 \big( \sigma_n(0) + \sigma_n(L)\big)\right]^2 \right\} \\
&= \frac{1}{\Zcal} \exp\left\{ \frac{1}{2} \sum_{i,j=1}^{N_p} \tilde \hscal_i \tilde \hscal_j C_\phi(R_i, R_j) + \tilde h_1 \sum_{j=1}^{N_p} \tilde \hscal_j \left[ C_\phi(R_j,0) + C_\phi(R_j,L) \right] \right\} ,
\end{split}\label{eq_Pss_h_act_h1_Np_base}\eeq 
with $\tilde \hscal_k \equiv \hscal_k/T_R$.
In the final result, we omitted all terms independent of $R_k$, as they are canceled by the normalization.
While the sums in \cref{eq_Pss_h_act_h1_Np_base} can be calculated for any number of tracers, we focus in the following on $N_p=2$ and couplings of equal magnitude $|\hscal_j|=h$. 
In the case of Dirichlet \bcs on $\phi$, the contributions associated with the boundary field $h_1$ identically vanish and one is left with 
\beq \bar P_s(R_1, R_2)\big|_{c=0} =
\begin{cases} \displaystyle
\frac{1}{\Zcal} \exp\left\{ -\onehalf T_\phi L \tilde h^2 \left[ (\rho_1 + \rho_2)^2 + |\rho_1 - \rho_2| \right]\right\}, \qquad \text{(D$++$D)} \\ \displaystyle
\frac{1}{\Zcal} \exp\left\{ -\onehalf T_\phi L \tilde h^2 \left[ (\rho_1-\rho_2)^2 - |\rho_1 - \rho_2| \right]\right\}, \qquad \text{(D$+-$D)} 
\end{cases}
\label{eq_Pss_h_act_D_twop}\eeq 
with $\rho_k \equiv R_k/L$, $\tilde h = h/T_R$, and a normalization factor $\Zcal$ that can be calculated analytically.
In the case of capillary \bcs on $\phi$ (which, as before, requires taking the $\sigma_n$ to be Neumann modes), the joint probability distribution of the two tracers is given by
\beq \bar P_s(R_1, R_2)\big|_{c=0} = 
\begin{cases}
 \frac{1}{\Zcal} \exp\left\{ T_\phi L \left[ \tilde h\tilde h_1 \left(  - \rho_1 + \rho_1^2 - \rho_2 + \rho_2^2\right) + \tilde h^2 \left(  - \rho_1 + \rho_1^2 - \rho_2 + \rho_2^2 - \frac{1}{2} |\rho_1 - \rho_2|\right)  \right]\right\}, \quad \text{($+$$+$$+$$+$)} \\
 \frac{1}{\Zcal} \exp\left\{ T_\phi L  \left[ \tilde h \tilde h_1 \left( - \rho_1 + \rho_1^2 + \rho_2 - \rho_2^2\right) + \onehalf \tilde h^2 |\rho_1 - \rho_2| \right]\right\}. \qquad \text{($+$$+$$-$$+$)}  
\end{cases}
\label{eq_Pss_h_act_h1_twop}\eeq 
When the OP obeys Neumann \bcs, the distribution follows by setting $h_1=0$ in this expression.

\Cref{fig_Pss_act_twop} illustrates the probability distribution of two linearly coupled reactive tracers in confinement. 
The essential features can be readily understood based on the behavior of a single tracer [see \cref{fig_Pss_act_h,fig_Pss_act_h1_h}]. 
To this end, we first note that two tracers having couplings of the same (opposite) sign attract (repel) each other. This behavior directly reflects the interactions of a single tracer with a boundary field [see \cref{fig_Pss_act_h1_h}] and is well-known in the context of critical Casimir forces (see \cref{sec_discuss_react}).
Accordingly, two tracers with $(++)$-type couplings tend to occupy the same region in the system, which, in the case of Dirichlet \bcs, is in the center [\cref{fig_Pss_act_twop}(a)], or, in the case of Neumann or capillary \bcs of the same sign, at the boundaries [\cref{fig_Pss_act_twop}(c,e)]. 
For the latter two \bcs, due to the attractive interactions between tracer and boundary, the tracers also have a nonzero probability to be located at opposite walls [in the case (N$++$N), the effect is suppressed except for small $h$, and is hence not visible in \cref{fig_Pss_act_twop}(c)].
By contrast, two tracers coupled linearly with opposite signs [$(+-)$] to an OP subject to Neumann \bcs likely reside at different boundaries of the system [\cref{fig_Pss_act_twop}(d)].
If the interactions between boundary and tracer are (partly) repulsive [as for $(D\pm)$, see \cref{fig_Pss_act_h}], one tracer tends to be located near the center of the system, while the other resides at a boundary [see \cref{fig_Pss_act_twop}(b,f)].
The above calculations can be readily extended to tracers with quadratic or mixed couplings.

\subsection{Dynamics}
\label{sec_react_dynamics}

We discuss in the following the effective dynamics of a reactive tracer in the adiabatic regime, i.e., assuming the OP field $\phi$ to be a fast variable. This regime is typically realized in experiments on colloidal particles in critical solvents~\cite{magazzu_controlling_2019}.

\subsubsection{Linear tracer-field coupling}

\label{sec_act_lin_adiab}

We first consider a tracer linearly coupled to a critical fluctuating OP field. This situation approximately describes a colloidal particle subject to critical adsorption.
In order to perform the adiabatic elimination of $\phi$ in the Langevin equations in \cref{eq_langevin_1d}, we apply the method of Refs.\ \cite{theiss_systematic_1985,theiss_remarks_1985}, which renders the following FPE for the effective tracer distribution $\bar P$ (see \cref{app_adiab_elim}):
\beq \pd_t \bar P(R,t) =  - \pd_R \left[\mu(R)  \bar P(R,t) \right] + \pd_R^2 \left[ D(R) \bar P(R,t) \right],
\label{eq_FP_act_h_adiab}\eeq 
with the drift and diffusion coefficients given by [see \cref{eq_adiab_drift_diffus_r}] 
\begin{subequations}
\begin{align}
\mu(R) &\equiv  \mu_0(R) + D'(R), \qquad \mu_0(R) \equiv -\left[ 1- \chit \kappa_h m(R) \right] \Ucal'(R)\label{eq_FP_act_h_drift} \\
D(R) &\equiv T \left[ 1 - \chit \kappa_h m(R)\right], \label{eq_FP_act_h_diff}
\end{align}\label{eq_FP_act_h_coeffs}
\end{subequations}
\hspace{-0.13cm}and $T=T_R=T_\phi$.
Furthermore, the coupling constants $\chit$ and $\kappa_h$ are defined in \cref{eq_adiab_param,eq_effcoupl_h},  $m(R)$ is reported in \cref{eq_m_func}, and 
\beq \Ucal(R) \equiv -\left[ \frac{h^2}{2 T}  V_\phi(R) + h \bra\phi(R)\ket_{h_1} \right]
\label{eq_FP_act_h_effpot}\eeq 
is an effective potential [see \cref{eq_phiR_var,eq_avg_prof_h1}]. 
The contribution $-h^2 V_\phi/(2T)$ in $\Ucal$ stems from the polarization of the OP by the tracer (representing critical adsorption) and is not present in the passive case [see \cref{eq_FP_pass_h_drift}]. 

The term $1- \chit \kappa_h m(R)$ in \cref{eq_FP_act_h_drift} represents an effective mobility [in units of the bare mobility $\gamma_R$; see \cref{eq_tracer_dyn_bare}], which is generally reduced compared to an uncoupled or a passive tracer [see \cref{eq_FP_pass_h_coeffs}].
Relative to the boundaries, the effective mobility is reduced (enhanced) in the center of the system for Neumann (Dirichlet) \bcs.
A positive effective mobility requires $\chit\kappa_h\ll 1$, which also defines the regime of validity of the adiabatic approximation [see \cref{eq_adiab_valid}].

As one easily checks, the steady-state solution $\bar P_s$ of \cref{eq_FP_act_h_adiab} is given by 
\beq \bar P_s(R) = \frac{1}{\Zcal} \exp(-\Ucal(R)/T),
\label{eq_Pss_act_h_adiab}\eeq 
which coincides with the equilibrium distribution for non-symmetry breaking and capillary \bcs, \cref{eq_Pss_h_act,eq_Pss_h_act_h1_base}, respectively.
For periodic \bcs, \cref{eq_FP_act_h_coeffs} reduces to $\mu=0$ and $D=T$, such that the tracer behaves as a simple Brownian particle in confinement, for which $\bar P_s=1/L$ [see \cref{eq_Pss_h_act}].

The Langevin equation associated with \cref{eq_FP_act_h_adiab} takes its simplest form in the ``\emph{iso}thermal'' convention (also known as anti-Ito or H\"{a}nggi-Klimontovich convention \cite{lau_state-dependent_2007,haenggi_stochastic_1978,klimontovich_ito_1990,volpe_effective_2016}, see \cref{app_spurious}):
\beq \pd_t R = \mu_0(R) + \sqrt{D(R)} \overset{\mathrm{iso}}{\circ} \theta,\qquad \bra \theta(t)\theta(t')\ket=2\delta(t-t'),
\label{eq_lang_react_h}\eeq 
where $\theta$ is a Gaussian white noise.
We remark that, using any other convention for the noise requires adding a spurious drift term to the Langevin equation in \cref{eq_lang_react_h} in order to recover the correct form of the drift in \cref{eq_FP_act_h_drift} (see \cref{app_spurious}). 
It has been previously noted that the isothermal convention is indeed a natural choice for the Langevin description of particles in an equilibrium system \cite{lau_state-dependent_2007,volpe_influence_2010}.  
Here, we have rigorously derived the underlying FPE from a system of stochastic differential equations with additive noise [\cref{eq_langevin_1d}], for which there is no ambiguity in its interpretation.

\subsubsection{Quadratic tracer-field coupling}
\label{sec_act_quadr}

We now determine the effective dynamics of a point-like colloidal particle quadratically coupled to a critical medium. In the strong coupling limit, $c\to\infty$, the particle imposes Dirichlet \bcs on the OP at $R$. 
However, in the present approach we are concerned with the opposite limit of a weak coupling.
We focus on the dynamics to $\Ocal(c)$, which is obtained (analogously to the passive case, see \cref{sec_passive_quadr_ad}) by inserting the adiabatic weak-coupling solutions for $\phi$ reported in \cref{eq_P2_c_act_phi1} into the Langevin equation in \cref{eq_tracer_dyn_simpl}. 
Applying the adiabatic elimination procedure of Refs.\ \cite{stratonovich_topics_1963,gardiner_stochastic_2009} \footnote{We remark that the adiabatic elimination procedure presented in Refs.\ \cite{theiss_systematic_1985,theiss_remarks_1985} applies only to a linear tracer-OP coupling.}, using the $\Ocal(c^0)$-expression $\bra\phi\izero(R)^2\ket = V_\phi(R) + \bra\phi(R)\ket_{h_1}^2 = (2/c) U_c(R)$ [see \cref{eq_avg_prof_h1,eq_phiR_var}], renders the linear Langevin equation 
\beq \dot R(t) =  -\pd_R U_c(R(t))  + \eta(t),
\label{eq_langevin_act_c}\eeq 
with the potential $U_c$ reported in \cref{eq_Pi_c_effpot}.
The associated FPE is given by
\beq \pd_t \bar P(R,t) = \pd_R \left[U_c'(R) \bar P(R,t) \right] + \pd_R^2 \left[ T_R \bar P(R,t) \right] ,
\label{eq_FP_act_c_adiab}\eeq 
which coincides with the one for a passive tracer [see \cref{eq_FPE_pass_c_coeffs}] at $\Ocal(c)$. 
Accordingly, the associated steady-state distribution is given by \cref{eq_Pss_pass_c_ad}, i.e.,
\beq \bar P_s(R) = \frac{1}{\Zcal} \exp\left[-\sfrac{U_c(R)}{T_R} \right],
\label{eq_Pss_act_c_ad}\eeq 
which agrees at $\Ocal(c)$ with \cref{eq_Pss_pass_c} as well as with the exact equilibrium distributions in \cref{eq_Pss_c_act_det,eq_Pss_c_act_h1}.
Numerical simulations (see \cref{sec_sim}) indeed confirm that, for a quadratically coupled tracer, steady-state distributions in the passive and reactive cases are similar \footnote{An exception occurs for OP fields subject to periodic \bcs, in which case $\bar P_s=1/L$ for a reactive tracer, while $\bar P_s$ is non-uniform for a passive tracer (see also the discussion in \cref{sec_discuss_pass_quad}).}.

\section{Simulation results \& Discussion}
\label{sec_sim}

In the next subsection, we present results for the stationary distribution and the mean squared displacement of the tracer obtained from numerical Langevin simulations of \cref{eq_langevin_1d}. Furthermore, we discuss the generic mechanism underlying the observed behaviors (\cref{sec_discuss_pass}) and place our results in the general context of boundary critical phenomena (\cref{sec_discuss_react}).

\subsection{Simulations}

\begin{figure}[t]\centering
    \includegraphics[width=0.65\linewidth]{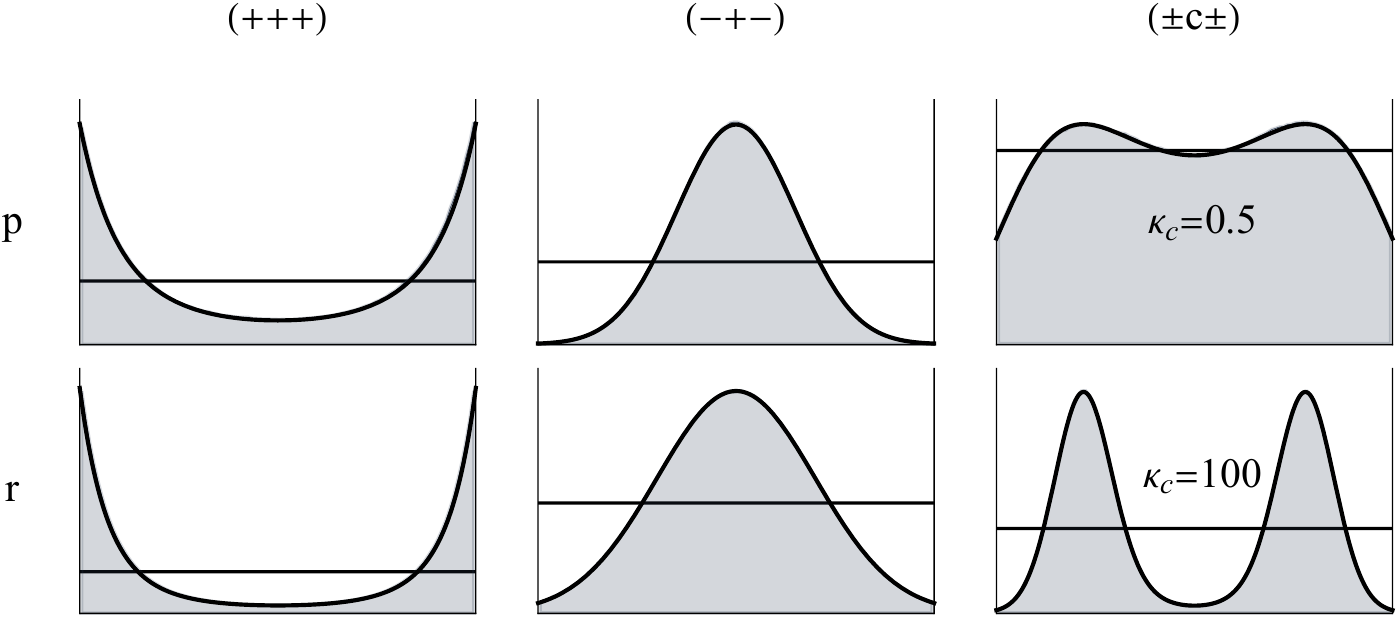} 
    \caption{Steady-state distribution $\bar P_s(R)$ of a tracer confined to the interval $[0,L]$, numerically determined from simulations of \cref{eq_langevin_1d} for non-vanishing boundary fields $h_1\neq 0$. The tracer is either \emph{p}assive (top row) or \emph{r}eactive (bottom row) and is coupled either linearly (left and middle panel) or quadratically (right panel) to the OP field. The sign of the fields $h$ and $h_1$ is indicated by the labels, e.g., ($-$+$-$) corresponds to $h_1<0$ and $h>0$. In the case of a quadratic coupling, $\bar P_s$ is independent of the sign of $h_1$ [see \cref{eq_Pss_pass_c_ad,eq_Pss_c_act_h1}]. Simulation results (thin curves with adjacent filling) are compared to the analytically determined distributions (thick cures) reported in \cref{eq_Pss_h_pass_adiab,eq_Pss_pass_c_avg} (passive tracer) and \cref{eq_Pss_h_act_h1_ppp,eq_Pss_h_act_h1_mpm,eq_Pss_c_act_h1} (reactive tracer). The location of the two peaks in the right panels is described by \cref{eq_Pss_c_pass_h1_peakpos,eq_Pss_c_act_h1_peakpos}. The dimensionless couplings in \cref{eq_effcoupl_h,eq_effcoupl_h1} take the values $|\kappa_h|=H_1^2=30$ (left), 50 (middle), and $H_1^2=100$ (right panel), while $\chit=0.02$ (passive) and $\chit=1$ (reactive); the value of $\kappa_c$ [\cref{eq_effcoupl_c}] is indicated. [Note that the adiabatic approximation also requires $\kappa_c\ll 1$ in order to be applicable to a quadratically coupled passive tracer (see \cref{sec_eff_Lang,sec_passive_quadr_ad}).] The horizontal lines represent $\bar P_s=1$.
    }
    \label{fig_Pss_grid_h1}
\end{figure}

Numerical results for the tracer position are generated by solving the Langevin \cref{eq_tracer_dyn} in real space (for a system size of $L=100$ in simulation units) and \cref{eq_field_dyn} in mode space using \cref{eq_phi_mode_eqn} (with a total number of 100 modes) \cite{press_numerical_2007}.
For both equations a standard Euler forward integration scheme is employed \cite{gardiner_stochastic_2009} with time steps $\Delta t\sim \Ocal(0.1)$ in the non-conserved [$a=0$ in \cref{eq_field_dyn}] and $\Delta t\sim \Ocal(0.01)$ in the conserved ($a=1$) case. 
All other parameters are set to unity in the simulations, unless otherwise indicated. In particular, the effective coupling constants [see \cref{eq_effcoupl_h,eq_effcoupl_c}] take the values $\kappa_h = h^2 L T_\phi/T_R^2 = 100$ and $\kappa_c=c L T_\phi/T_R = 100$, while the adiabaticity parameter [see \cref{eq_adiab_param}] $\chit=1$ (except if $T_R/T_\phi\neq 1$ and in parts of \cref{fig_Pss_grid_h1}). We remark that, since standard Dirichlet \bcs violate global OP conservation (see discussion in \cref{sec_BCs}), we consider in this case only non-conserved dynamics. For all other \bcs, we typically study both conserved as well as non-conserved dynamics. Simulation data is recorded after an initial transient period [of approximate duration $\chi(L/\pi)^{2+2a}$] required to equilibrate the OP field.

\begin{figure}[t]\centering
    \includegraphics[width=0.88\linewidth]{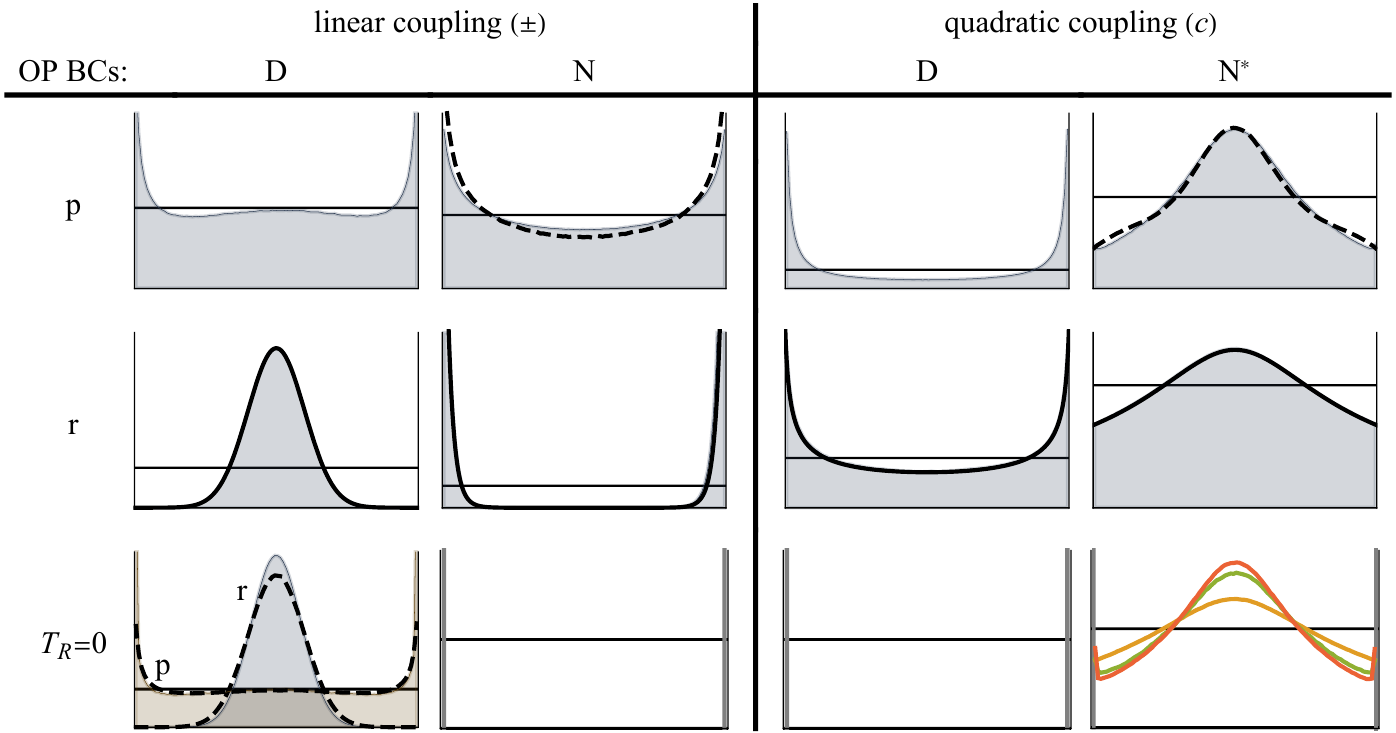} 
    \caption{Steady-state distribution $\bar P_s(R)$ of a tracer confined to the interval $[0,L]$, numerically determined from simulations of \cref{eq_langevin_1d} for $h_1=0$. The tracer is either coupled linearly (coupling $h$, denoted as $\pm$) or quadratically (coupling $c$) to a fluctuating OP field $\phi$, the latter obeying either \emph{D}irichlet or \emph{N}eumann \bcs (N$^*$ denotes Neumann \bcs without a zero mode). The dimensionless couplings [see \cref{eq_effcoupl_h,eq_effcoupl_c}] take the values $\kappa_h=\kappa_c=100$, while the adiabaticity parameter [see \cref{eq_adiab_param}] $\chit=1$ (except for the last row). 
    The first two rows show $\bar P_s$ for a \emph{p}assive ($\zeta=0$) and a \emph{r}eactive tracer ($\zeta=1$), while the thin curves (with adjacent filling) pertain to non-conserved dynamics and the dashed curves to conserved dynamics. 
    The numerically determined distributions of a reactive tracer (second row) are accurately captured by the analytical equilibrium distributions in \cref{eq_Pss_h_act,eq_Pss_c_act} (thick black curves).
    The last row illustrates the steady-state distributions obtained for vanishing tracer noise $T_R=0$, both in the reactive and passive cases, with the gray bars at the boundaries representing Dirac-$\delta$ distributions. In the last row, the dashed curves in the case (D$\pm$D) pertain to $T_R=T_\phi$, while the various solid curves in the case (N$^*$$c$N$^*$) correspond to $T_R/T_\phi = 1, 10^{-1} , 10^{-4}$ (center bottom to top). The horizontal lines represent $\bar P_s=1$.}
    \label{fig_Pss_grid}
\end{figure}

\begin{figure}[b]\centering
    \subfigure[]{\includegraphics[width=0.32\linewidth]{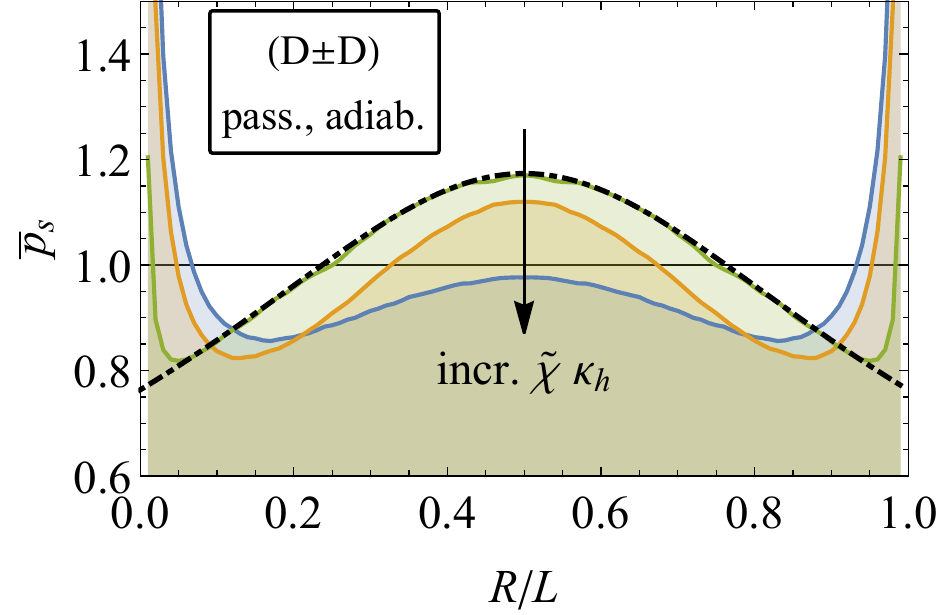} \label{fig_Pss_adiab_Dh}}\qquad
    \subfigure[]{\includegraphics[width=0.32\linewidth]{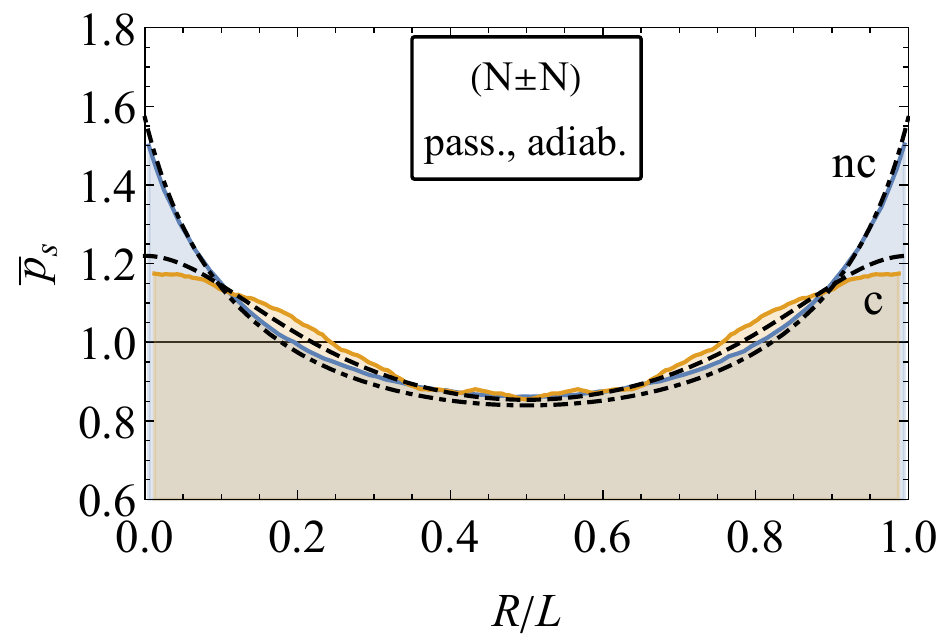} }
    \caption{Steady-state distribution $\bar p_s = L \bar P_s$ of a passive tracer in the adiabatic regime ($\chit\ll 1$), linearly coupled to an OP field subject to (a) Dirichlet and (b) Neumann \bcs. The various solid curves (with adjacent filling) represent simulation results obtained for (a) $\chit\kappa_h = 10,100,1000$ (in the direction of the arrow) and (b) $\chit\kappa_h=10$ (50) in the \emph{n}on-\emph{c}onserved (\emph{c}onserved) case. The dashed-dotted (dashed) curves represent the expression in \cref{eq_Pss_h_pass_adiab} for $\chit\kappa_h=10$ [$\chit\kappa_h=50$ in the conserved case in (b)] and $h_1=0$. }
    \label{fig_Pss_adiab}
\end{figure}

\subsubsection{Statics}

In \cref{fig_Pss_grid}, the numerically determined steady-state distributions $\bar P_s(R)$ of a tracer are summarized, focusing on OP fields obeying Dirichlet or Neumann \bcs.
A passive tracer (first row) is genuinely out of thermal equilibrium and thus $\bar P_s$ depends, in principle, on the specific dynamics and the conservation law. 
However, for Neumann \bcs, simulations reveal only a marginal difference between conserved (dashed curves) and dissipative dynamics (thin curves with filling) in the considered parameter regime. 
Since, for a quadratically coupled tracer, the presence of a zero mode would lead to a uniform distribution $\bar P_s=1/L$ [see \cref{eq_Pss_c_act}], we consider in that case only Neumann \bcs without a zero mode (N$^*c$N$^*$, last column). While this is easily arranged in a simulation with dissipative dynamics, it requires conserved dynamics in an actual physical setup.

In the case of a reactive tracer (second row), the equilibrium distributions in \cref{eq_Pss_h_act,eq_Pss_c_act} match the numerical results essentially exactly, both for dissipative and conserved dynamics.
For a passive tracer, the distributions obtained within the weak-coupling approximation in \cref{eq_Pss_h_pass_dom,eq_Pss_pass_c} [see also \cref{fig_Pss_pass_h,fig_P_pass_quadr}] still capture the simulations qualitatively -- in particular, the attraction of the tracer towards a boundary or the center of the system. 
This is remarkable insofar as the numerical results are obtained using effective couplings of significantly larger magnitude than typically permitted in a perturbative solution. This points to the generic character of the phenomenon (see \cref{sec_discuss_pass} below).
Note that, if Dirichlet or Neumann \bcs act on the OP, the sign of the coupling $h$ is irrelevant, since $\bar P_s$ depends on $h^2$ [see \cref{eq_effcoupl_h}].

The local maximum of $\bar P_s$ at $R=L/2$ observed in \cref{fig_Pss_grid} in the case of a passive tracer for (D$\pm$D) is predicted by the adiabatic approximation [see \cref{eq_Pss_h_pass_adiab,fig_Pss_pass_h_ad}].
This is detailed in \cref{fig_Pss_adiab}(a), which shows $\bar P_s$ determined from simulations in the adiabatic regime for (D$\pm$D), non-conserved OP dynamics, and various values of $\chit \kappa_h$. As the latter parameter is decreased, the central maximum of $\bar P_s$ grows at the expense of the maxima at the boundaries. 
\Cref{fig_Pss_adiab}(b) illustrates the corresponding behavior of $\bar P_s$ in the case (N$\pm$N) for conserved and non-conserved OP dynamics \footnote{In order to increase the statistical accuracy of the data for conserved dynamics shown in \cref{fig_Pss_adiab}(b), we have taken advantage of the mirror symmetry of $\bar P_s(R)$ around $R=L/2$.}.
Notably, for $\chit\kappa_h\lesssim \Ocal(1)$, the adiabatic result in \cref{eq_Pss_h_pass_adiab} (broken curves in \cref{fig_Pss_adiab}) agrees well with the numerically determined steady-state distributions.
The adiabatic approximation breaks down for larger values of $\chit \kappa_h$ [see \cref{eq_adiab_valid}], where non-Markovian effects instead dominate the behavior of a passive tracer, leading to an enhancement of the probability near the boundaries (see \cref{sec_discuss_pass} below).

The last row in \cref{fig_Pss_grid} shows the steady-state distributions resulting for vanishing tracer noise intensity [$T_R=0$, see \cref{eq_tracer_noise}], in which case the only source of noise stems from the coupling to the OP [see \cref{eq_tracer_dyn_simpl}]. 
In the case (D$\pm$D), the  resulting distributions (thin curves with filling) are close to the ones obtained for $T_R=T_\phi$ (dashed curves). For all other couplings and \bcs, one obtains two Dirac-$\delta$-like distributions located at the boundaries (represented by thin vertical bars in the plot). This behavior can be readily understood by noting that, in \cref{eq_tracer_dyn_simpl} the coupling terms $\pd_R\phi$ and $\phi$ vanish at the boundaries for Neumann and Dirichlet \bcs, respectively.
Interestingly, in the case (N$^*$$c$N$^*$), the transition from $T_R=T_\phi$ to $T_R=0$ proceeds by a growth of the central peak of $\bar P_s$ and a rather sharp increase of the probability at the boundaries.

\Cref{fig_Pss_grid_h1} illustrates the steady-state distribution of a tracer in the presence of non-vanishing boundary fields $h_1$. 
In the passive case (top row), $\bar P_s$ is dominated by the deterministic potentials $U_h$ and $U_c$ [see \cref{eq_Pss_h_pass_adiab0,eq_Pss_pass_c_avg}] and is thus inhomogeneous even for $\chit=0$, in contrast to the case with non-symmetry-breaking \bcs ($h_1=0$). 
We find that, for $\chit\ll 1$, the simulation results (thin curves with filling) are accurately captured by the expressions obtained based on the adiabatic approximation in \cref{eq_Pss_h_pass_adiab,eq_Pss_pass_c_avg} (thick curves). 
As was the case for $h_1=0$, the analytically determined equilibrium distributions for a reactive tracer in \cref{eq_Pss_h_act_h1_ppp,eq_Pss_h_act_h1_mpm,eq_Pss_c_act_h1} (thick curves, bottom row) exactly match the numerical results.

\subsubsection{Dynamics}

\begin{figure}[t]\centering
    \subfigure[]{\includegraphics[width=0.33\linewidth]{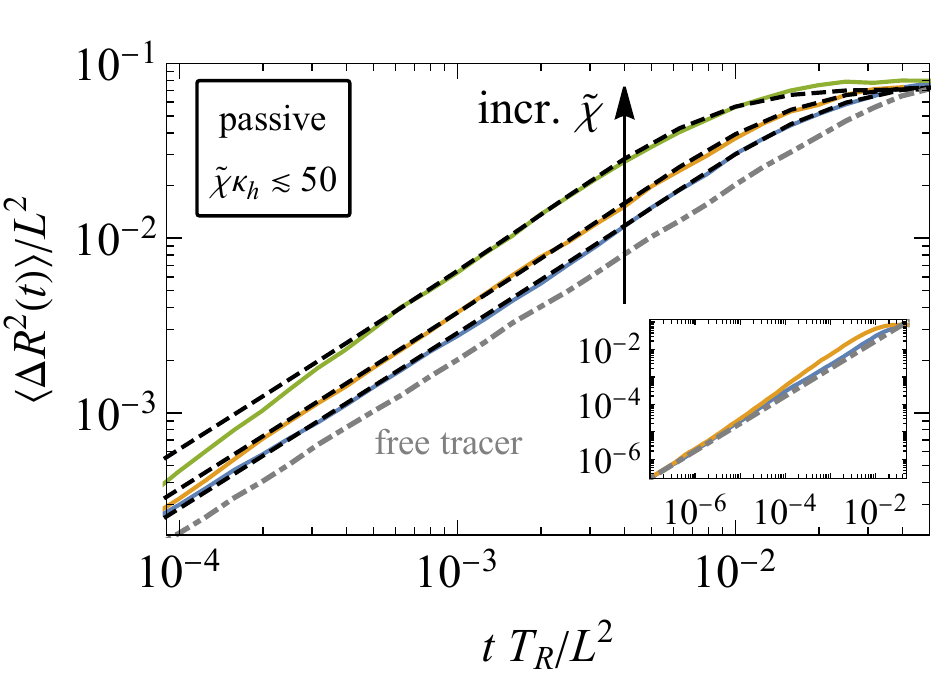} }\qquad
    \subfigure[]{\includegraphics[width=0.33\linewidth]{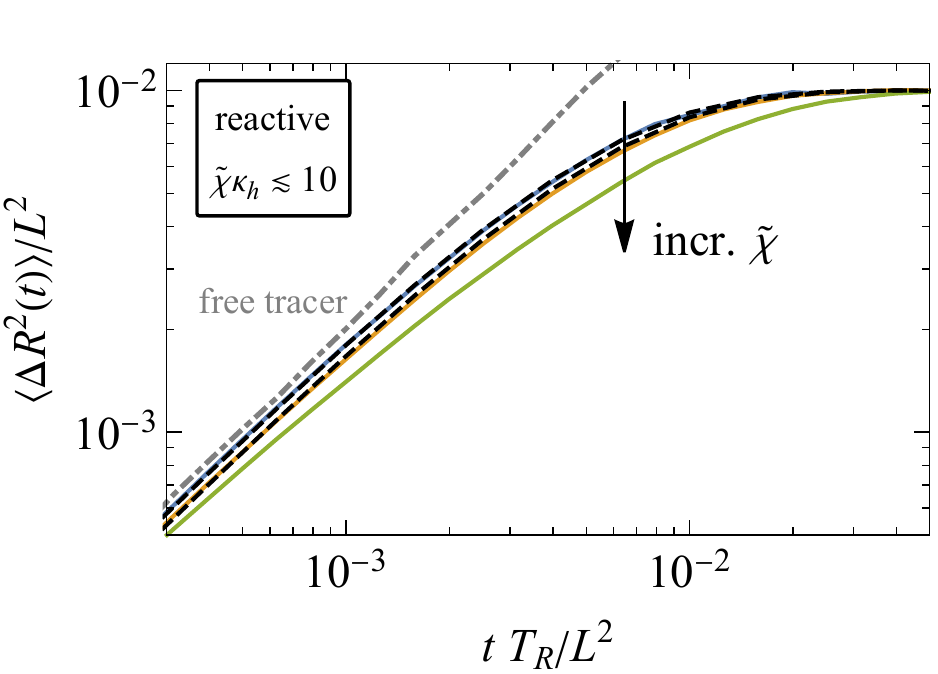} } 
    \caption{MSD [\cref{eq_MSD}] of (a) a passive and (b) a reactive tracer in the interval $[0,L]$, coupled linearly to an OP field with Dirichlet \bcs and dissipative dynamics in the adiabatic regime. Results of Langevin simulations (solid curves) are compared to the predictions obtained numerically from the FPEs in \cref{eq_FP_pass_h_adiab,eq_FP_act_h_adiab} (dashed curves) for $R_0=L/2$. The MSD obtained from the FPE grows $\propto t$ [see \cref{eq_MSD_shorttime}] up to a cross-over time [\cref{eq_OU_relax}]. For large $\chit \kappa_h$ [see \cref{eq_adiab_valid}], the adiabatic approximation ceases to hold and non-Markovian effects become important at intermediate times. At short times, the corresponding MSD obtained from the simulations approaches the one of a free ($h=0$) tracer (dashed-dotted curve, see also inset) and hence deviates from the predictions of the (Markovian) FPE. The coupling parameters $\kappa_h$ [\cref{eq_effcoupl_h}] and $\chit$ [\cref{eq_adiab_param}] used in the simulations take the following values (in the direction of the arrows): (a) $\kappa_h=10^4$, $\chit=10^{-3}, 0.01, 0.025$, (b) $\kappa_h=10^2$, $\chit=10^{-3}, 0.02, 0.1$. In the reactive case, no solution of the FPE in \cref{eq_FP_act_h_adiab} exists for $\chit\kappa_h\gtrsim 1$.}
    \label{fig_msd}
\end{figure}

In order to assess the tracer dynamics, we consider the mean-squared displacement (MSD) of the tracer location, 
\beq \bra\Delta R(t)^2 \ket \equiv \bra [ R(t)-R_0 ]^2 \ket = \int_0^L \d R\, (R-R_0)^2 \bar P(R,R_0,t),
\label{eq_MSD}\eeq 
where we take $R_0\equiv R(0)=L/2$ as initial position.
\Cref{fig_msd} illustrates the MSD of a tracer coupled linearly to a (non-conserved) OP field with Dirichlet \bcs in the adiabatic regime.
In simulation, the average in \cref{eq_MSD} is obtained over multiple stochastic realizations of the noise, whereas the theoretical prediction is determined by inserting in \cref{eq_MSD} for $\bar P$ the (numerically calculated) solution of the FPEs reported in \cref{eq_FP_pass_h_adiab,eq_FP_act_h_adiab}.
Since these FPEs describe a Markovian process with time-independent drift and diffusion coefficients, the resulting MSD depends linearly on time $t$ up to a cross-over time, at which the MSD attains its steady-state value determined by $\bar P_s$.
The crossover time can be estimated by the characteristic relaxation time $t_R$ of the stochastic process [see \cref{eq_OU_relax}].
A short-time solution of the FPE in \cref{eq_FP_pass_h_adiab,eq_FP_act_h_adiab} yields \cite{risken_fokker-planck_1989}
\beq \bra\Delta R(t)^2\ket \simeq 2 D(R_0) t + \Ocal(t^2), \qquad (t\ll t_R)
\label{eq_MSD_shorttime}\eeq 
with the diffusivity $D(R)$ given in \cref{eq_FP_pass_h_diff,eq_FP_act_h_diff}.
For a passive tracer within the adiabatic regime, the OP acts as an additional Markovian noise source [cf.\ \cref{eq_tracer_dyn_simpl}], causing the diffusivity to surpass the free one $T_R$. In contrast, a reactive tracer polarizes the surrounding medium, which hinders displacement and consequently reduces the diffusivity relative to the free one.
These behaviors are consistent with the results of Ref.\ \cite{demery_perturbative_2011} and we emphasize that the trends observed when varying $\chit$ apply only to $\chit\ll 1$.
Remarkably, already for $\chit \kappa_h\lesssim \Ocal(10)$, our simulation results (solid curves) are accurately captured by the analytical predictions (dashed curves). 
At shorter times as well as for larger $\chit\kappa_h$, the non-Markovian character of the OP fluctuations becomes prominent, causing the simulations to increasingly deviate from the adiabatic approximation [see \cref{eq_adiab_valid}]. 
In fact, for $t\to 0$, the MSD obtained from the simulations approaches the one of a free ($h=0$) tracer, $\bra\Delta R(t\ll t_R)^2\ket = 2 T_R t$ [dashed-dotted curve, see inset in \cref{fig_msd}(a)].
We remark that, even in the Markovian regime, the spatially heterogeneous character of the diffusivity $D(R)$ can lead to non-Brownian diffusion \cite{lau_state-dependent_2007,cherstvy_anomalous_2013,leibovich_infinite_2019}.
A more detailed analysis of the diffusivity will be performed in a separate study.

\subsection{Discussion}

\subsubsection{Passive tracer}
\label{sec_discuss_pass}

\begin{figure}[t]\centering
    \subfigure[]{\includegraphics[width=0.275\linewidth]{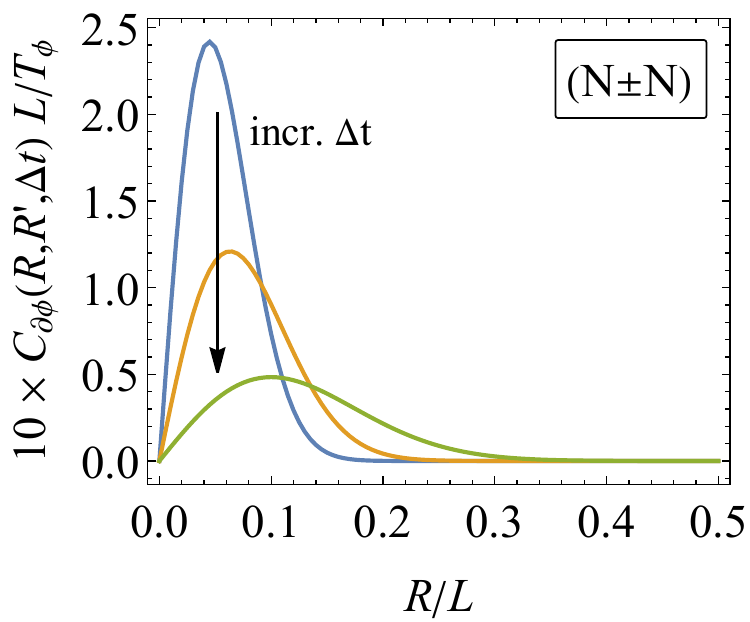} }\qquad
    \subfigure[]{\includegraphics[width=0.27\linewidth]{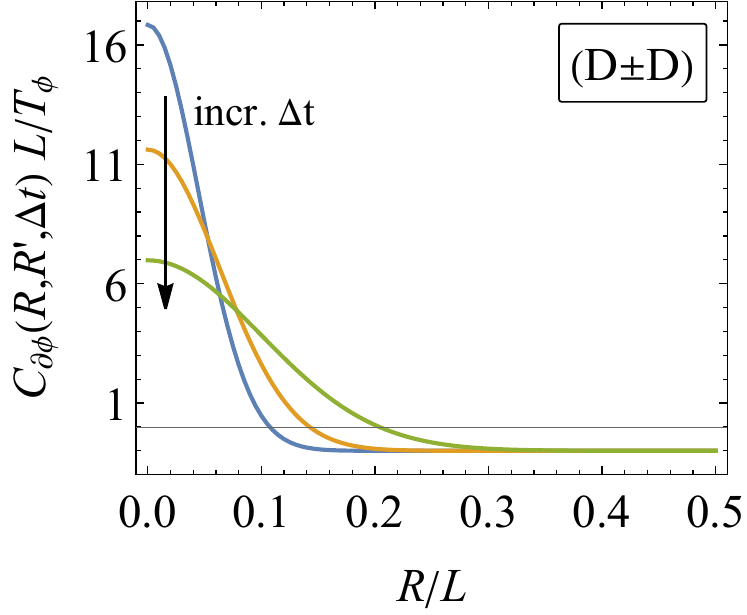} } \qquad
    \caption{Correlations of the effective random force $\Pi_h$ [see \cref{eq_Pi_h_correl,eq_C_phi_deriv}] imposed by the OP onto a linearly coupled passive tracer [see \cref{eq_tracer_dyn_simpl,eq_Pi_h_effnoise}]. The different curves in the plot correspond to dimensionless times $\Delta t\, T_R/(L^2 10^{-3})=1,2,5$ (from top to bottom left). The positive temporal correlation of $\Pi_h$ leads to an enhancement of the occupation probability of a confined passive tracer near the system boundaries, as observed in simulations (see the first two panels in \cref{fig_Pss_grid}).}
    \label{fig_effnoise}
\end{figure}

The steady-state distributions of a passive tracer (first row in \cref{fig_Pss_grid,fig_Pss_grid_h1}) can be understood by inspecting the forcing terms $\Xi_{h,c}$ in the Langevin equation in \cref{eq_tracer_dyn_simpl}. 
In the presence of boundary fields $h_1\neq 0$, the tracer dynamics is essentially controlled by the non-vanishing mean OP profile $\bra\phi(z)\ket$, which gives rise to the deterministic potentials $U_{c,h}(z)$ [see \cref{eq_effpot,eq_Pss_h_pass_adiab,eq_Pss_pass_c_ad}].
For $h_1=0$, instead, the OP fluctuations, as encoded in the effective noises $\Pi_{h,c}$ [\cref{eq_effnoise}], determine the behavior of the tracer.
In the a linearly coupled case, the correlations of the effective noise $\Pi_h$ are proportional to $C_{\pd\phi}(R,R',\Delta t)$ [see \cref{eq_Pi_h_correl,eq_C_phi_deriv}], which is illustrated in \cref{fig_effnoise} as a function of $R$ and $\Delta t$ around some fixed location $R'\simeq 0$ near one boundary \footnote{The behavior is, in fact, similar in the bulk, where $C_{\pd\phi}$ as a function of $R$ is symmetric around $R'$.}. 
Note first that $C_{\pd\phi}(R,R',\Delta t)$ is positive if $R$ is near $R'$. 
Assume now that the tracer is located near a boundary and receives a ``kick'' from the noise which, in the absence of a boundary, would move the tracer beyond it. Due to the boundary condition, however, the tracer is reflected back to a position, where, owing to the positive temporal correlation of the noise, it is likely to again be kicked towards the boundary in the next time step. 
As a consequence, a passive tracer linearly coupled to a uniform OP field has an enhanced probability to reside near a boundary, as observed in \cref{fig_Pss_grid}.
This effect is, in fact, generically expected for a confined stochastic process driven by a temporally correlated noise \cite{guggenberger_fractional_2019,vojta_probability_2019}.
The dynamics of a quadratically coupled tracer, by contrast, is dominated by the non-vanishing mean of the forcing term $\Xi_c$ [\cref{eq_Pi_c_effpot}], as described by the deterministic potential $U_c(R)$ [see \cref{fig_effpot,eq_Pss_pass_c_ad,eq_Pss_act_c_ad}].

\subsubsection{Reactive tracer}
\label{sec_discuss_react}

The equilibrium distribution $\bar P_s$ of a reactive tracer (second row in \cref{fig_Pss_grid} and \cref{fig_Pss_grid_h1}) encodes the fluctuation-induced interactions between inclusions in a critical medium \cite{diehl_field-theoretical_1986,krech_casimir_1994,brankov_theory_2000}.
Accordingly, it is informative to connect the present results to previous studies of the critical Casimir force (CCF) acting on a spatially extended spherical particle in front of a planar wall \cite{burkhardt_casimir_1995,eisenriegler_casimir_1995,hanke_critical_1998}.
To this end, the Hamiltonian coupling of the point-like tracer [see \cref{eq_Hamilt}] must be mapped to a boundary condition for the OP $\phi$ at the particle surface: a linear coupling ($h\neq 0$, $c=0$) corresponds to a $+$ (or $-$) boundary condition, while a quadratic coupling ($c\neq 0$, $h=0$) corresponds to a Dirichlet boundary condition (provided that $|h|$ and $|c|$ are sufficiently large).
In the limit where the particle-wall distance $R$ is large compared to the particle radius $\varrho$, a small-sphere expansion renders the Casimir (excess) free energy \cite{burkhardt_casimir_1995,eisenriegler_casimir_1995,hanke_critical_1998} 
\beq \Fcal\st{cas} \simeq - T_c \frac{A_a^\psi A_b^\psi}{B_\psi} \left(\frac{\varrho}{2R}\right)^{x_\psi}.
\label{eq_excess_FE}\eeq 
Here, $\psi= \phi$ and $x_\phi = \beta/\nu$, if both the wall and the particle impose symmetry-breaking \bcs [i.e., ($\pm\pm$)], whereas $\psi= \phi^2$, $x_{\phi^2} = d-\nu^{-1}$, if neither particle nor wall, or only one, impose symmetry-breaking \bcs [corresponding to the cases ($\pm$, D/N), (D,D/N), (NN)]. (Here, $\beta$ and $\nu$ denote the standard bulk critical exponents.)
The amplitudes $A_a^\psi$ and $B^\psi$ and the exponent $x_\psi$ are defined via the associated bulk correlation function, $\bra \psi(\rv) \psi(\rv')\ket\st{bulk} = B_\psi r^{-2 x_\psi}$, and the profile in the half-space with boundary condition $a$, $\bra \psi(r_\perp)\ket\st{half-space}^a = A_a^\psi (2r_\perp)^{-x_\psi}$.
The excess free energy contributes to the Casimir potential $\Ucal(R) = \Fcal\st{cas}(R) + \Ucal\st{add}(R)$, which enters a Boltzmann-like probability distribution for the tracer \cite{maciolek_collective_2018},
\beq \bar P_s(R) \propto \exp(-\Ucal(R)/T).
\label{eq_Boltzmann}\eeq 
The potential $\Ucal\st{add}(R)$ accounts for additional interactions, such as those stemming from van der Waals forces, which are relevant at short distances $R$ and regularize a possible divergence of $\Fcal\st{cas}$ for $R\to 0$ \cite{gambassi_critical_2009,valchev_critical_2015}. We do not consider these here.
The CCF acting between particle and wall follows from \cref{eq_excess_FE} as $\Kcal = -\d \Fcal\st{cas}/\d R = x_\psi  \Fcal\st{cas} / R$.

Since $\Fcal\st{cas}$ in \cref{eq_excess_FE} is singular for a point-like particle ($\varrho\to 0$) and, moreover, does not include the effect of the second, distant wall, we do not expected quantitative agreement with the results obtained in our study. 
We thus focus instead on the sign of the CCF and asymptotic behavior of $\Fcal\st{cas}$: 
for the \bcs considered here, \cref{eq_excess_FE} implies an \emph{attractive} CCF for the combinations $(a,b)$ = (D,D), (N,$+$), ($+$,$+$), and a \emph{repulsive} one for (D,N), (D,$+$), ($+$,$-$) (see also Ref.\ \cite{eisenriegler_casimir_1995}).
These predictions are consistent with the behaviors observed in \cref{fig_Pss_grid,fig_Pss_grid_h1}.
For the Gaussian model in $d=1$ dimensions, the exponent $x_\psi$ takes the value $x_{\phi^2} = -1$ if either particle or wall (or both) have non-symmetry breaking \bcs. This implies that $\Fcal\st{cas}\propto R$ as $R\to 0$, in agreement with the asymptotic behavior resulting from (the negative logarithm of) the expressions in \cref{eq_Pss_h_act,eq_Pss_c_act,eq_Pss_c_act_h1} \footnote{For symmetry-breaking \bcs both on particle and wall, the prediction $x_\phi = 1$ does not describe the asymptotics of \cref{eq_Pss_h_act_h1_ppp,eq_Pss_h_act_h1_mpm}, which might be due to the specifics of the present model.}. 
In $d=3$ dimensions (see \cref{app_steadyst_3d}), we have $x_\phi = x_{\phi^2} = 1$ within the Gaussian model, which agrees in the cases (D/N$,\pm,$D/N) and ($\pm$$\pm$$\pm$) with the asymptotics reported in \cref{eq_pot_3d_h_asympt}.

\section{Summary}
\label{sec_sum}

We have investigated in this study the behavior of a confined point-like tracer particle coupled to a fluctuating order parameter (OP) field $\phi$ within the Gaussian approximation. 
The OP field represents a critical fluid medium in equilibrium and follows either dissipative or conservative dynamics (model A/B \cite{hohenberg_theory_1977}). 
The tracer is governed by a Langevin equation [see \cref{eq_dynamics_genD}] and is subject to reflective \bcs.
We have considered passive as well as reactive types of tracers.
The former is out of equilibrium, since the coupling to the fluctuating fluid represents an energy source that is unbalanced by dissipation. In this sense, a passive tracer bears resemblance to an ``active'' particle \cite{farage_effective_2015,maggi_multidimensional_2015,rein_applicability_2016,fodor_statistical_2018}. 
By contrast, a reactive tracer interacts with the fluid in accordance with the fluctuation-dissipation theorem, such that its steady state obeys equilibrium statistical mechanics. A reactive tracer can be viewed as a simplified model of a colloidal particle in a critical fluid.
The action of a reactive tracer on the fluid is described either by a local chemical potential (linear coupling) or a locally altered correlation length (quadratic coupling) [see \cref{eq_Hamilt}]. A linear coupling enhances the OP around the tracer, as is typically observed for colloids \cite{hertlein_direct_2008,gambassi_critical_2009}.
We have also considered non-vanishing boundary fields ($h_1\neq 0$). These induce a non-uniform OP profile, which manifests as a deterministic force acting on the tracer [see \cref{eq_effpot}]. 

The central quantity in our study is the probability distribution $\bar P(R,t)$ of the tracer position $R$. 
While previous studies of tracers in fluctuating media focused mostly on bulk systems \cite{dean_diffusion_2011,demery_perturbative_2011,demery_diffusion_2013}, we have addressed here the effect of spatial confinement by considering the fluctuation-induced interactions of the tracer with two fixed boundaries (at $R=0$ and $L$).
This complements investigations of the critical Casimir force for (spatially extended) colloidal particles in a half-space \cite{burkhardt_casimir_1995,eisenriegler_casimir_1995,hanke_critical_1998} and in strong confinement \cite{vasilyev_nonadditive_2018,kondrat_probing_2018,vasilyev_bridging_2020}. 
In order to make analytical progress, we have focused on $d=1$ spatial dimensions and employed adiabatic as well as weak-coupling approximations. 
In the adiabatic regime, the dynamics of the OP is fast compared to the one of the tracer, such that the effect of the OP can be approximated as a (spatially correlated) Markovian noise.
This enables a description of the tracer dynamics in terms of a Fokker-Planck equation with (spatially dependent) drift and diffusion coefficients.
In the case of a (linearly coupled) passive tracer, the nonlinear multiplicative noise in the associated Langevin equation can be interpreted in the Stratonovich sense, whereas, in the reactive case, an ``isothermal'' interpretation (also called anti-Ito or H\"anggi-Klimontovich prescription \cite{lau_state-dependent_2007,haenggi_stochastic_1978,klimontovich_ito_1990}) emerges naturally. 
Note that the adiabatic approximation applies only to systems which do not involve a zero mode, such that the otherwise diverging relaxation time of a critical fluid \cite{onuki_phase_2002} is cut off.  
One of our main results is given by \cref{eq_FP_act_h_adiab}, which describes the effective (adiabatic) equilibrium dynamics of a point-like colloidal particle in a confined critical medium in the presence of critical adsorption, i.e., a local enhancement of the OP around the particle. 

We have validated our analytical results by numerically solving the associated Langevin equations [\cref{eq_langevin_1d}] to obtain the steady-state distribution [see \cref{fig_Pss_grid,fig_Pss_adiab,fig_Pss_grid_h1}] as well as the mean-squared displacement [see \cref{fig_msd}].
While we focused a one-dimensional system, analytical calculations of the equilibrium distribution of a reactive tracer in three dimensions revel that they are qualitatively similar to the one-dimensional case (see \cref{app_steadyst_3d}).
A reactive tracer typically obeys a Boltzmann-like equilibrium distribution, $\bar P_s(R)\sim \exp(-\Ucal(R)/T)$, with an effective potential $\Ucal(R)$.
The latter is a consequence of the coupling to the OP field [see \cref{sec_react_dynamics}] and encodes the critical Casimir interactions between tracer and boundaries [see \cref{sec_discuss_react}]. 
In a homogeneous medium ($h_1=0$) and for a linearly coupled tracer (coupling $h$), one has $\Ucal(R) \propto -h^2 \bra\phi(R)^2\ket$ [see \cref{eq_Pss_act_h_adiab,eq_Pss_h_act}], while $\Ucal(R)\propto c\, \bra\phi(R)^2\ket$ in the case of a quadratic coupling ($c$) [see \cref{eq_Pss_act_c_ad}], where $\bra\phi(R)^2\ket = V_\phi(R)$ is the variance of $\phi$ [see \cref{eq_phiR_var}].
(For a quadratically coupled tracer, these expressions apply only in the weak-coupling approximation, as the exact equilibrium distribution is given in \cref{eq_Pss_c_act_det}.)
Non-uniform OP profiles $\bra\phi(R)\ket$, caused by boundary fields $h_1\neq 0$ [see \cref{eq_avg_prof_h1}], render contributions to $\Ucal$ of the form $-h \bra\phi(R)\ket$ and $c\bra\phi(R)\ket^2$, respectively.

The effective potential $\Ucal$ determines also the dynamics of the tracer at the leading order in the adiabatic and the weak-coupling approximations [see \cref{eq_FP_pass_h_adiab,eq_FP_pass_c_adiab,eq_FP_act_h_adiab,eq_FP_act_c_adiab}].
Since, within the Gaussian model considered here, $\bra\delta\phi(R)^2\ket$ and $\bra\phi(R)\ket$ are quadratic functions of $R$, the tracer can in certain cases be effectively described by a confined Ornstein-Uhlenbeck process \footnote{This strictly applies only when the quadratic potential has its minimum in the center of the system (see \cref{fig_effpot}). We recall furthermore that a tracer coupled quadratically to an OP subject to boundary fields $h_1$ experiences a quartic potential.}.
Beyond leading order, deviations from this simple behavior arise because the mobility acquires a spatial dependence [see discussion after \cref{eq_FP_pass_h_adiab,eq_FP_act_h_adiab}].
Note that the adiabatic dynamics of a quadratically coupled reactive tracer has been considered here only to $\Ocal(c)$ and the inclusion of higher-order corrections is reserved for a future study. In the adiabatic regime, the mean-squared displacement of the tracer grows linearly in time, with an effective diffusivity that increases (decreases) with the adiabaticity parameter $\chit\kappa_h$ [see \cref{eq_adiab_valid}] in the passive (reactive) case (see \cref{fig_msd}).

A linearly coupled passive tracer has a higher probability to be located near the boundaries than in the center of the system. 
This is a generic effect resulting from the interplay between confinement and a temporally correlated noise \cite{guggenberger_fractional_2019,vojta_probability_2019}. 
It also arises in the case of active matter, where it gives rise to the accumulation of active particles at surfaces and plays a role for motility-induced phase separation \cite{farage_effective_2015,maggi_multidimensional_2015,rein_applicability_2016}.
The behavior of a quadratically coupled (passive or reactive) tracer, instead, is dominated by the effective potential $\propto c\, \bra\phi(R)^2\ket$, stemming from the nonzero average of the OP-related noise $\sim \phi^2$ [see \cref{eq_tracer_dyn_simpl,fig_phiVar}].
Interestingly, in the parameter regimes considered here, the steady-state distributions of a passive and a reactive tracer are qualitatively similar (see \cref{fig_Pss_grid,fig_Pss_grid_h1}). 
An exception is a tracer coupled linearly to an OP field subject to Dirichlet \bcs (see \cref{fig_Pss_grid}), in which case the steady-state distribution exhibits a crossover behavior controlled by the non-Markovian character of the dynamics [see \cref{fig_Pss_adiab_Dh}].
We finally remark that the steady-state distribution of a passive tracer is similar for dissipative and conservative OP dynamics.

The present study opens up various possibilities to investigate colloidal dynamics in a critical medium within an analytical approach.
In particular, our model can be readily extended to more than one tracer (see \cref{sec_equil_manypart}), which could be utilized to address many-body critical Casimir interactions \cite{mattos_many-body_2013,mattos_three-body_2015,hobrecht_many-body_2015,edison_critical_2015,paladugu_nonadditivity_2016,vasilyev_nonadditive_2018}. 
Furthermore, while we assumed the OP to remain in equilibrium at all times, non-equilibrium scenarios such as OP quenches \cite{gambassi_critical_2006,gross_surface-induced_2018,gross_dynamics_2019,rohwer_correlations_2019} appear to be a rewarding topic.
It is also pertinent to extend the present work towards two and three spatial dimensions, which are the relevant cases for membranes \cite{reister_lateral_2005,honerkamp-smith_introduction_2009,reister-gottfried_diffusing_2010,honerkamp-smith_experimental_2012,camley_contributions_2012}, interfaces \cite{lehle_effective_2006,lehle_importance_2007,oettel_colloidal_2008}, and colloidal suspensions \cite{maciolek_collective_2018}.
Moreover, effects of off-criticality as well as hydrodynamics \cite{bleibel_hydrodynamic_2014,fujitani_fluctuation_2016} could be taken into account in the future.
The Gaussian and weak-coupling approximations employed here provide the leading order contributions of a perturbation expansion of a $\phi^4$-theory \cite{diehl_field-theoretical_1986}. In fact, the couplings are expected to flow under a renormalization group and attain fixed-point values which, depending on the dimension, are not necessarily small. This should be addressed in a future study.
Further attention should also be devoted to non-Markovian effects in the dynamics, which become relevant at large coupling strengths.

\appendix

\section{Dimensional considerations}
\label{app_dimensions}
From the fact that the argument of the exponential in \cref{eq_Pss_joint_T} must be dimensionless, one infers the following dimensions of the field and the static couplings: $[\phi] = [L]^{1-d/2} [T_\phi]^{1/2}$, $[h] = [L]^{d/2-1} [T_R] [T_\phi]^{-1/2}$, $[c]= [L]^{d-2} [T_R]/[T_\phi]$, and $[h_1] = [L]^{-d/2} [T_\phi]^{1/2}$.
The dimensions of the dynamical couplings follow from \cref{eq_tracer_dyn_bare,eq_field_dyn_bare}: $[\gamma_\phi] = [L]^{2+2a} [t]^{-1}$, $[\gamma_R] = [L]^{2} [T_R]^{-1} [t]^{-1}$, $[\chi] = [L]^{-2a} [T_R]^{-1}$, where $[R_\alpha] = [L]$. Although $[T_R] = [T_\phi]$, we distinguish here the two temperatures for clarity. Note that $t$ refers to the unrescaled time [see \cref{eq_dynamics_genD}].

\section{Equilibrium correlation functions of the OP}
\label{app_profile_correl}

Here, we determine the equilibrium OP profile and correlation function within the Gaussian model as defined by \cref{eq_Pss_joint_T,eq_Hamilt}.

\subsection{System with a linearly coupled tracer or boundary fields}
\label{app_profile_correl_lin}

We consider an OP field $\phi$, which may be subject to boundary fields $h_1$ as well as to a bulk field $h$ (at location $R$, representing a linearly coupled tracer).
Connected correlation functions of $\phi$ can be determined in the usual fashion \cite{le_bellac_quantum_1991} from the generating functional
\beq \Zcal[J] \equiv \int\Dcal\phi\, e^{-\frac{1}{T_\phi} \hat\Hcal[\phi,J]},
\label{eq_Z_Hprime}\eeq 
which is obtained by introducing a position-dependent auxiliary field $J(z)$ into the Hamiltonian in \cref{eq_Hamilt}:
\beq \hat \Hcal[\phi,J] \equiv \int\d z \left[ \onehalf (\pd_z\phi)^2 - J(z)\phi(z) \right] - h_1\left[\phi(0)+ \phi(L)\right] - h \phi(R).
\eeq 
We first consider the averaged profile
\beq \bra\phi(z)\ket = \frac{1}{\Zcal} \int\Dcal\phi \, \phi(z) e^{-\frac{1}{T_\phi} \hat\Hcal[\phi,J=0] } =  T_\phi \frac{\delta}{\delta J(z)} \ln \Zcal[J] \Big|_{J=0}.
\label{eq_avg_profile}\eeq 
Switching to mode space, we write $J(z) = \sum_n \sigma_n(z) J_n$ and use \cref{eq_phi_expand,eq_eigenf_ortho} to bring the Hamiltonian into the form
\beq \hat\Hcal = \sum_n \left[ \onehalf k_n^2 |\phi_n|^2 - J_n^* \phi_n -  \hat\tau_n \phi_n \right],
\label{eq_Hamilt_modesp_J}\eeq 
with $\hat\tau_n \equiv h_1[\sigma_n(0) + \sigma_n(L)] + h \sigma_n(R)$. 
In order to regularize a possible zero mode $k_0=0$, we set $k_0 =\epsilon> 0$ and perform the limit $\epsilon\to 0$ at the end of the calculation.
After a Gaussian integration using \cref{eq_gaussint}, the generating functional takes the form \footnote{In contrast to standard field theoretical approaches (see Ref. \cite{gross_statistical_2017} and references therein), we do not separate here the mean profile $\bra\phi(z)\ket$ from the fluctuating part beforehand.}
\beq \Zcal[J] = \int \Dcal\phi_n e^{-\frac{1}{T_\phi} \hat\Hcal[\phi_n, J_n]} \propto \exp\left( \frac{1}{2 T_\phi} \sum_n \frac{| J_n^* + \hat\tau_n |^2}{k_n^2} \right),
\label{eq_Z_J_modes}\eeq 
where we omitted an unimportant normalization factor.
The mode-space expression for the averaged profile resulting from \cref{eq_avg_profile} follows as 
\beq \bra\phi(z)\ket = \sum_n \sigma_n(z) \bra\phi_n\ket =  T_\phi\sum_n \sigma_n(z) \frac{\pd \ln \Zcal[J]}{\pd J_n}\Big|_{J=0} =   \sum_n \frac{\sigma_n(z) \hat\tau_n^*}{k_n^2}.
\label{eq_avg_prof_series}\eeq 
The zero mode $k_0 = \epsilon\to 0$ renders a divergent mean profile in equilibrium if $h\neq 0$ or $h_1\neq 0$. 
In the absence of a zero mode (which applies, in particular, to conserved dynamics), we take Neumann modes for $\sigma_n$ [see \cref{eq_eigenf_Nbc} \footnote{The contribution of boundary fields vanishes when representing $\phi$ in terms of Dirichlet modes, which can be readily seen by using $\sigma_n(0)=0=\sigma_n(L)$ in the expressions in \cref{eq_Pss_h_act_h1_base,eq_Pss_c_act_h1_mat}.}] and use standard Fourier-series relations \cite{gradshteyn_table_2014} to evaluate \cref{eq_avg_prof_series} with $h=0$: 
\beq \bra\phi(z)\ket_{h_1}  \equiv \bra\phi(z)\ket\big|_{h=0} = h_1 L \left[ \frac{1}{6} - \frac{z}{L} + \pfrac{z}{L}^2 \right].
\label{eq2_avg_prof_h1}\eeq 
This profile agrees with the one obtained within linear MFT [see Eq.~(38) in \cite{gross_critical_2016}] and accordingly fulfills, in an averaged sense, (++) capillary \bcs of critical adsorption, 
\beq \pd_z \bra\phi(z)\ket_{h_1}\big|_{z\in\{0,L\}} = \mp h_1,
\label{eq2_cap_bcs}\eeq 
as well as $\int_0^L \d z \bra\phi(z)\ket = 0$.
If, instead, $h_1=0$ and the OP field is only subject to a bulk-like field $h\neq 0$ at location $R$, \cref{eq_avg_prof_series} renders
\beq 
\bra\phi(z)\ket\big|_{h_1=0} =  \begin{cases} \displaystyle
	  h L \left[ \frac{1}{L} \min(R, z) - \frac{z  R}{L^2}   \right], & \text{(D)} \\ \displaystyle
	  h L \left[ \frac{1}{3} - \frac{1}{L} \max(R, z) + \frac{1}{2L^2}\left(R^2 + z^2\right) \right]. \qquad & \text{(N$^*$)} 
  \end{cases}
\label{eq_avg_prof_h}\eeq 

In an analogous way, we obtain from \cref{eq_Z_J_modes} the connected static correlation function [see \cref{eq_correl_def,eq_C_phi}]: 
\beq C_\phi(x,y) = \bra\delta\phi(x) \delta\phi(y)\ket = \sum_{n,m} \sigma_n(x) \sigma_m(y) \bra \delta\phi_n \delta\phi_m\ket =  T_\phi^2 \sum_{n,m} \sigma_n(x)\sigma_m(y) \frac{\pd^2 \ln \Zcal}{\pd J_n \pd J_m}\Big|_{J=0} =  T_\phi \sum_n \frac{\sigma_n(x)\sigma_n^*(y) }{k_n^2},
\label{eq_phi_correl_genfunc}\eeq 
where $\delta\phi \equiv \phi - \bra\phi\ket$ denotes the fluctuation part of $\phi$ and we used $J_n^* = J_{-n}$ (which follows from $J(z)$ being real-valued) and $[\sigma\pbc_n(z)]^* = \sigma\pbc_{-n}(z)$.
Note that $C_\phi$ is independent of $h$ and $h_1$ and thus applies irrespective of the presence of boundary fields or linearly coupled tracers. Explicit expressions for $C_\phi$ are provided in \cref{eq_phi_correl}. It is useful to note that the profile in \cref{eq2_avg_prof_h1} can be written as 
\beq 
\bra\phi(z)\ket_{h_1} = \frac{h_1}{T_\phi}\left[ C_\phi\Nbc(0,z) + C_\phi\Nbc(L,z)\right],
\label{eq_avg_prof_h1_correl}\eeq 
which readily follows from \cref{eq_avg_prof_series,eq_phi_correl_genfunc}.

\subsection{System with a quadratically coupled tracer}
\label{sec_phivar_quadcoupl}

\begin{figure}[t]\centering
    \includegraphics[width=0.32\linewidth]{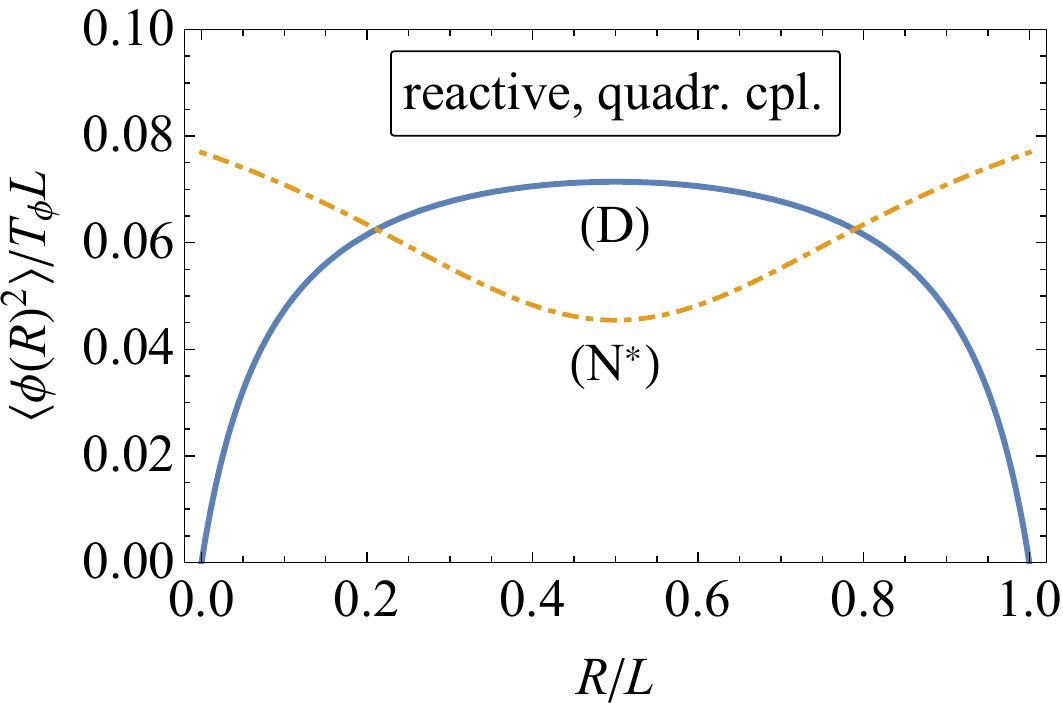} 
    \caption{Static variance $\bra\phi(R)^2\ket$ [\cref{eq_phiR_var_c_act}] of an OP field $\phi$ subject to Dirichlet or Neumann \bcs (without a zero mode) and quadratically coupled to a tracer at location $R$. A value $\kappa_c=10$ is used for illustrative purposes, noting that $\bra\phi(R)^2\ket=V_\phi(R)$ for $\kappa_c=0$ [see \cref{eq_phiR_var,fig_phiVar}].
    }
    \label{fig_phiVar_c_act}
\end{figure}

In the case of a quadratically coupled tracer, correlations can be determined analogously to \cref{eq_phi_correl_genfunc} by defining $\Zcal$ [\cref{eq_Z_Hprime}] in terms of the Hamiltonian $\hat \Hcal[\phi,J] = \int\d z \left[ \onehalf (\pd_z\phi)^2 - J(z)\phi(z) \right] +\onehalf c \phi(R)^2 = \onehalf \sum_{n,m} \phi_n \Gamma_{nm} \phi_m^* - \sum_n J^*_n\phi_n$, with $\boldsymbol\Gamma$ given in \cref{eq_Gamma_mat} [see also \cref{eq_Gamma_mat_alt}].
We focus here on the variance $\bra\phi(R)^2\ket$ of the field at the location of the tracer, which can be obtained from \cref{eq_Pss_pathint} as
\beq \bra\phi(R)^2\ket = -2T_R \pd_c \ln \Qcal(R),\qquad  \Qcal(R) \equiv \int \Dcal\phi \exp\left\{ - \int_V \d z  \frac{1}{2 T_\phi} \left(\pd_z \phi\right)^2 - \frac{c}{2 T_R} \phi(R)^2  \right\} = [\det \boldsymbol \Gamma(R)]^{-1/2},
\label{eq_phiR_var_basic}\eeq 
where we have used \cref{eq_gaussint,eq_Pss_c_act_det} and omitted an (infinite) numerical prefactor which cancels out in \cref{eq_phiR_var_basic}. (The same result follows from $\delta^2 \Zcal/\delta J(R)^2$.)
Upon using the expression for $\det\boldsymbol\Gamma$ stated in \cref{eq_actC_det_res,eq_actC_det_zeromode}, one obtains (see \cref{fig_phiVar_c_act})
\beq \bra\phi(R)^2\ket = \frac{1}{\displaystyle \frac{\kappa_c}{T_\phi L} + \frac{1}{V_\phi}} =  \begin{cases} 
   \frac{T_\phi L}{\kappa_c + 12}, \qquad &\text{(p$^*$)} \smallskip \\ 
   \frac{T_\phi L }{\kappa_c}, &\text{(p,N)} \smallskip \\ 
    \frac{T_\phi L}{\kappa_c + \left(\rho-\rho^2\right)^{-1}}, & \text{(D)} \smallskip \\ 
    \frac{T_\phi L}{\kappa_c + \left(\frac{1}{3}-\rho + \rho^2\right)^{-1} }, & \text{(N$^*$)}
  \end{cases}
\label{eq_phiR_var_c_act}\eeq 
with $\rho\equiv R/L$, $\kappa_c$ given in \cref{eq_effcoupl_c} and $V_\phi$ in \cref{eq_phiR_var}.
In the absence of a coupling to the tracer, i.e., for $\kappa_c= 0$, \cref{eq_phiR_var_c_act} reduces to the expression in the passive case, \cref{eq_phiR_var}.
In the limit $\kappa_c\to\infty$, corresponding to Dirichlet \bcs at $R$, $\bra\phi(R)^2\ket$ vanishes.

\section{Adiabatic limit of the OP dynamics}
\label{app_adiab_OP}

In order to determine an approximation to the solution $\phi(z,t)$ [\cref{eq_phi_mode_sol}] in the adiabatic limit ($\chit\ll 1$), we substitute in \cref{eq_phi_mode_sol_n} the integration variable $s=t-u \chi/( k_n^2 k_n^{2a})$ and obtain 
\begin{align}
\phi_n(t) = \frac{\chi}{ k_n^2 k_n^{2a}} \int_0^{\infty} \d u\, e^{-u} \left\{ \chi^{-1} k_n^{2a} \left[ \zeta h \sigma_n^*\left(R\left(t - \frac{u \chi}{ k_n^2 k_n^{2a}}\right)\right) + h_1\tau_n^*\right] + \chi^{-1/2}\xi_n\left(t - \frac{u \chi}{ k_n^2 k_n^{2a}}\right)\right\}. \quad (n\neq 0) \label{eq_phi_h_adiab_expand_n_base} 
\end{align}
The dependence on $\chit$ can be made explicit by rescaling time accordingly, see the discussion around \cref{eq_dynamics_genD}.
Due to the exponential, the integrand gives substantial contributions only if $u\lesssim \Ocal(1)$. 
Accordingly, for $\chit \ll 1$ it is justified to Taylor expand the terms in the square bracket in \cref{eq_phi_h_adiab_expand_n_base} up to first order in $u \chi/( k^2 k^{2a})$, rendering
\beq\begin{split} 
\phi_n(t) &\simeq \int_0^\infty \d u\ e^{- u} \left[ \frac{\zeta h}{k_n^2} \sigma_n^*(R(t)) + \frac{h_1}{k_n^2} \tau_n^* - \frac{\zeta h \chi}{ k_n^4 k_n^{2a}} u \dot R(t) \pd_R\sigma_n^*(R(t))  + \frac{\sqrt{\chi}}{ k_n^2 k_n^{2a}} \xi_n(t)\right] \\
&= \frac{\zeta h}{k_n^2} \sigma_n^*(R(t)) + \frac{h_1}{k_n^2} \tau_n^* - \frac{\zeta h \chi}{ k_n^4 k_n^{2a}} \dot R(t) \pd_R \sigma_n^*(R(t)) + \frac{\sqrt{\chi}}{ k_n^2 k_n^{2a}} \xi_n(t). \qquad (n\neq 0)
\end{split}\label{eq_phi_h_adiab_expand_n}\eeq 
It is not feasible here to expand beyond $\Ocal(u)$ or, correspondingly $\Ocal(\chit)$, since this would generate derivatives of the noise $\xi_n(t)$.
We remark that, alternatively to deriving an equation of motion, the adiabatic approximation can also be applied on the level of the correlation function $C_\phi$ in \cref{eq_C_phi}.

\section{Langevin and Fokker-Planck equations: spurious drift and steady-state}
\label{app_spurious}

Given the FPE 
\beq \pd_t P(R,t) = - \pd_R \left[\mu(R) P(R,t) \right] + \pd_R^2 \left[ D(R) P(R,t) \right] ,
\label{eq_app_FPE}\eeq 
the form of the associated Langevin equation depends on the chosen integration rule for the noise \cite{gardiner_stochastic_2009}. 
Possible rules can be parametrized by a quantity $\alpha$, with $0\leq \alpha\leq 1$, such that \cref{eq_app_FPE} maps to \cite{lau_state-dependent_2007,volpe_effective_2016}
\beq \pd_t R = \mu(R) - \alpha D'(R) + \sqrt{D(R)} \theta,\qquad \bra\theta(t)\theta(t')\ket = 2\delta(t-t'),
\label{eq_app_Lang}\eeq 
with a Gaussian white noise $\theta$.
The parameter $\alpha$, in fact, determines the point $t_i^* = t_{i-1}+\alpha(t_i-t_{i-1})$ in the interval $[t_i, t_{i-1}]$ at which the noise is evaluated in a discretization of \cref{eq_app_Lang}. Common choices in the literature are $\alpha=0$ (Ito convention), $\alpha=1/2$ (Stratonovich convention), and $\alpha=1$ (called anti-Ito, isothermal, or H\"{a}nggi-Klimontovich convention) \cite{lau_state-dependent_2007,haenggi_stochastic_1978,klimontovich_ito_1990,volpe_effective_2016}.
Conversely, when starting from a Langevin equation of the form $\pd_t R = \tilde \mu(R) + \sqrt{D(R)} \theta$, the drift term $\mu(R)$ in the FPE in \cref{eq_app_FPE} would be replaced by $\tilde\mu(R) + \alpha D'(R)$.
The contribution $\alpha D'(R)$ denotes a ``spurious'' drift and appears in either the Langevin or the FPE when $\alpha\neq 0$.

The steady-state solution $P_s(R)$ of \cref{eq_app_FPE} is given by
\beq P_s(R) = \frac{1}{\Zcal} \frac{1}{D(R)} \exp\left[ \int_{0}^R \d z \frac{\mu(z)}{D(z)} \right]
\label{eq_app_Pss}\eeq 
with a normalization constant $\Zcal$.
The drift typically takes the following generic form:
\beq \mu(z) = -\frac{D(z)}{T} \Ucal'(z) + \alpha D'(z),
\label{eq_app_drift}\eeq 
where $\Ucal(z)$ is a potential, $D(z)/T$ represents a mobility and $\alpha D'(z)$ is the ``spurious'' drift.
Using \cref{eq_app_drift} in \cref{eq_app_Pss} renders
\beq P_s(R) = \frac{1}{\Zcal} \frac{1}{D(R)^{1-\alpha}} \exp\left[ -\Ucal(R) \right].
\label{eq_app_Boltzm}\eeq 
Accordingly, in order to recover the  standard Boltzmann equilibrium distribution from the FPE in \cref{eq_app_FPE}, the drift in \cref{eq_app_drift} must involve a ``spurious'' contribution with $\alpha=1$ (isothermal convention) \cite{lau_state-dependent_2007}.

\section{Adiabatic elimination for a linearly coupled tracer}
\label{app_adiab_elim}

\begin{figure}[t]\centering
    \subfigure[]{\includegraphics[width=0.3\linewidth]{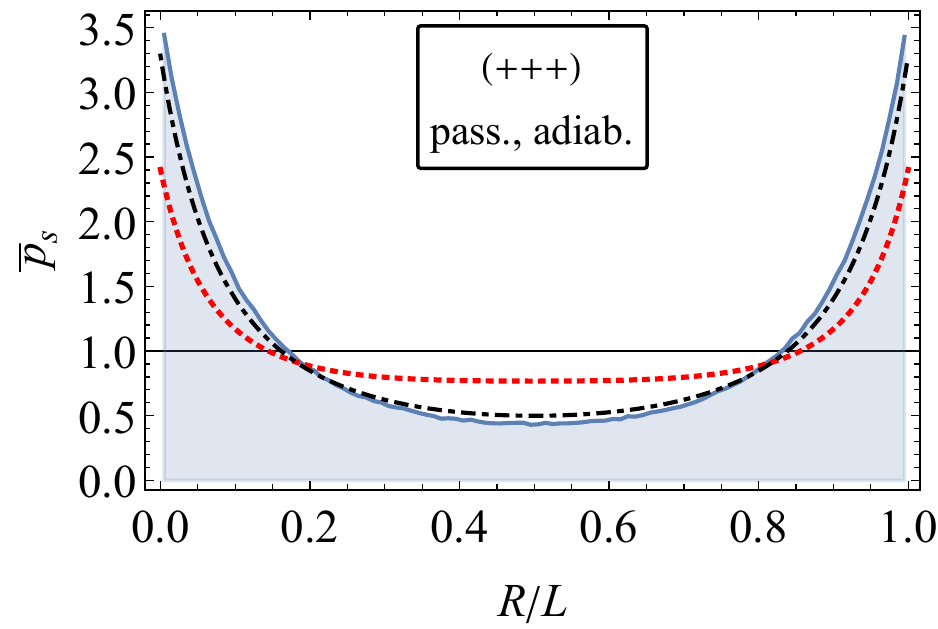} }\qquad
    \subfigure[]{\includegraphics[width=0.3\linewidth]{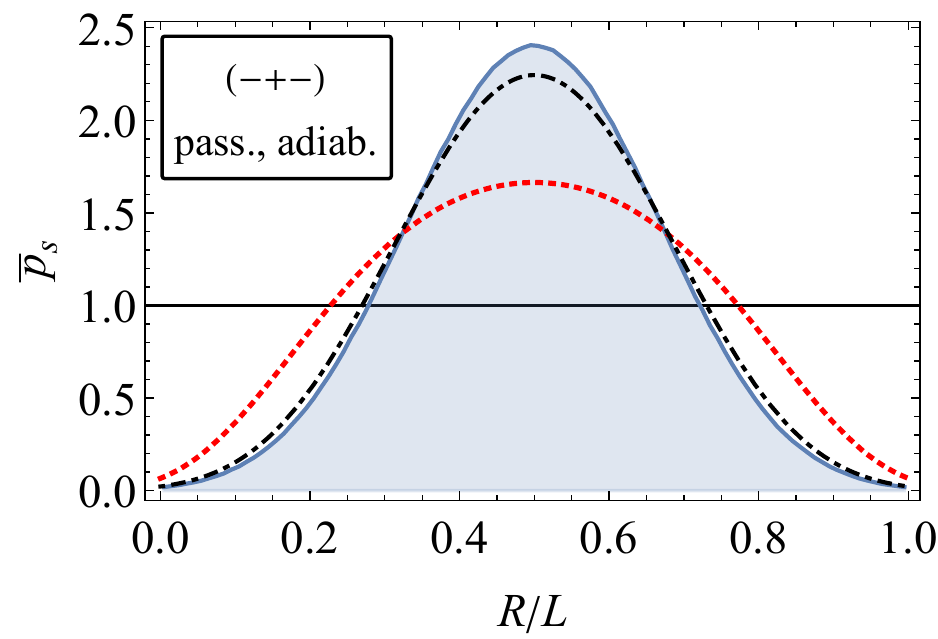} }
    \caption{Assessment of the predictions for the steady-state distribution $\bar p_s = L \bar P_s$ of a passive tracer ($\zeta=0$) determined from the two adiabatic elimination schemes in \cref{eq_adiab_drift_diffus_r,eq_adiab_drift_diffus_p}. The tracer is linearly coupled to an OP field subject to capillary \bcs with (a) $h h_1>0$ and (b) $h h_1<0$. The solid curves (with filling) represent simulation results (see \cref{sec_sim}) obtained for $\kappa_{h_1}\simeq 25$ and (a) $\chit = 0.5$, (b) $\chit=0.1$. The dashed-dotted curve represents $\bar p_s$ [\cref{eq_app_Pss}] obtained from \cref{eq_adiab_drift_diffus_p}, which accurately describes the simulation results even for large values of $\chit\kappa_{h_1}$ [see \cref{eq_adiab_valid}]. In comparison, the dotted curve shows $\bar p_s$ determined from \cref{eq_adiab_drift_diffus_r} (with $\zeta=0$). }
    \label{fig_Pss_adiab_comp}
\end{figure}

We focus on a linearly coupled tracer and recall that \cref{eq_tracer_dyn,eq_phi_mode_eqn} form a system of Langevin equations:
\begin{subequations}
\begin{align}
\dot R(t) &= h \sum_n k_n \tilde\sigma_n(R(t)) \phi_n(t) + \eta(t), \label{eq_adiab_langevin_R} \\
\pd_t \phi_n &= \chi^{-1} k_n^{2a} \left\{ -k_n^2 \phi_n + \zeta h   \sigma_n^*(R(t)) + h_1\tau_n^* \right\} + \chi^{-1/2} \xi_n, 
\end{align}\label{eq_adiab_langevin}
\end{subequations}
\hspace{-0.1cm}where $\tau_n \equiv \sigma_n(0)+\sigma_n(L)$.
We seek to determine an effective Markovian Langevin equation for the tracer position $R$, assuming the field $\phi$ to be a fast variable, i.e., $\chit\ll 1$.
In the literature, a number of adiabatic elimination schemes for multidimensional Langevin equations (linear in the fast variable) have been developed, see, e.g., Refs.\ \cite{stratonovich_topics_1963,gardiner_stochastic_2009,pavliotis_stochastic_2014,mori_contraction_1980,morita_contraction_1980,kaneko_adiabatic_1981,theiss_systematic_1985}.
For the specific system in \cref{eq_adiab_langevin}, the Chapman-Enskog based approach of Ref.\ \cite{theiss_systematic_1985,theiss_remarks_1985} results in a FPE which renders a steady-state distribution in accord with equilibrium statistical mechanics [see \cref{eq_Pss_h_act,eq_Pss_h_act_h1_base}].

\Cref{eq_adiab_langevin} can be matched onto the formalism in Refs.\ \cite{theiss_systematic_1985,theiss_remarks_1985} by identifying $c_1=R$, $b_n=\phi_n$, $\alpha_{1 i}=-h k_i \tilde\sigma(R)$, $\gamma_{ij} = \chi^{-1} k_i^{2+2a}$, $\beta_i = \zeta \chi^{-1} h k_i^{2a} \sigma_i^*(R) + \chi^{-1} h_1 k_i^{2a} \tau_i $, $V_{ij} = (T_\phi/k_i^2) \delta_{ij}$, where $V_{ij} = \bra \delta\phi_i \delta\phi_j\ket$ is the equilibrium correlation function of the field modes, which can be inferred from \cref{eq_phi_correl_genfunc}. In the case of periodic \bcs, one has $V_{ij}=(T_\phi/k_i^2)\delta_{i,-j}$, instead.

A \emph{reactive} tracer ($\zeta=1$) affects the local value of $\phi$, implying that the scheme stated in Eq.~(2.9) of Ref.\ \cite{theiss_remarks_1985} is appropriate.
This renders the following FPE for the tracer distribution $\bar P$:
\beq \pd_t \bar P(R,t) = - \pd_R \left[\mu(R)  \bar P(R,t) \right] + \pd_R^2 \left[ D(R) \bar P(R,t) \right] ,
\label{eq_adiab_FPE}\eeq 
with the drift and the diffusion coefficient
\begin{subequations}
\begin{align}
\mu(R) &=  \Acal(R) \left[ 1- \chi h^2 L^{1+2a} m(R) \right]  - \left(\frac{3}{2}\zeta T_R - \onehalf T_\phi\right) \chi h^2 L^{1+2a} m'(R), \label{eq_adiab_drift_r}  \\
D(R) &= T_R - \chi h^2 L^{1+2a} m(R)\left(2\zeta T_R - T_\phi\right).
\end{align}\label{eq_adiab_drift_diffus_r}
\end{subequations}
\hspace{-0.1cm}Here, $m(R)$ is defined in \cref{eq_m_func} and 
\beq \Acal(R) \equiv  \sump_n \frac{1}{k_n}\left[ \zeta h^2 \tilde \sigma_n(R) \sigma_n^*(R) + h h_1 \tilde \sigma_n(R) \tau_n^* \right] 
= \frac{1}{T_\phi} \pd_R \left\{ \onehalf \zeta h^2  V_\phi(R) + h h_1 \left[ C_\phi(0,R) + C_\phi(L,R) \right] \right\}
\label{eq_adiab_drift_force}
\eeq 
is an effective forcing term [recall \cref{eq_avg_prof_h1_correl}], which is, in fact, independent of $T_\phi$. For clarity, we have made the presence of $\zeta$ explicit.
Note that the steady-state solution of \cref{eq_adiab_FPE}, which is given by \cref{eq_app_Pss} with $\Acal(R)=\Ucal'(R)$, attains the correct Boltzmann-like equilibrium form [see \cref{eq_Pss_h_act,eq_Pss_h_act_h1_base,eq_app_Boltzm}] only for $T_\phi=T_R$.

By contrast, if the tracer is \emph{passive} ($\zeta=0$), the appropriate scheme is given by Eq.~(2.10) of Ref.\ \cite{theiss_remarks_1985}, which yields the FPE in \cref{eq_adiab_FPE} with the following drift and diffusion coefficients:
\begin{subequations}
\begin{align}
\mu(R) &=  \Acal(R)  + \onehalf T_\phi \chi h^2 L^{1+2a} m'(R), \label{eq_adiab_drift_p} \\
D(R) &= T_R + T_\phi\chi h^2 L^{1+2a} m(R).
\end{align}\label{eq_adiab_drift_diffus_p}
\end{subequations}
\hspace{-0.1cm}For $\zeta=0$, these expressions are identical to the ones in \cref{eq_adiab_drift_diffus_r}, except for the prefactor of $\Acal$ (mobility), which is unity in \cref{eq_adiab_drift_p}.

In order to demonstrate that the drift in \cref{eq_adiab_drift_p} rather than the one in \cref{eq_adiab_drift_r} (for $\zeta=0$) provides the correct description of a passive tracer, we compare in \cref{fig_Pss_adiab_comp} the steady-state distributions $\bar P_s$ resulting from \cref{eq_app_Pss} to the ones obtained from simulation (see \cref{sec_sim}). (Note that this requires us to use $h_1\neq 0$, as otherwise $\Acal$ vanishes.) As illustrated in \cref{fig_Pss_adiab_comp}, the expressions in \cref{eq_adiab_drift_diffus_p} indeed provide a more accurate description of $\bar P_s$ than the ones in \cref{eq_adiab_drift_diffus_r}, in particular also for larger values of the adiabaticity parameter $\chit$.

It is useful to compare the above result to a ``naive'' adiabatic elimination based on the adiabatic approximation of the OP $\phi$ given in \cref{eq_phi_h_adiab_expand_n}. Inserting this into \cref{eq_tracer_dyn} renders 
\beq \Mcal(R(t)) \dot R(t) = \Acal(R(t)) + \Pi_h(R(t),t) + \eta(t),
\label{eq_tracer_dyn_active_ad}\eeq  
with the noise $\Pi_h$ reported in \cref{eq_tracer_pass_ad_noise}.
Notably, the coupling to the OP ($\zeta = 1$) has generated an effective damping term
\beq \Mcal(R) \equiv 1 + \zeta h^2 \sump_n \frac{\chi |\tilde \sigma_n(R)|^2}{ k_n^2 k_n^{2a}}  = 1 + \frac{\zeta \chit h^2 L}{T_\phi} m(R) .
\label{eq_M_term}\eeq
Upon dividing \cref{eq_tracer_dyn_active_ad} by $\Mcal$ and expanding to $\Ocal(\chi)$, the associated FPE takes, for $\zeta=1$, the form given in \cref{eq_adiab_FPE,eq_adiab_drift_diffus_r} and, for $\zeta=0$, the form in \cref{eq_adiab_FPE,eq_adiab_drift_diffus_p}, except for the spurious drift term.
The latter cannot be determined from the Langevin approach, as it leaves the stochastic interpretation of the multiplicative noise unclear. 

In order to quantify the regime of validity of the adiabatic approximation, we estimate [noting that $\Acal(R)$ in \cref{eq_adiab_drift_force} is linear in $R$] the characteristic relaxation time $t_R$ of the process described by \cref{eq_adiab_FPE} at $\Ocal(\chi^0)$ by the one of an Ornstein-Uhlenbeck process \cite{gardiner_stochastic_2009}:
\beq t_R\sim L(\zeta h^2 + 2h h_1)^{-1}.
\label{eq_OU_relax}\eeq 
Note that, for small $|h|$ and $|h_1|$, the relaxation time is bounded by the one of a free (uncoupled, $h=0$) tracer in confinement, $t_{R,\text{free}}\sim L^2/T_R$.
On the other hand, the typical OP relaxation time is given by $t_\phi \sim \chi L^{2+2a}/\pi^{2+2a}$ [see \cref{eq_C_phi}].
The adiabatic approximation holds if the ratio of the characteristic relaxation times of the OP and the tracer is small, i.e., 
\beq t_\phi/t_R \ll 1.
\eeq 
Using \cref{eq_adiab_param,eq_effcoupl_h,eq_effcoupl_hh1} (with $T_R=T_\phi$), this implies, in particular,
\beq \chit\kappa_h \ll 1,\qquad \chit \kappa_{h_1}\ll 1.
\label{eq_adiab_valid}\eeq

\section{Noise contribution in the adiabatic limit for a quadratically coupled passive tracer}
\label{sec_pass_c_adiab}

Since $C_\phi$ and its derivatives appear quadratically in the correlation function in \cref{eq_Pi_c_correl}, one cannot simply use \cref{eq_C_phi_adiab,eq_C_phi_deriv_ad} [or, equivalently, \cref{eq_phi_mode_adiab}] in order to obtain the adiabatic limit, since this gives rise to terms $\propto \delta^2(t-t')$ (see also Refs.\ \cite{miguel_theory_1981,luczka_non-Markovian_1995} for related discussions).
In the standard adiabatic elimination procedure \cite{stratonovich_topics_1963,gardiner_stochastic_2009}, $\Pi_c$ is approximated by a zero-mean white noise  $\tilde \Pi_c$ with correlations
\beq \bra\tilde \Pi_c(z,t) \tilde \Pi_c(z,t')\ket = 2\Pcal(z) \delta(t-t'),
\eeq 
where $\tilde \Pi_c$ and $\Pi_c$ have the same fluctuation amplitude, 
\beq \int_{-\infty}^\infty \d t\, \bra\Pi_c(z,t) \Pi_c(z,t')\ket = \int_{-\infty}^\infty \d t\, \bra \tilde \Pi_c(z,t) \tilde \Pi_c(z,t')\ket = \Pcal(z),
\label{eq_correl_adiab_int}\eeq 
assuming time translation invariance.
In order to calculate the amplitude $\Pcal(z)$, we insert the mode representations stated in \cref{eq_C_phi,eq_C_phi_deriv} into \cref{eq_Pi_c_correl} and obtain (in the case $h_1=0$)
\beq 
\Pcal(z) = \chi c^2 T_\phi^2 \sump_{n,m} \left[\frac{|\sigma_n(z)|^2  |\tilde\sigma_m(z)|^2}{k_n^2( k_n^{2+2a} + k_m^{2+2a})} + \frac{\sigma_n(z) \tilde\sigma_n^*(z)\sigma_m(z) \tilde\sigma_m^*(z)}{k_n k_m (k_n^{2+2a} +  k_m^{2+2a})}\right].
\label{eq_correl_adiab_modesum}\eeq 
Note that, due to the adiabatic limit, $n=0$ or $m=0$ are excluded from the sum.

\section{Equilibrium distributions and matrix calculations}
\label{sec_matrix}

In order to obtain the equilibrium distributions in \cref{eq_Pss_c_act_det,eq_Pss_c_act_h1_mat}, the determinant and the inverse of the matrix $\boldsymbol \Gamma$ [\cref{eq_Gamma_mat}] entering the multivariate Gaussian integral in \cref{eq_gaussint} must be evaluated.
To this end, we define a diagonal matrix $\Av$ and a vector $\uv$,
\beq A_{nm} = \frac{k_n^2}{T_\phi} \delta_{n,m},\qquad u_n = \sqrt{\frac{c}{T_R}}\, \sigma_n(R),
\label{eq_matrix_A}\eeq 
where the eigenfunctions $\sigma_n$ and indices for the various \bcs are specified in \cref{eq_eigenspec}.
For the zero mode $n=m=0$, we replace the vanishing entry $A_{00}$ with 
\beq A_{00} = \frac{\varepsilon}{T_\phi} >0,
\label{eq_matrix_zeromode_regul}\eeq
where $\varepsilon$ is regularization parameter which set to zero at the end of the calculation.
According to \cref{eq_matrix_A,eq_matrix_zeromode_regul}, we can write \cref{eq_Gamma_mat} as
\beq \boldsymbol \Gamma = \Av + \uv \uv^\dagger,
\label{eq_Gamma_mat_alt}\eeq 
where $^\dagger$ denotes transposition and complex conjugation (the latter being relevant only in the case of periodic \bcs).

\subsection{Determinant}
\label{sec_det}

The form of \cref{eq_Gamma_mat_alt} allows us to apply the matrix determinant lemma: 
\beq \det \boldsymbol \Gamma =  (1+\uv^\dagger \Av^{-1} \uv)\det(\Av).
\label{eq_det_lemma}\eeq 
For the purpose of regularization, we keep the mode number $M$ finite and let $M\to \infty$ only at the end of the calculation. 
Considering first \bcs without a zero mode, it follows from the expression for the variance $V_\phi(R)$ in \cref{eq_phiR_var} that 
\beq \uv^\dagger \Av^{-1} \uv \overset{M\to\infty}{=} \frac{c}{T_R} V_\phi(R) .
\eeq 
Accordingly, \cref{eq_det_lemma} evaluates to 
\beq \det \boldsymbol \Gamma = \Ncal \left[1+\frac{c}{T_R} V_\phi(R) \right] = \begin{cases}
   \Ncal\pbc \left[1 + \frac{1}{12}\kappa_c\right], \qquad &\text{(p$^*$)} \\
   \Ncal\Dbc \left[ 1 + \kappa_c \left( \rho - \rho^2 \right) \right], & \text{(D)} \\
   \Ncal\Nbc \left[ 1 + \kappa_c \left( \frac{1}{3} - \rho + \rho^2 \right) \right], & \text{(N$^*$)}
  \end{cases}
\label{eq_actC_det_res}\eeq 
where $\rho\equiv R/L$ and $\kappa_c=c L T_\phi/T_R$ [see \cref{eq_effcoupl_c}].
While the quantity $\Ncal \equiv \det(\Av)$ formally diverges for $M\to\infty$, it is independent of $R$ and $c$ and thus canceled by the normalization of the distribution [see, e.g., \cref{eq_Pss_c_act_det}].
We obtain $\Ncal\ut{(D,N$^*$)} =  \frac{\pi^{2M} \prod_{n=1}^M n^2}{T_\phi^M L^{2M}}$ and $\Ncal\pbc = \left(2^{2M} \Ncal\NbcNZM\right)^2$.
For \bcs with a zero mode, \cref{eq_phi_correl} implies $1 +\uv^\dagger \Av^{-1} \uv \simeq \frac{\kappa_c }{L^2 \varepsilon}$ in the limit $\varepsilon\to 0$, such that
\beq \det \boldsymbol \Gamma = \frac{\kappa_c}{L^2} \Ncal, \qquad \text{(p,N)}
\label{eq_actC_det_zeromode}\eeq 
which is independent of $R$.

\subsection{Inverse}
\label{sec_inv}

According to the Sherman-Morrison formula \cite{press_numerical_2007}, the inverse of $\boldsymbol{\Gamma}$ in \cref{eq_Gamma_mat_alt} is given by
\beq \boldsymbol\Gamma^{-1} = \Av^{-1} - \frac{\mathbf{G}}{1+\uv^\dagger \Av^{-1} \uv},
\label{eq_sherman_morrison}\eeq 
with $(\uv \uv^\dagger)_{nm}=u_n u^*_m$ and 
\beq G_{nm}\equiv \left(\Av^{-1} \uv \uv^\dagger \Av^{-1}\right)_{nm} = \left(\Av^{-1} \uv\right)_n \left(\Av^{-1} \uv\right)_m^* ,
\eeq
where the last equation follows from the fact that $\Av$ is a real symmetric matrix.
Upon introducing the static correlation function $C_\phi$ via \cref{eq_phi_correl}, one obtains:
\beq \sum_{n,m} \sigma_n(x) \Gamma^{-1}_{nm}(R) \sigma^*_m(y) = C_\phi(x,y) - \frac{c}{T_R} \frac{C_\phi(x,R) C_\phi(R,y)}{1 + \frac{c}{T_R} V_\phi(R)}.
\label{eq_actC_Gamma_correl}\eeq 
In the specific case of Neumann \bcs with a zero mode we also need the following expression:
\beq \sum_{n,m=0}^\infty  [\sigma_n(0)+\sigma_n(L)] \Gamma_{nm}^{-1} [\sigma_n(0) + \sigma_n(L)] = L T_\phi + 4 \frac{T_R}{c} ,\qquad \text{(N)}
\label{eq_sherman_morrison_inv_zeromode}\eeq
which is obtained by inserting $C_\phi\Nbc$ [\cref{eq_phi_correl}] into  \cref{eq_actC_Gamma_correl} and performing the limit $\varepsilon\to 0$.

\section{Equilibrium distributions of a reactive tracer in $d=3$ dimensions}
\label{app_steadyst_3d}

The calculation of the equilibrium distributions for $d=1$ in \cref{sec_react_equil} is extended here to a film geometry in $d>1$ dimensions. Where necessary, we specialize the calculation to $d=3$.
The film is assumed to have boundaries at $z=0,L$ and is macroscopically extended (transverse area $A$) in the other directions. 
For the purpose of regularization, we first consider a finite $A$ and perform the thin film limit $A\to\infty$ at the end of the calculation.

Inserting the standard eigenmode expansion of the OP,
\beq  \phi(\rvp,z) = \frac{1}{\sqrt{A}} \sum_{n,\pv} \sigma_n(z) e^{\im \pv\cdot\rvp} \phi_n(\pv),
\label{eq_phi_mode_expansion}\eeq 
into the Hamiltonian in \cref{eq_Hamilt} renders
\beq \Hcal = \sum_{n,\pv} \left\{ \onehalf \sum_{m,\qv} \phi_n(\pv)\tilde\Gamma_{n,\pv,m,\qv} \phi_m^*(\qv) - \frac{h}{T_R \sqrt{A}} \sigma_n(R_z) e^{\im\pv\cdot \Rvp} \phi_n(\pv) - \frac{\sqrt{A} h_1}{T_\phi} [\sigma_n(0) + \sigma_n(L)] \delta_{\pv,0} \phi_n(\pv) \right\}, \label{eq_Hamilt_ddim_base}
\eeq 
with
\beq 
\tilde \Gamma_{n,\pv,m,\qv}(R_z) = \frac{1}{T_\phi} (k_n^2 + \pv^2) \delta_{n,m}\delta_{\pv,\qv} + \frac{c}{A T_R} e^{\im\pv\cdot\Rvp} \sigma_n(R_z) e^{-\im\qv\cdot\Rvp} \sigma_{-m}(R_z), \label{eq_Gamma_tilde}
\eeq 
wherein $R_z$ denotes the $z$-coordinate of the tracer position.
In writing the last term in $\boldsymbol{\tilde\Gamma}$, we used the fact that $\phi_{-n}(-\pv) = \phi_n^*(\pv)$ for a real-valued $\phi(\rv)$, such that one recovers the expected contribution $\frac{c}{2A} \sum_{m,n\atop \pv,\qv} e^{\im\pv\cdot\Rvp + \im \qv\cdot\Rvp} \times \\ \sigma_n(R_z)\sigma_m(R_z) \phi_n(\pv)\phi_m(\qv)$ in the Hamiltonian.
We now separately discuss the case of a linearly and a quadratically coupled tracer.

\subsection{Linearly coupled tracer}

\begin{figure}[t]\centering
    \subfigure[]{\includegraphics[width=0.31\linewidth]{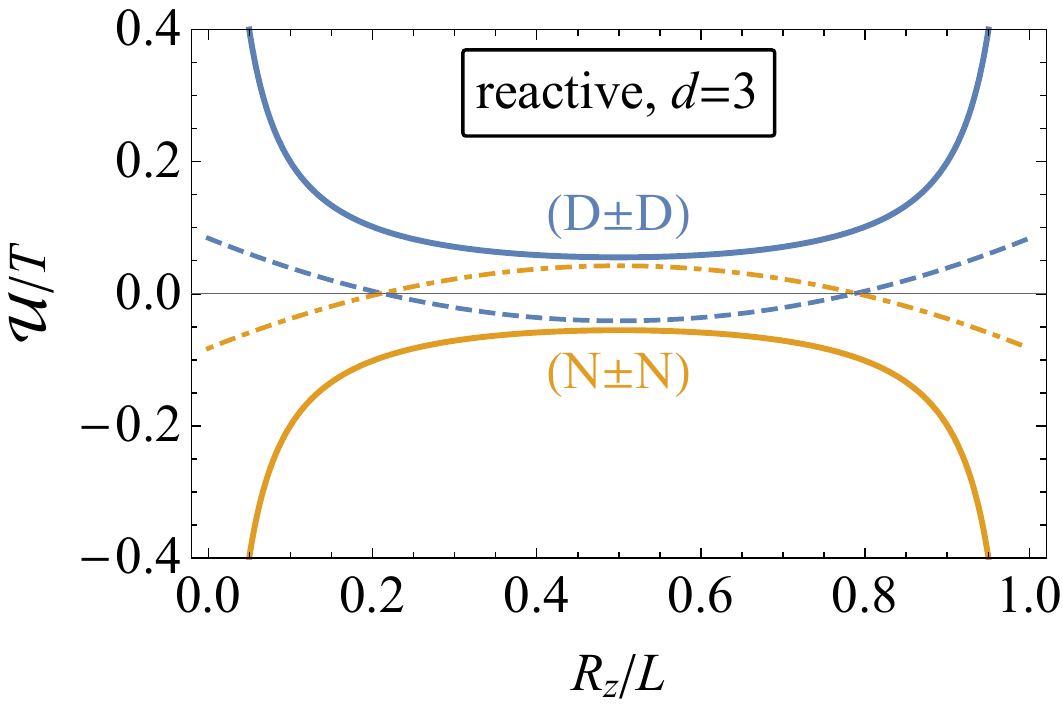} }\quad
    \subfigure[]{\includegraphics[width=0.31\linewidth]{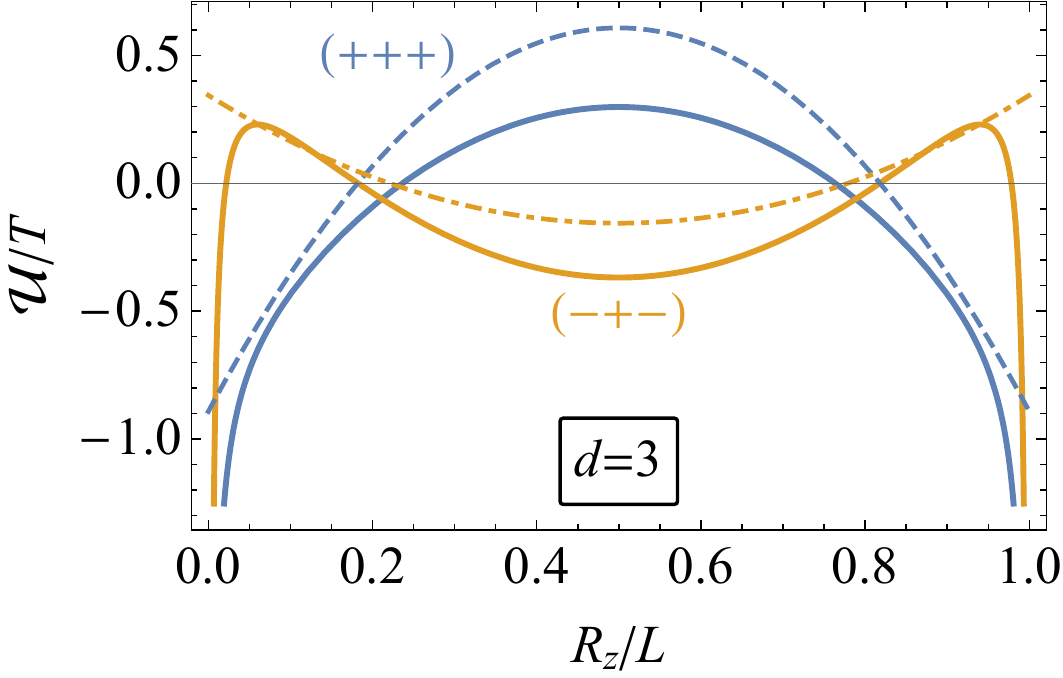}} \quad
    \subfigure[]{\includegraphics[width=0.31\linewidth]{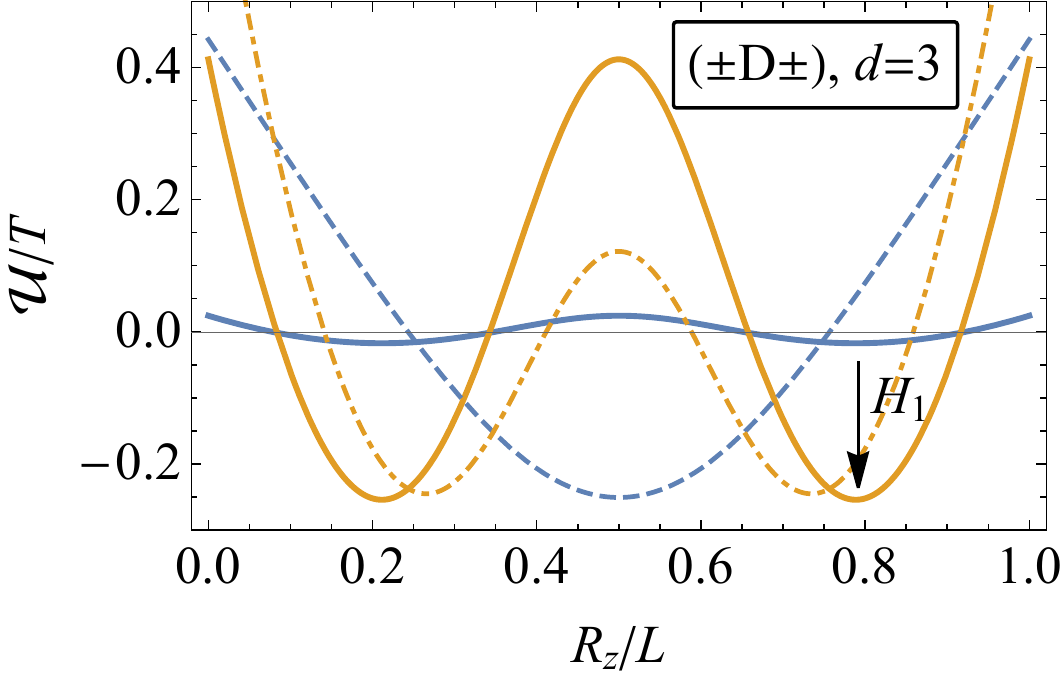}} 
    \caption{Effective potential $\Ucal(R_z) = -\hat\Phi(R_z)$ [see \cref{eq_Boltzmann}, solid lines] of a reactive tracer in a finite interval, where $\hat\Phi$ is given by (a) \cref{eq_pot_3d_h}, (b) \cref{eq_pot_3d_hh1}, and (c) \cref{eq_Pss_c_act_h1_largeC_3d} [with $\hat\Phi = -T \log(L\bar P_s)$]. For comparison, in (a,b) the dashed and dashed-dotted lines represent the potential $\Ucal(R_z) =  - T \log(L\bar P_s)$ in $d=1$ dimensions, as obtained from \cref{eq_Pss_h_act,eq_Pss_h_act_h1_ppp,eq_Pss_h_act_h1_mpm} [the dashed line corresponding to the case (D$\pm$D) and (+++), respectively]. In (c), the two solid (broken) curves represent \cref{eq_pot_3d_hh1} [\cref{eq_Pss_c_act_h1_largeC}] for increasing values of $H_1$ (in the direction of the arrow). The potentials have similar shapes in the three- and one-dimensional case, with noticeable differences being in (b) the additional attractive part of $\Ucal$ for $d=3$ near the boundaries in the case $(-+-)$ [which stems from the first term in \cref{eq_pot_3d_hh1}], and in (c) the fact that the bimodal shape of $\Ucal$ persists in $d=3$ even for small values of $H_1$.}
    \label{fig_Pss_act_3d}
\end{figure}

We assume $h\neq 0$, $h_1=c=0$ and perform the Gaussian integration over the OP field [see \cref{eq_P_margin,eq_gaussint}], which yields the equilibrium distribution of $R_z$:
\beq \bar P_s(R_z)\big|_{c=0} = \frac{1}{\Zcal} \exp\left[ \Phi(R_z) \right],\qquad \Phi(R_z) = \frac{h^2 T_\phi K_{d-1}}{2 T_R^2}  \int_0^\infty\d p\, p^{d-2} \sum_n \frac{|\sigma_n(R_z)|^2}{k_n^2 + p^2} ,
\label{eq_Pss_ddim_base}\eeq 
with
\beq K_d \equiv \Omega_d/(2\pi)^{d},\qquad \Omega_d = \frac{2\pi^{d/2}}{\Gamma(d/2)},
\label{eq_unitball}\eeq 
where $\Omega_d$ is the surface area of the $d$-dimensional unit sphere.
In \cref{eq_Pss_ddim_base}, we have performed the thin film limit $A\to\infty$ by applying the rule
\beq \frac{1}{A}\sum_\pv \to \frac{1}{(2\pi)^{(d-1)}} \int_\lambda^\Lambda \d^{d-1}p,
\label{eq_inf_area_repl}\eeq 
where $\lambda$ and $\Lambda$ are low- and high-wavenumber cutoffs introduced for the purpose of regularization.
Note that the expression for $\Phi$ in \cref{eq_Pss_ddim_base} essentially corresponds to the variance of the OP $\phi(\Rvp,R_z)$ in $d$ dimensions, i.e., $\Phi(R_z) = \frac{h^2}{2 T_R^2} V_\phi(\Rvp=\bv0,R_z)$ [cf. \cref{eq_phiR_var}], rendering \cref{eq_Pss_ddim_base} fully analogous to \cref{eq_Pss_h_act}.
In order to further evaluate \cref{eq_Pss_ddim_base}, we make use of the fact that [see \cref{eq_eigenspec}]
\beq |\sigma_n(R_z)|^2 = 
  \begin{cases} 
    \frac{1}{L}, \qquad & \text{(p)} \\
    \frac{1}{L}\left(1 - \cos(2 \pi n R_z/L) \right),\qquad & \text{(D)} \\
    \frac{1}{L}\left(1 + \cos(2 \pi n R_z/L) \right), & \text{(N, $n\geq  1$)}  \\
    \frac{1}{L}. & \text{(N, $n= 0$)} 
  \end{cases}\label{eq_sigma_trig}
\eeq 
In the case of periodic \bcs, we accordingly obtain a spatially uniform distribution $\bar P_s$. 
For the other \bcs, we get
\beq  \Phi(R_z) =  \frac{h^2 T_\phi K_{d-1}}{2 T_R^2 L} \sum_{n=1}^\infty \int_\lambda^\Lambda \d p \, p^{d-2} \frac{1 \mp \cos(2\pi n R_z /L)}{(\pi n/L)^2 + p^2}, \qquad \begin{matrix}\text{(D)} \\ \text{(N)}\end{matrix}
\label{eq_pot_d_h_1}\eeq 
where the $-$ ($+$) sign applies to Dirichlet (Neumann) \bcs.
In order to calculate \cref{eq_pot_d_h_1} in $d=3$, we first determine the sum over $n$ (see \S 1.445 in \cite{gradshteyn_table_2014}) and subsequently integrate over $p$.
The result can be written in the form $\Phi(R_z) = \hat\Phi(R_z) + \Phi_0$, where $\Phi_0$ is a $R_z$-independent term that diverges $\sim \lambda^{d-3}$ ($\sim \log\lambda$ in $d=3$) and $\sim \Lambda^{d-1}$, while the $R_z$-dependent part $\hat\Phi(R_z)$ is independent of the cutoffs.
Since all $R_z$-independent parts are canceled by the normalization in \cref{eq_Pss_ddim_base}, one may omit them in $\hat\Phi$ and accordingly obtain
\beq \hat \Phi(R_z)  = \pm \frac{\kappa}{8} \left[ \psi(1-R_z/L) + \psi(R_z/L) \right], \qquad \begin{matrix}\text{(D$\pm$D)} \\ \text{(N$\pm$N)}\end{matrix}, \qquad (d=3)
\label{eq_pot_3d_h}\eeq 
where $\psi(z) = \Gamma'(z) / \Gamma(z)$ is the digamma function \cite{olver_nist_2010} and 
\beq \kappa \equiv \frac{h^2 T_\phi K_{2}}{T_R^2 L}
\eeq 
is a dimensionless coupling [cf. \cref{eq_effcoupl_h}].
The normalization factor $\Zcal$ in \cref{eq_Pss_ddim_base} has to be computed numerically.
The function $\hat\Phi$ diverges at the boundaries: 
\beq \hat\Phi(R_z\to 0)\simeq \mp \frac{\kappa}{8} \frac{L}{R_z}
\label{eq_pot_3d_h_asympt}\eeq 
(analogously for $R_z\to L$), rendering $\bar P_s$ non-normalizable for Neumann \bcs.  
This divergence of $\hat\Phi$ is cut off if $\Lambda$ is finite.
In a physical system, van der Waals interactions regularize the singular Casimir potential near the boundary \cite{maciolek_collective_2018}.
Instead of $\bar P_s$, we thus show in \cref{fig_Pss_act_3d} the quantity $\hat \Phi$, which essentially corresponds to the negative Casimir potential defined in \cref{eq_Boltzmann}, $\Ucal(R_z) = - T \Phi(R_z)$ (setting $T=T_R=T_\phi$).

For $h_1\neq 0$ (but still $c=0$), we take Neumann modes for $\sigma_n$ and obtain from \cref{eq_Hamilt_ddim_base}:
\beq\begin{split} 
\Phi(R_z) &= \frac{T_\phi }{2} \sum_{n,\pv} \frac{ \left| A^{-1/2}\tilde h \sigma_n(R_z) e^{\im\Rvp\cdot\pv} + A^{1/2} \tilde h_1 [\sigma_n(0) + \sigma_n(L)] \delta_{\pv,0} \right|^2 }{k_n^2 + \pv^2} \\
&= \frac{T_\phi }{2 } \sum_n \left\{ K_{d-1} \tilde h^2 \int\d p\, p^{d-2}  \frac{\sigma_n^2(R_z)}{k_n^2 + \pv^2} + 2 \tilde h \tilde h_1 \frac{\sigma_n(R_z)[\sigma_n(0) + \sigma_n(L)]}{k_n^2} \right\},
\end{split}\eeq 
with $\tilde h\equiv h/T_R$, $\tilde h_1 \equiv h_1/T_\phi$ and where, in the last step, we performed the continuum limit $A\to\infty$.
As before, we generally omit any constants independent of $R_z$. 
The required expressions have already been determined in \cref{eq_Pss_h_act_h1_ppp,eq_pot_3d_h}, resulting (for $d=3$) in the exponent 
\beq \hat \Phi(R_z) = -\frac{\kappa }{8}  [\psi(1-R_z/L) + \psi(R_z/L)] + L T_\phi\tilde h \tilde h_1 \left[ \frac{1}{6} - \pfrac{R_z}{L} + \pfrac{R_z}{L}^2 \right],  \qquad \text{($\pm$$\pm$$\pm$)}
\label{eq_pot_3d_hh1}\eeq 
which is illustrated in \cref{fig_Pss_act_3d}(a,b).

\subsection{Quadratic coupling}

For $c\neq 0$ and $h=h_1=0$, a Gaussian integration over the field modes using \cref{eq_Hamilt_ddim_base} yields [cf.\ \cref{eq_Pss_c_act_det}]
\beq \bar P_s(R_z)\big|_{h=0} = \frac{1}{\Zcal} \left[\det \boldsymbol{\tilde\Gamma}(R_z) \right]^{-1/2}  .
\label{eq_Pss_c_act_det_3d}\eeq 
Upon expressing $\boldsymbol{\tilde\Gamma}$ in terms of $(\uv)_{n\pv} = \left[\sqrt{c/(A T_R)} e^{\im\pv\cdot\Rvp} \sigma_n(R_z)\right]_{n\pv}$ and $A_{n\pv,m\qv} = \delta_{n,m}\delta_{\pv,\qv} (k_n^2 + \pv^2)/T_\phi$, the determinant can be calculated using the formalism in \cref{sec_matrix}. In the thin film limit, this results in [see \cref{eq_det_lemma}]
\beq \uv^\dagger \Av^{-1}\uv = \frac{c T_\phi}{(2\pi)^{d-1} T_R} \sum_n \int\d^{d-1}p \frac{|\sigma_n(R_z)|^2}{k_n^2 + \pv^2},
\eeq 
which has essentially been calculated after \cref{eq_Pss_ddim_base}.
Accordingly, for periodic \bcs, it is independent of $R_z$, whereas for Neumann or Dirichlet \bcs, the analysis around \cref{eq_pot_d_h_1} implies that it is dominated by an $R_z$-independent contribution which diverges $\propto \Lambda^{d-1}$ in the continuum limit.
Specifically in $d=3$, \cref{eq_Pss_c_act_det_3d} can be expressed as $\bar P_s = \Zcal^{-1} [1+ c \alpha \Lambda + c f(R_z) ]^{-1/2}$, where $\alpha$ is a constant and $f$ is a function describing the $R_z$-dependent part [in fact, $f\propto \hat\Phi$, see \cref{eq_pot_3d_h}].
In the limit $\Lambda\to\infty$, one obtains $\bar P_s \simeq [1+ f(R_z)/(\alpha \Lambda)]/L \to 1/L$. 

Finally, in the case $c\neq 0$ and $h_1\neq 0$ (but $h=0$), one finds (for finite $A$)
\beq \bar P_s(R_z)\big|_{h=0} = \frac{1}{\Zcal} \frac{1}{\left[\det \boldsymbol{\tilde\Gamma}(R_z) \right]^{1/2} }  \exp\left\{ \onehalf A \tilde h_1^2 \sum_{n,m} [\sigma_n(0) + \sigma_n(L)]  \tilde \Gamma(R_z)_{nm,\pv=0=\qv}^{-1} [\sigma_m(0)+\sigma_m(L)] \right\}.
\label{eq_Pss_c_act_h1_3d_fin}\eeq 
As discussed above [see \cref{eq_Pss_c_act_det_3d}], in the continuum limit $\Lambda\to\infty$, the contribution of the determinant is dominated by a divergent term and is thus canceled by the normalization factor $\Zcal$. 
The exponent in \cref{eq_Pss_c_act_h1_3d_fin} coincides (apart from the prefactor $A$) with its one-dimensional counterpart in \cref{eq_Pss_c_act_h1_mat} and, taking Neumann modes for $\sigma_n$ and using \cref{sec_matrix}, we obtain (with $\rho\equiv R_z/L$) 
\beq \bar P_s(R_z)\big|_{h=0} = \frac{1}{\Zcal} \exp\left\{ -\frac{1}{2}H_1^2 \kappa_c  \frac{\left( \frac{1}{6} - \rho + \rho^2 \right)^2}{1 + \kappa_c \left( \frac{1}{3} - \rho + \rho^2 \right)} \right\}. \qquad \text{($\pm c \pm$)}
\label{eq_Pss_c_act_h1_3d}\eeq 
The (dimensionless) effective couplings $H_1\equiv h_1\sqrt{\sfrac{L^d}{T_\phi}}$ and $\kappa_c\equiv c L^{2-d} T_\phi/T_R$ generalize the ones in \cref{eq_effcoupl_h1,eq_effcoupl_c} to $d$ dimensions. 
The expression in \cref{eq_Pss_c_act_h1_3d} differs from \cref{eq_Pss_c_act_h1} by the absence of the $R_z$-dependent prefactor stemming from the determinant.
In the limit $\kappa_c \to\infty$, the probability distribution is given by
\beq \bar P_s(R_z)\big|_{h=0,\kappa_c\to\infty} = \frac{1}{\Zcal} \exp\left\{ -\frac{1}{2}H_1^2  \frac{\left( \frac{1}{6} - \rho + \rho^2 \right)^2}{ \frac{1}{3} - \rho + \rho^2 }  \right\}, \qquad \text{($\pm$D$\pm$)} 
\label{eq_Pss_c_act_h1_largeC_3d}\eeq
which is independent of $\kappa_c$.
The normalization factors $\Zcal$ in \cref{eq_Pss_c_act_h1_3d,eq_Pss_c_act_h1_largeC_3d} have to be determined numerically.
Analogously to the one-dimensional case [see \cref{eq_Pss_c_act_h1_largeC}], the most likely position of the tracer is at a certain distance from the nearest boundary. 
In the limit $H_1\to\infty$, \cref{eq_Pss_c_act_h1_largeC_3d} reduces to two $\delta$-functions:
\beq \bar P_s(R_z)\big|_{h=0,\kappa_c\to\infty} = \frac{1}{2} \left[ \delta\left(R_z- R_- \right) + \delta\left(R_z - R_+\right)\right], \qquad (H_1\to\infty)
\eeq 
with $R_\pm=L \left(\frac{1}{2} \pm \frac{1}{2\sqrt{3}}\right)$, identically to \cref{eq_Pss_c_act_h1_peakpos}.
For Neumann \bcs with a zero mode, \cref{eq_sherman_morrison_inv_zeromode} implies $\bar P_s(R_z)|_{h=0} = 1/L$.



%

\end{document}